\documentclass[aps,pra,nofootinbib,floatfix,superscriptaddress,twocolumn]{revtex4-2}

\newcommand{\doctitle}{Characterization of Overparameterization in Simulation of Realistic Quantum Systems}
\newcommand{\docauthor}{Matthew Duschenes}
\newcommand{\docauthorone}{Juan Carrasquilla}
\newcommand{\docauthortwo}{Raymond Laflamme}
\newcommand{\docaffil}{Department of Physics \& Astronomy, University of Waterloo, Ontario, N2L 3G1, Canada}
\newcommand{\docaffilone}{Institute for Quantum Computing, University of Waterloo, Ontario, N2L 3G1, Canada}
\newcommand{\docaffiltwo}{Vector Institute, Toronto, Ontario, M5G 1M1, Canada}
\newcommand{\docaffilthree}{Perimeter Institute for Theoretical Physics, Waterloo, Ontario, N2L 2Y5, Canada}

\newcommand{\docaffilfour}{Institute for Theoretical Physics, ETH Zürich, 8093, Switzerland}
\newcommand{\docemail}{mduschen@uwaterloo.ca}

\newcommand{\pdftitle}{\doctitle}

\usepackage{main}

\begin{document}

\preprint{APS/123-QED}

\title{\doctitle}

\author{\docauthor}
 \email[Contact author: ]{\docemail}
\affiliation{\docaffil}
\affiliation{\docaffilone}
\affiliation{\docaffiltwo}
\affiliation{\docaffilthree}
\author{\docauthorone}
\affiliation{\docaffilfour}
\affiliation{\docaffil}
\affiliation{\docaffiltwo}
\author{\docauthortwo}
\affiliation{\docaffil}
\affiliation{\docaffilone}
\affiliation{\docaffilthree}

\date{\today}

\begin{abstract}
Quantum computing devices require exceptional control of their experimental parameters to prepare quantum states and simulate other quantum systems. Classical optimization procedures used to find such optimal control parameters, have further been shown in idealized settings to exhibit different regimes of learning. Of interest in this work is the overparameterization regime, where for systems with a sufficient number of parameters, global optima for prepared state and compiled unitary fidelities may potentially be reached exponentially quickly. Here, we study the robustness of overparameterization phenomena in the presence of experimental constraints on the controls, such as bounding or sharing parameters across operators, as well as in the presence of noise inherent to experimental setups. We observe that overparameterization phenomena are resilient in these realistic settings at short times, however fidelities decay to zero past a critical simulation duration due to accumulation of either quantum or classical noise. This critical depth is found to be logarithmic in the scale of noise, and optimal fidelities initially increase exponentially with depth, before decreasing polynomially with depth, and with noise. Our results demonstrate that parameterized ansatze can mitigate entropic effects from their environment, offering tantalizing opportunities for their application and experimental realization in near-term quantum devices.
\end{abstract}

\maketitle

\section{Introduction} \label{sec:introduction}
There exist many active avenues and experimental platforms for the quantum information community to explore and advance quantum technologies, including trapped ions \cite{Bruzewicz2019,Low2020}, superconducting qubits \cite{Kim2023,Werninghaus2021,Wendin2016}, neutral atoms \cite{Bluvstein2023,Bluvstein2021,Wintersperger2023,Levine2018}, nuclear magnetic resonance \cite{Peterson2020a,Rao2013}, and several other intriguing approaches. To harness these technologies' full potential for tasks of interest in quantum information \cite{Preskill2018,Ge2022, Wang2021a}, in particular state preparation or unitary compilation \cite{Sharma2020}, it is imperative to exercise precise control through the manipulation of experimental parameters. This complex, high-dimensional quantum control problem arises in numerous applications \cite{Dalgaard2022,Peterson2020,Feng2018,Magann2021,Pan2014} and is addressed through classical simulation and parameter optimization. Insight into such procedures, in experimentally relevant quantum settings, is therefore necessary to further our ability to rigorously control such systems.

Quantum control shares striking similarities with the field of classical deep learning, where large parameterized models are used to discover and represent complex patterns in large amounts of data. Beyond the resemblance of being variational algorithms concerned with optimizing high dimensional systems for technological advancement, a series of observations \cite{Cerezo2021,Liu2020,Schuld2015} has lead to direct equivalencies in learning phenomena between variational quantum algorithms and deep learning. A striking example is overparameterization and lazy training \cite{Bahri2020,Chizat2018}. In classical systems, excessively parameterized models can learn efficiently with negligible adjustments to their parameters, leading to improved generalization performance and training efficiency. A similar phenomenon has been anticipated in ideal settings for noise-free variational quantum algorithms \cite{Kiani2020,Wiersema2020,Larocca2021}. It is observed that in the overparameterized regime, the optimization landscape becomes almost free of sub-optimal minima and optimization may converge exponentially quickly.

Here we explore overparameterization via the classical simulation of quantum systems, within experimentally feasible settings. We investigate these phenomena within the quantum optimal control paradigm, where systems evolve under continuous time evolution \cite{Dalgaard2022,Magann2021,Peterson2020a}, and within the variational quantum algorithms paradigm, where systems evolve under discretized sets of operations \cite{Cerezo2021,Benedetti2021,Sharma2020}. 

We find that overparameterization phenomena are robust under realistic settings, including constrained parametrizations, and imposing noisy non-unitary ansatz. For a given periodic ansatz, where quantum circuit depth dictates the number of model parameters, we find that inclusion of parameter constraints shifts the overparameterization depth boundary by a system size dependent factor. However, the dominant overparameterization phenomenon, of exponential convergence of optimization routines with depth, persists. In noisy settings, we observe that there are different regimes of optimization convergence. For depths beyond the overparameterization depth, but before a noise-induced critical depth, exponential convergence of optimization routines with depth, still occurs. However, beyond this critical depth, an excessive amount of noise accumulates, and the optimization diverges polynomially with depth, and with noise. To complement these numerical findings, analytical investigations into the noise and depth scaling of the infidelity objectives, and other metrics including the entropy and purity of the parameterized states, provide an explanation for these behaviors. Overall, these findings suggest overparameterization is resilient when imposing experimental feasibility, offering opportunities for this phenomenon to be exploited in future simulated and existing quantum experiments.

The work is structured as follows. In \cref{sec:background}, we define general parameterized quantum channels, and objective tasks of interest, namely noisy infidelities for state preparation and unitary compilation. In \cref{sec:methods}, we interpret the form of noisy parameterized channels as expectation values of $k$-error channels, and we perform an analysis on the scaling of infidelity objectives with respect to noise and depth. In \cref{sec:results}, we demonstrate overparameterization and other learning phenomena in constrained and noisy parameterized quantum systems. From numerical studies, we quantify the relationships between noise and depth at optimality. Finally in \cref{sec:discussion} and \cref{sec:conclusion}, we discuss the implications of these results and the resulting compromises that occur between numerical and experimental feasibility.

\section{Background}\label{sec:background}
This work aims to understand the abilities of parameterized quantum systems in realistic, experimentally feasible settings. Critically, noise, resulting from systems interacting with their environment, is well known to be detrimental to quantum computation \cite{Ravi2022,Sharma2020,Botelho2022}. Example effects include noise-induced symmetry breaking \cite{Fontana2020}, fundamental differences in sampling and annealing trajectories \cite{Franca2020}, and noise-induced barren plateaus \cite{Wang2020}. Initial numerical investigations by Fontana \etal \cite{Fontana2021} demonstrate that noise, represented by noise scales, or probabilities $\gamma$, accumulates with depth $M$. There are also well known compromises between expressiveness, i.e., how much of the desired space of solutions can be represented by an ansatz via increased depth \cite{Holmes2021}, and trainability, i.e., the ability of an ansatz to be optimized. However, these precise relationships in noisy settings have yet to be confirmed quantitatively. 

The parameterized systems of interest simulated in this work are representative of various noisy intermediate scale quantum devices (NISQ) \cite{Kim2023,Levine2018,Wendin2016}. These systems consist of $N$ qubits, each with local space dimension $D=2$, and total space dimension $d = D^{N}$. These qubits may have fixed inter-qubit couplings, however they can be manipulated with external, time-dependent fields over a time $T$, or equivalently depth of simulation $M$. Please refer to \cref{app:background} for a complete description of the parameterized ansatze studied in this work. Such experimental parameters are generally constrained due to feasibility \cite{Pan2014,Ravi2022}, and this work seeks to quantify the amount with which constraints affect the capabilities of parameterized quantum systems. We assume there are generally on the order of $P = O(\textrm{poly}(N)M)$ variable parameters in the system, where generally the system size $N$ dependence is held fixed. Thus changes in parameter counts are reflected in the simulation depth $M$, and any notion of overparameterization is discussed in the context of depth. We take as an example in this work, nuclear magnetic resonance systems where nuclei, acting as $D=2$ level qubits, are manipulated by time-dependent magnetic pulses \cite{Laflamme2002,Cory2000,Rao2013,Pan2014}.

Underlying this analysis, are principles from learning theory, based on studies of overparameterization in ideal quantum settings, including unitary compilation \cite{Kiani2020}, variational quantum eigensolvers \cite{Wiersema2020,You2022}, and general quantum circuits \cite{Larocca2021}. These works have subsequently been followed up by initial theoretical studies on the effects of noise. Within an information theoretic context, the quantum Fisher information \cite{Garcia-Martin2023} is used as a metric to determine whether a quantum system is overparameterized. Within a general optimization context to complement neural-tangent kernel approaches for asymptotic learning dynamics \cite{Liu2020}, Riemannian gradient flow dynamics \cite{You2022} are used to assert the convergence of overparameterized systems with bounded gradient noise. For our purposes, overparameterization refers to when there are an adequate number of parameters $P > \tilde{P}$, or depths $M > \tilde{M}$, such that the full space of $G$-dimensional solutions to tasks is spanned by the parameterized ansatz. Within this regime, we investigate the resulting overparameterization phenomena of possible exponential convergence of noisy optimization procedures with depth. Imposing constraints on the optimization related to experimental feasibility, including restricting individual control of qubits, is also known to decrease convexity in the objective landscape \cite{Moore2012,Song2022,Ge2022}, and it requires many optimization heuristics. Please refer to \cref{app:learning_phenomena} for a more complete description of overparameterization, including quantitative bounds on the overparameterization depth related to the quantum Fisher information.

Based on these studies, we hypothesize that there exists a critical evolution time or depth $M_{\gamma} > \tilde{M}$, where overparameterization has occurred, however too much noise has also accumulated. This noise is expected to prevent parameterized systems from accomplishing fidelity-based tasks with arbitrary precision. To confirm these predictions, we will consider the average behavior of infidelities, optimized independently over a distribution of tasks. We therefore conjecture that there are convergent and divergent phases of the optimization
\begin{align}
	\textrm{Optimization} \sim \left\{ \begin{array}{cc} \textrm{Plateau} & M < \tilde{M} \\
	\textrm{Convergent} & \tilde{M} < M < M_{\gamma} \\
	\textrm{Divergent} & M > M_{\gamma} \end{array} \right.~.
\end{align}
This work aims to confirm these conjectures of depth-dependent regimes of learning phenomenon in non-ideal settings.

To quantify the effects of noise on the evolution and abilities of parameterized quantum systems to perform tasks of interest, we define parameterized quantum channels as
\begin{align}
	\Lambda_{\theta\gamma} = &~ \mathcal{N}_{\gamma} \circ \mathcal{U}_{\theta}~,
\end{align}
with a unitary channel $\mathcal{U}_{\theta}$ parameterized by variable parameters $\theta$, and a non-unitary channel $\mathcal{N}_{\gamma}$ parameterized by constant noise scales $\gamma$.

For the unitary channel, we assume the Hamiltonian driving the evolution
\begin{align}
	H_{\theta}^{(t)} = \sum_{\mu} H_{\mu}^{(t)} ~:~ H_{\mu}^{(t)} =&~ \theta_{\mu}^{(t)} G_{\mu}
\end{align}
at a continuous time $t \in [0,T]$. The Hamiltonian is defined by a set of generators $\{G_{\mu}\}$, which are generally assumed to be acting on at most $k$ qubits. Assuming the evolution is approximately piecewise constant over $M$ time steps $\tau = T/M$, allows for first-order temporal Trotterization of the resulting unitary operator
\begin{align}
	U_{\theta} =&~ \Tau e^{-i \int_{0}^{T} dt ~H_{\theta}^{(t)}} = \prod_{m}^{M} U_{\theta}^{(m)} ~+~ O(\tau^2)~.\\
	\intertext{$Q$ order spatial Trotterization of the operator across the $N$ qubits at time index $m$ is also possible. Here Trotterization is represented as a product of a function of lower-order Trotterizations, denoted by the $Q$ superscript,}
	U_{\theta}^{(m)} =&~ e^{-i\tau H_{\theta}^{(m)}} = \prod_{\mu}^{(Q)}U_{\mu}^{(m,Q)} ~+~ O(\tau^{Q+1})~.
\end{align}
The final first-order temporally localized unitary channel, with $Q$-order spatial Trotterization is
\begin{align}
	\mathcal{U}_{\theta} =&~ \circ_{m}^{M} \mathcal{U}_{\theta}^{(m,Q)} ~:~ \mathcal{U}_{\theta}^{(m,Q)} = \circ_{\mu} \mathcal{U}_{\mu}^{(m,Q)} ~,
\end{align}
with resulting gate operators related to the Hamiltonian generators
\begin{align}
	U_{\mu}^{(m,Q)} =&~ e^{-i\tau H_{\mu}^{(m,Q)}}~.
\end{align}
Please refer to \cref{app:background} for a complete description of these schemes.

\begin{figure}[h]
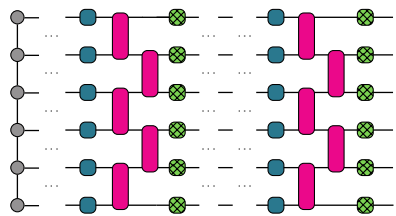

	\centering
	\tikzfig{0.5}{circuit}
	\captionsetup{justification=raggedright}	
	\caption{Parameterized quantum channel with layers of trotterized local (blue squares) and two-body (pink rectangles) unitary operators, followed by local noise channels (hatched green squares) after each layer, with an initial state (left-most gray circles).}
	\label{fig:circuit}
\end{figure}

For the non-unitary channel, we consider temporally and spatially local, independent noise acting on the $K = NM$ sites $(m,i)$ of possible errors
\begin{align}
	\mathcal{N}_{\gamma} =&~ \circ_{m}^{M} \left(\circ_{i}^{N}\mathcal{N}_{\gamma_{i}}^{(m)}\right)~.
\end{align}
For our purposes, we decompose each local noise channel into a convex combination of an identity component, and what we refer to as a non-identity error component,
\begin{align}
	\mathcal{N}_{\gamma_{i}}^{(m)} \equiv (1-\gamma)\mathcal{I}_{i} ~+~ \gamma\mathcal{K}_{\gamma_{i}}^{(m)}~.
\end{align}
The forms of the non-trivial $\mathcal{K}_{\gamma}$ depend on the specific noise model of interest. Noise models considered in this work include local dephasing, amplitude damping, and depolarization noise. Such local noise models are known to be relevant in several quantum computing implementations \cite{Georgopoulos2021,Bharti2022}. Our analytical and numerical approaches are easily transferable to spatially correlated noise models across multiple qubits, however such studies concerning any non-trivial effect of correlated noise are left for future work. Please refer to \cref{app:behavior_of_multiple_layer_noise_channels} for a complete description of each noise channel, and to \cref{app:classical_and_quantum_error_analysis} for an analytical treatment of noise-induced effects. Given the temporal and spatial locality of the Trotterized unitary and non-unitary channel, we finally reach the explicit ansatz form of interlaced noise and unitary evolution, 
\begin{align}
	\Lambda_{\theta\gamma} =&~ \circ_{m}^{M}\left(\mathcal{N}_{\gamma}^{(m)} \circ \mathcal{U}_{\theta}^{(m)}\right)~,
\end{align}
leading to our overall channel circuit diagram in \cref{fig:circuit}.

In this work, we take as our parameterized unitary ansatz, evolution generated by the nuclear magnetic resonance (NMR) Hamiltonian consisting of $N$, $D=2$ qubit Pauli operators
\begin{align}
		H_{\theta}^{(t)} =&~ \sum_{i}{\theta_{i}^{x}}^{(t)}X_{i} ~+~ \sum_{i}{\theta_{i}^{y}}^{(t)} Y_{i} ~+~ \\
		&~\sum_{i}h_{i}Z_{i} ~+~ \sum_{i<j} J_{ij} Z_{i}Z_{j}~. \nonumber
\end{align}
Generally in such systems, we have control over the variable time-dependent local transverse $X$ and $Y$ fields at qubit $i$, with additional constant time-independent longitudinal local $Z$ at qubit $i$ and non-local $ZZ$ fields at qubits $i \neq j$. Such Hamiltonians allow for universal control over qubits, with the local and non-local gates allowing entangling gates to be implemented. We use experimentally relevant scales for our ansatze in \cref{tab:constants}, and details of the ansatz are discussed in \cref{app:nuclear_magnetic_resonance_ansatz}.

To compare our choice of NMR ansatz to other implementations, we collect rough estimates \cite{Linke2017} from recent literature of the $1$-qubit $T_{U1}$ and $2$-qubit $T_{U2}$ gate times, and decoherence times $T_{\gamma}$, for NMR \cite{Peterson2020a,Peterson2020,Majidy_Wilson_Laflamme_2024}, trapped ion \cite{Bruzewicz2019,Low2020,Majidy_Wilson_Laflamme_2024}, superconducting qubit \cite{Kim2023,Werninghaus2021,Majidy_Wilson_Laflamme_2024}, and neutral atom \cite{Wintersperger2023,Bluvstein2021,Bluvstein2023} quantum computing experiments in \cref{tab:implementations} of \cref{app:nuclear_magnetic_resonance_ansatz}. NMR systems are limited by experimental feasibility. Their small non-local coupling constants $J$ limit how much qubits can be correlated at each time step $T_{U1} = O(\tau)$, increasing significantly the necessary $2$-qubit gate times $T_{U2} = O(1/J)$. This  translates into NMR systems having the largest effective depth $T_{U2}/T_{U1} \sim O(100)$ required for each $2$-qubit gate, and having the smallest effective maximum depth $T_{\gamma}/T_{U2} \sim O(10^{2})$ before coherence, amongst considered implementations. We therefore note that any conclusions drawn from this work regarding the explicit scale of noise or depth of models where phenomena occurs, are specific to this NMR ansatz. However, we believe other similarly universal ansatz should exhibit comparable behavior, at their specific ansatz-dependent scales.
\begin{table}[h]
\captionsetup{justification=raggedright}	
\caption{Experimentally relevant constants for constrained NMR ansatz.}
\label{tab:constants}
\begin{tabular}{c|l}
	$N$      & Number of qubits $\sim 1-4$           \\
	$M$      & Number of time steps $\sim O(10^0-10^4)$  \\
	$\tau$ & Trotterization time step $\sim O(75-100~\si{\micro\s})$          \\
	$T$   & Evolution time $ = M\tau \sim O(375~\si{\micro\s}-500~\si{\milli\s})$  \\
	$Q$ & Spatial Trotterization order $ = 2$ \\	
	$P$ & Number of parameters $\sim O(\textrm{poly}(N)M)$ \\
	$J$      & Constant longitudinal coupling $\sim O(\pi/2 \times 10^{2}~\si{\hertz})$ \\
	$h$      & Constant longitudinal field $\sim O(\pi/2 ~\si{\kilo\hertz})$   \\
	$\theta$ & Variable transverse field $\sim O(\pi/2 ~\si{\mega\hertz})$ \\
	$\gamma$ & Noise scale $\sim O(10^{-14} - 10^{-1})$    
\end{tabular}
\end{table}\\
The choice of a universal ansatz spanning the full space of $G = O(d^{2})$ unitaries, also simplifies transferability to other universal ansatze, and avoids any bias by restricting evolution to being within a subspace. Techniques used other implementations \cite{Kim2023} that also generally have origins in NMR techniques and their proof of principle quantum algorithm experiments \cite{Cory2000,Feng2018}, including zero-noise extrapolation \cite{Bharti2022}, dynamical decoupling \cite{Lidar2014}, and refocusing procedures \cite{Cory2000}. These approaches are highly relevant to interpretations of the phenomena observed.

Given this parameterized ansatz, we wish to assess its ability to represent targets of interest, such as operator compilation, where sequences of operators are optimized to approach a target operator, and state preparation, where initial states transformed by sequences of operators, are optimized to approach a target state \cite{Ashhab2022}. Such tasks arise often in quantum algorithms, and they depend crucially on targets being within span of the ansatz.

Here we focus on unitary compilation and pure state preparation, given a parameterized ansatz with universal control over the full space of unitary operators. The parameters $\theta$ of the ansatze are optimized via optimization routines, and given an initial pure state $\sigma$, and target unitaries $U$ or target pure states $\rho$, we assess the abilities of the respective parameterized unitaries and states,
\begin{align}
	U_{\theta\gamma} \approx &~ U \\ 
	\rho_{\theta\gamma} = \Lambda_{\theta\gamma}(\sigma) \approx&~ \rho = \mathcal{U}(\sigma)~.
\end{align}
Objective metrics of infidelities with respect to the given unitary compilation and state preparation tasks \cite{Glaser1998} are chosen to quantify these ansatze through optimization,
\begin{align}
	\mathcal{L}_{\theta\gamma}^{U} =&~ 1 - (1/d^{2})\abs{\trace{U^{\dagger}U_{\theta\gamma}}}^2 \\
	\mathcal{L}_{\theta\gamma}^{\rho} =&~ 1 - \trace{\rho\rho_{\theta\gamma}}~.
\end{align}
We also define the impurity, von-Neumann entropy, and relative entropy divergence, relative to a pure state $\rho$ as
\begin{align}
	\mathcal{I}_{\theta\gamma} =&~1- \trace{\rho_{\theta\gamma}^{2}} \\
	\mathcal{S}_{\theta\gamma} =&~- \trace{\rho_{\theta\gamma}\log{\rho_{\theta\gamma}}}/\log{d} \\
	\mathcal{D}_{\theta\gamma}^{\rho} =&~- \trace{\rho\log{\rho_{\theta\gamma}}}/\log{d}~,
\end{align}
for later interpretations of noise-induced phenomena.

In general, the optimization of objectives $\mathcal{L}_{\theta\gamma} \to \mathcal{L}_{\theta\gamma}^{*}$, particularly in noisy settings to determine the optimal parameterization $\theta^{*}_{\gamma}$, has no closed forms \cite{Nocedal2006}. Whether there are similarities between optimal noisy and noiseless quantities, such as infidelities $\mathcal{L}_{\theta\gamma}^{*} \approx \mathcal{L}_{\theta}^{*}$, or even more strongly, between parameters $\theta^{*}_{\gamma} \approx \theta^{*}$, remains an open question \cite{Fontana2020}. In our subsequent analysis, in addition to numerical studies, we derive analytically the leading-order scalings of the discussed quantities of interest, given our variables of the depth $M$ and the noise scale $\gamma$.

We also note that all plotted statistics in this work reflect average behaviors of the ansatz across independent optimizations with respect to Haar random initial pure states, and Haar random targets. The minimum infidelity reached for each independent optimization for a given fixed depth and noise scale, is used for statistics across samples. Error bars and shaded regions represent one standard deviation from the mean of samples. Lower error bars are plotted equal in length to the upper error bar on a log scale, and non-visible error bars can be considered to be equal in scale to any plot markers. Error bars in this work generally appear to be relatively orders of magnitude smaller than their corresponding average values, and $S = 50 \leq O(100)$ samples are deemed adequate to capture all behaviors well in practice for $N \leq 4$ qubits. We note however that optimization hyperparameters, as per \cref{app:classical_simulation_and_optimization}, in particular the use of a modified conjugate-gradient-based optimizer with a Wolfe condition line search and appropriate learning rates, must be carefully selected. Our initial studies indicated that at larger depths, and larger small noise scales, after an initially smooth convergence in infidelity, optimization routines can oscillate rapidly between local minima, as also observed in previous optimization studies \cite{Wiersema2020}.

A point concerning notation used in this work: quantities computed within a noisy context generally have $\gamma$ subscripts, such as parameters obtained in a noisy setting $\theta_{\gamma}$. Otherwise in noiseless settings, any noise subscripts are dropped. The Trotterization order $Q$ superscripts are also generally dropped for simplicity as $Q=2$ is held fixed. 

\section{Methods}\label{sec:methods}
We now develop formalisms to understand the scaling of channel-dependent quantities with respect to depth and noise scales. We first develop methods for comparing constrained versus unconstrained noiseless ansatze, as detailed in \cref{app:learning_phenomena}. We follow the formalism developed by Larocca \etal \cite{Larocca2021}, which sets bounds on the rank $R = R(P) \leq P$ of the quantum Fisher information $\mathcal{F}_{\theta_{\mu\nu}}$ to determine whether a quantum system with $P$ parameters is overparameterized. At an overparameterization limit $\tilde{P} = O(G)$, defined in terms of the dimension of the space spanned by the ansatze, this metric's rank is shown to transition from being full rank $R=P$ for underparameterized $P < \tilde{P}$, to saturating at this limit $R=\tilde{P}$ for adequately or overparameterized $P \geq \tilde{P}$.

Underlying these definitions of overparameterization is the quantum Fisher information's rank generally reflecting how many directions a parameterized ansatze may span in its space of solutions. We seek to generalize these intuitions from state-dependent ansatze to unitary-dependent ansatze, independent of the initial state being transformed. The conventional state Fisher information $\mathcal{F}_{\theta}^{\rho}$ is defined in terms of states $\rho_{\theta}$, given a state preparation objective $\mathcal{L}_{\delta}^{\rho} \sim \trace{\rho_{\theta}\rho_{\theta+\delta}}$. We define a generalized unitary Fisher information $\mathcal{F}_{\theta}^{U}$, developed concurrently by Haug \etal \cite{Haug2023}, that is strictly dependent on the unitary ansatz $U_{\theta}$. Given our unitary compilation objective $\mathcal{L}_{\delta}^{U} \sim \trace{U^{\dagger}_{\theta}U_{\theta+\delta}}$, we may derive the unitary Fisher information as the leading-order deviation, in the perturbing parameters $\delta \to 0$, of the objective
\begin{align}
	\mathcal{F}^{U}_{\theta_{\mu\nu}} =&~ \frac{1}{d^{2}} \textrm{Re}\left(d~\trace{\partial_{\mu} U_{\theta}^{\dagger}\partial_{\nu} U_{\theta}} - \right. \\
	&~~~~~~~~~~~~ \left.\trace{\partial_{\mu} U_{\theta}^{\dagger}U_{\theta}}\trace{U_{\theta}^{\dagger}\partial_{\nu} U_{\theta}}\right)~, \nonumber
\end{align}
which reduces to the state Fisher information definition in the limit of tracing over $d=1$-dimensional states
\begin{align}
	\mathcal{F}^{\rho}_{\theta_{\mu\nu}} =&~ \real{\bra{\partial_{\mu}\rho_{\theta}}\ket{\partial_{\nu}\rho_{\theta}} - \bra{\rho_{\theta}}\ket{\partial_{\mu}\rho_{\theta}}\bra{\partial_{\nu}\rho_{\theta}}\ket{\rho_{\theta}}}~.
\end{align}
The rank of the unitary quantum Fisher information subsequently offers insight into the capabilities of an ansatz to span a set of unitaries, and to potentially become overparameterized for compilation tasks.

We also may derive expressions for the resulting noise-dependent quantities that allow for an easier interpretation for noise-induced phenomena. Crucially, we differentiate states by their number of errors due to local noise,
\begin{align}
	\hspace{-0.1cm}
	\mathcal{N}_{\gamma} =&~ \sum_{k}^{K} \hspace{-0.2cm}\sum_{\substack{\chi_{k} \in [2]^{K}\\\abs{\chi_{k}=k}}} \hspace{-0.2cm} \circ_{m}^{M}\otimes_{i}^{N}\gamma^{{\chi_{k}}_{i}^{m}}(1-\gamma)^{1-{\chi_{k}}_{i}^{m}}\mathcal{K}_{\gamma_{i}}^{(m){\chi_{k}}_{i}^{m}}.
\end{align}
Here, we refer to an error as any single non-identity noise operation $\mathcal{K}_{\gamma}$ acting locally at any of $K=NM$ possible qubits and time indices, interlaced within a unitary ansatz $\mathcal{U}_{\theta}$. We then may define $k$-error channels $\Lambda_{\theta\gamma_{k}}$, that are convex combinations of channels consisting of all possible locations of $k \leq K$ errors, represented by multi-indices ${\chi_{k}} \in [2]^{K} ~:~ \abs{\chi_{k}} = k$, each with probability $\gamma^{k}$. The channels that generate at most $K$-error noisy states, can therefore be seen to have an interesting form of an expectation value over a distribution of $k$-error channels,
\begin{align}
	\Lambda_{\theta\gamma} =&~ \expval{\Lambda_{\theta\gamma_{k}}}_{k \sim p_{K\gamma}}~.
\end{align}
For noise models defined in terms of strictly identity or non-trivial errors, the exact distribution over the errors
\begin{align}
	p_{K\gamma}(k) =&~ \tbinom{K}{k}\gamma^{k}(1-\gamma)^{K-k}	
\end{align}
is the binomial distribution with mean $K \gamma$. As discussed in \cref{app:behavior_of_multiple_layer_noise_channels}, other interpretations for noise can arise due to the binomial distribution being equivalent to other distributions in various limits of $\gamma$ and $K$. Grouping the $K,\gamma$-dependent terms yields the leading-order scaling
\begin{align}
	\Lambda_{\theta\gamma} - \Lambda_{\theta} =&~ \sum_{k>0}^{K} \tbinom{K}{k}\gamma^{k} \Lambda_{\theta\gamma_{\leq k}}  \\
	\intertext{We denote $k$-error channels as the uniform convex combination of all possible error locations}
	\Lambda_{\theta\gamma_{k}} =&~ \frac{1}{\tbinom{K}{k}}\sum_{{\chi_{k}}} \Lambda_{\theta\gamma}^{{\chi_{k}}}~, \\
	\intertext{and we denote at-most-$k$-error operators as the non-unfiorm combination of all possible $l \leq k$ error locations}
	\Lambda_{\theta\gamma_{\leq k}} =&~ \sum_{l}^{k} (-1)^{k-l}\tbinom{k}{l} \Lambda_{\theta\gamma_{l}}~,
	\intertext{where specific $k$-error channels with errors at locations $\chi_{k}$ are denoted as}
	\Lambda_{\theta\gamma}^{{\chi_{k}}} =&~ \circ_{m}^{M} \left[\left[\circ_{i}^{N} {\mathcal{K}_{\gamma_{i}}^{(m)}}^{{\chi_{k}}_{i}^{m}}\right] \circ \mathcal{U}_{\theta}^{(m)}\right]~.
\end{align}
For example, the $k=1$-order deviation from the noiseless channel 
\begin{align}
	\Lambda_{\theta\gamma_{\leq 1}} =&~ \frac{1}{K}\sum_{m,i}^{M,N}\mathcal{U}_{\theta}^{(>m)} \circ ({\mathcal{K}_{\gamma_{i}}^{(m)}} - \mathcal{I}_{i})
	\circ \mathcal{U}_{\theta}^{(\leq m)}
\end{align}
depends on the non-triviality of the $K$ possible single $\mathcal{K}_{i}^{(m)} \neq \mathcal{I}_{i}$ local channels. Generalizing this interpretation, the $k$-order deviation depends on how the $\binom{k}{l}$ possible $l \leq k$-error channels combine, weighted by their binomial coefficients, to cancel or deviate from the trivial noiseless channel. Channels consisting of multiple $q$ types of errors at indices $\chi_{k} \in [q]^{K}$, may also be written as an expectation over a generalized multinomial distribution of channels.

As derived in \cref{app:behavior_of_multiple_layer_noise_channels}, gradients along parameter directions $\mu$ of objectives, which are linear in the parameterized channels with constant noise, follow parameter shift rules \cite{Wierichs2021}. We denote such gradients with perturbing parameter angles $\varphi$ and coefficients $\alpha_{\varphi}$, which are dependent on the spectrum of the ansatz generators
\begin{align}
	\partial_{\mu} \Lambda_{\theta\gamma} =&~ \sum_{\varphi}\alpha_{\varphi}^{\mu}\Lambda_{\theta+\varphi~\gamma}~.
\end{align}
These parameter shift rules indicate that the linear nature of the noise interlaced with the parameterized unitary ansatz perturbatively affects the noisy quantities.

Whether classical sources of noise impose similar noise-induced phenomena to the quantum noise sources studied, is a separate important question when determining the robustness of parameterized systems. As developed in \cref{app:classical_and_quantum_error_analysis}, we introduce the notion of classical noise due to numerical precision, or floating point error scale $\epsilon$, and we determine its similarity to the quantum noise scale $\gamma$. This error affects both the representation and numerical operations of floating point scalars, which we extend to matrices $A \in \mathcal{M}(d)$ and matrix multiplication
\begin{align}
	\prod_{\mu}^{k}A_{\mu} \to \sideset{}{_{\epsilon}}\prod_{\mu}^{k}A_{\mu} =&~ \sum_{\chi \in [2]^{k}}\prod_{\mu}^{k}A_{\mu} \Sigma^{\chi_{\mu}}~,
\end{align}
for error matrices $\Sigma \in \mathcal{M}(d)$, interlaced at locations $\chi$.

We then relate classical $\epsilon = \epsilon(d) = \norm{\Sigma}$ and quantum $\gamma = O(\norm{\mathcal{K}})$ noise phenomena, given the number of $k = O(K) = O(\textrm{poly}({N})M)$ matrix multiplication operations that are subject to classical error, are related to the number of $K$ possible quantum errors in the ansatz. We are thus able to derive exact bounds on the deviations of noisy infidelities from their noiseless counterparts
\begin{align}
	\abs{\mathcal{L}_{\theta\epsilon} - \mathcal{L}_{\theta}} \leq&~ \hspace{5pt}\abs{1-(1+\epsilon)^{2k}} \sim O(2k\epsilon) \\
	\abs{\mathcal{L}_{\theta\gamma} - \mathcal{L}_{\theta}} \leq&~ 2\abs{1 - (1-\gamma)^{K}} \sim O(2K\gamma)~,
\end{align}
with the use of various properties of norms such as Holder and von-Neumann's inequalities for Schatten matrix norms \cite{Mirsky1975}. Both classical and quantum noisy infidelities scale similarly: polynomially in the noise scale, exponentially in the number of errors, and identically in the limit of small local error scales. However, the constant classical error factors depend on the dimension of the space of interest, and the number of parameters, indicating that larger classically simulated systems experience potentially greater error.

This scaling of infidelities analysis is ansatz-independent and provides valuable insight into general deviations of noisy quantities from their noiseless counterparts. However tighter bounds are potentially possible if there are symmetries or relationships between the noise and dynamics, or if the optimization is analytically tractable, and if an exact form for $\mathcal{L}_{\theta^{*}_{\gamma}\gamma}$ may be derived. These upper bounds, as will be discussed in \cref{sec:results}, also relate to the critical point at which noise effects begin to dominate for greater numbers of operations, and optimization enters a divergent regime.

\section{Results}\label{sec:results}
Classical simulations and analytical calculations are performed to quantify changes in the behavior of parameterized systems and their optimization, with depth and with noise scale. NMR ansatz and simulation details are described in \cref{app:nuclear_magnetic_resonance_ansatz} and \cref{app:classical_simulation_and_optimization}, and all data can be found in a repository \cite{duschenes_2024_10295792}. 

\subsection{Unconstrained versus Constrained Optimization}\label{subsec:unconstrained_versus_constrained_optimization}
We first investigate the effects of constraints on noiseless Haar random unitary compilation, with respect to the number of optimization iterations and various depths $M$. The controllable transverse fields $\theta_{i}^{x,y (m)}$ at each time step $m \in [M]$ and qubit $i \in [N]$, are constrained due to it being difficult to exercise individual control over qubits, $\theta_{i}^{x,y (m)} = \theta_{}^{x,y (m)}$, pulse amplitudes are bounded $\abs{\theta_{i}^{x,y (m)}} \leq \bar{\theta}$, and are generally turned off at the start and end of experiments, $\theta_{i}^{x,y (0,M-1)}=0$.

\begin{figure}
	\centering
	\begin{subfigure}[t]{\columnwidth}
		\captionsetup{singlelinecheck = false, justification=raggedright,margin={0pt,0pt},skip=-10pt}
		\subcaption{}
		\includegraphics[width=0.8\columnwidth]{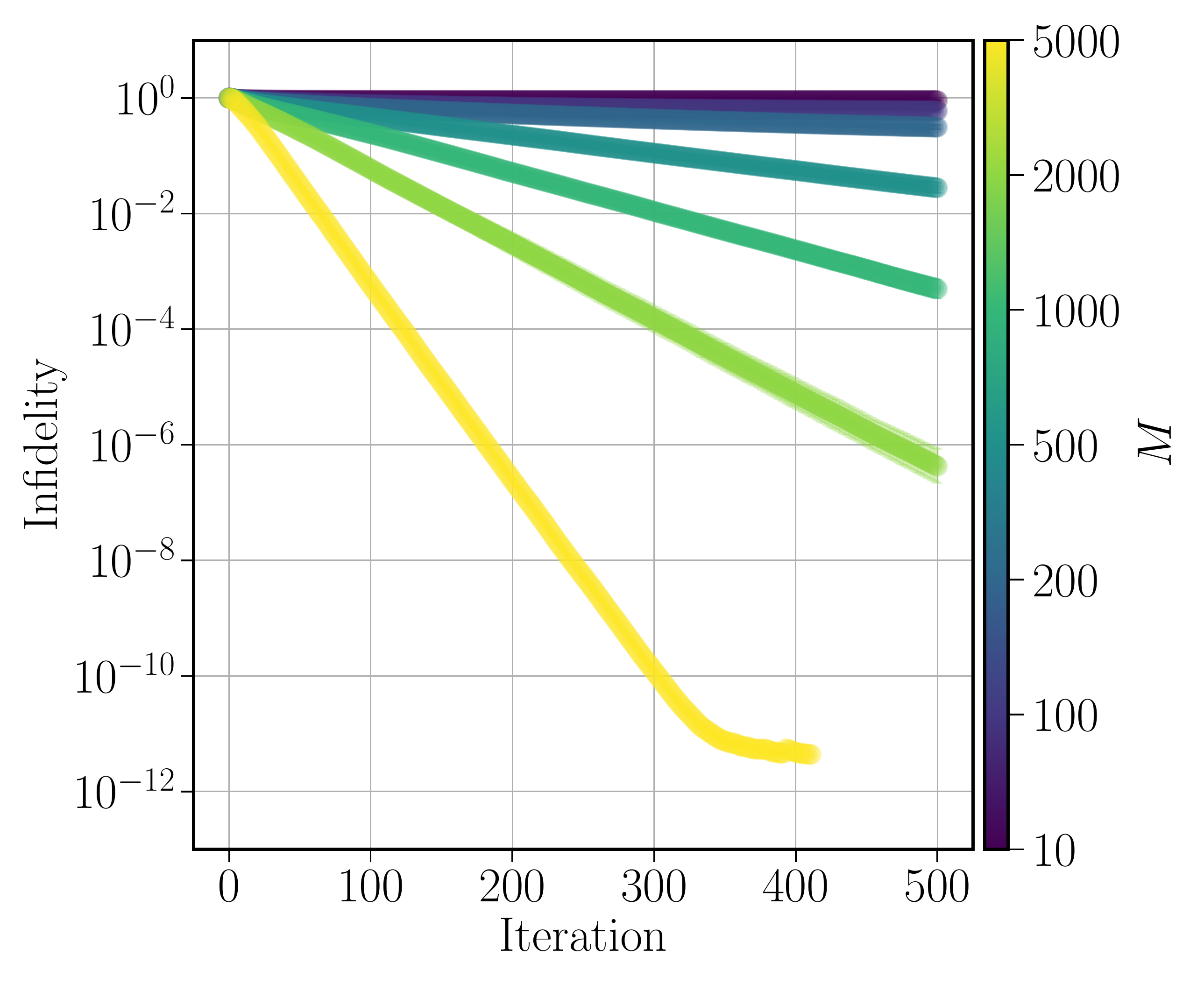}
		\label{fig:infidelity_unconstrained}
	\end{subfigure}
	\hfill
	\begin{subfigure}[t]{\columnwidth}
		\captionsetup{singlelinecheck = false, justification=raggedright,margin={0pt,0pt},skip=-10pt}
		\subcaption{}
		\includegraphics[width=0.8\columnwidth]{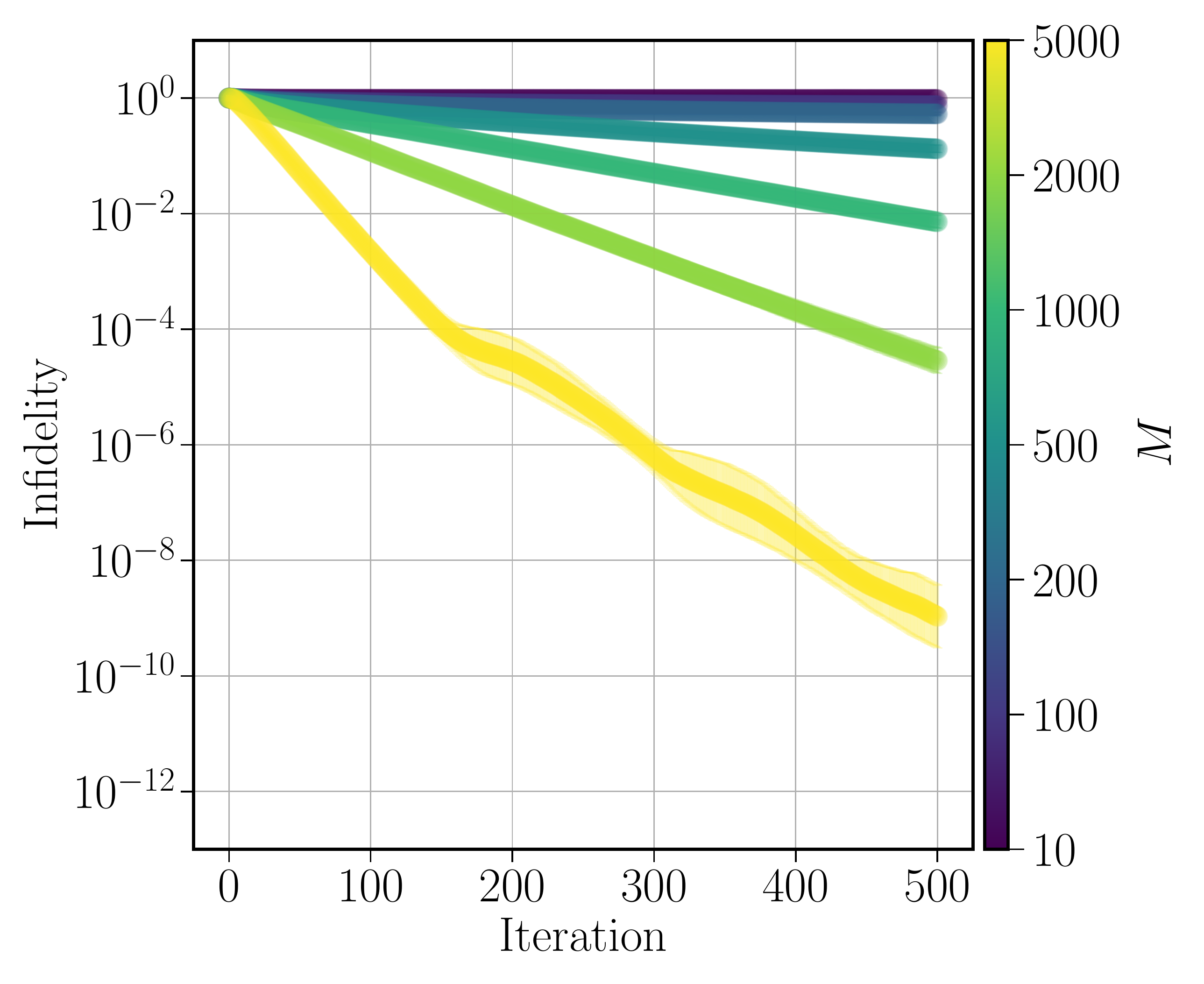}
		\label{fig:infidelity_constrained}
	\end{subfigure}
	\hfill
	\begin{subfigure}[t]{\columnwidth}
		\captionsetup{singlelinecheck = false, justification=raggedright,margin={0pt,0pt},skip=-10pt}
		\subcaption{}
		\hspace{-0.85cm}	
		\includegraphics[width=0.75\columnwidth]{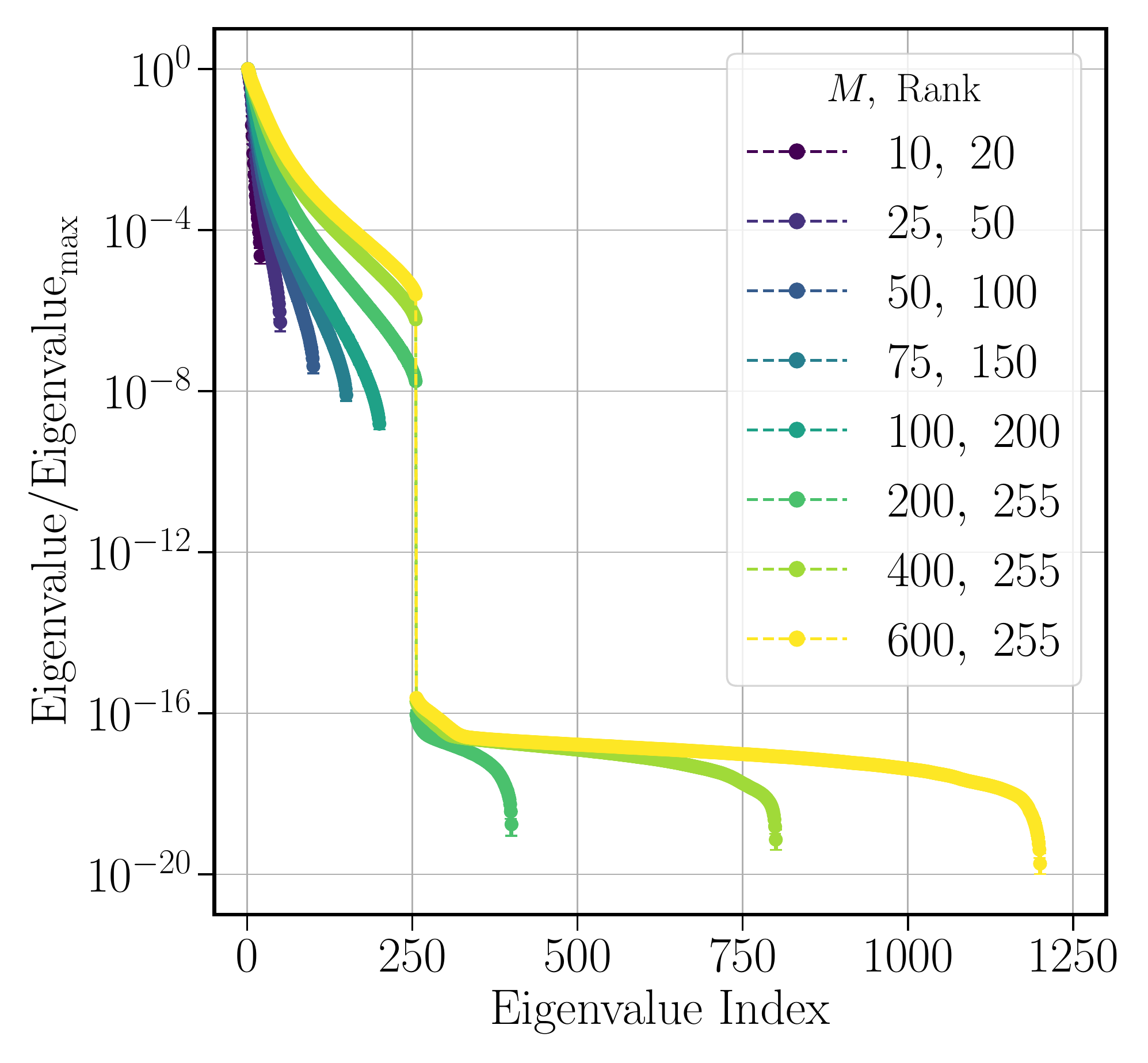}
		\label{fig:infidelity_fisher}
	\end{subfigure} 
	\captionsetup{justification=raggedright}	
	\caption{Convergence of unitary compilation infidelity with respect to optimization iteration and depth $M$ (colored/gradient) for the $N=4$ NMR ansatz. (a) Unconstrained parameterization with independent qubit parameters, and no boundary conditions. (b) Constrained parameterization with shared qubit parameters, and zero-field temporal boundary conditions. Constrained tasks require comparable iterations or depth, to converge comparably to unconstrained tasks, and exhibit exponential convergence beyond $M>\tilde{M} \approx O(d^{2}) = O(256)$. (c) Constrained quantum Fisher information eigenvalue spectrum at optimality. The spectrum is full-rank $R=P$ for $M < \tilde{M}$ before saturating at rank $R = \tilde{P} \sim O(d^{2}) \leq P$ for $M \geq \tilde{M}$, indicating overparameterization.}
	\label{fig:infidelity}
\end{figure}

Given the trends throughout optimization for the $N=4$ NMR ansatz and Haar random targets, as well as the rank saturation of the generalized unitary Fisher information metric in \cref{fig:infidelity}, we observe exponential infidelity convergence beyond a depth $M > \tilde{M} \approx G \sim O(256)$. We thus claim numerically that in noiseless settings, constrained optimization converges as
\begin{align}
	\mathcal{L}_{\theta} \sim e^{-\alpha M} ~:~ M > \tilde{M}
\end{align}
for some rate of convergence $\alpha$. In practice, $M \approx O(1000) \gg \tilde{M}$ is necessary for adequate convergence within a feasible number of optimization iterations. We also note that, in particular at larger depths, constrained tasks exhibit greater variance relative to unconstrained tasks, however both types of tasks have generally low relative variance, suggesting an appropriate choice of optimization hyperparameters, as per \cref{app:classical_simulation_and_optimization}.

We note that the non-local coupling $J \ll h,\theta$ is much smaller than other scales, and for experimentally realistic time steps, $\tau J \ll 1$. This limits how much of a non-local entangling gate can be implemented in a single time step, an essential part of generating Haar random unitaries. From these simulations, we conjecture that these constraints necessitate that the minimum depth where overparameterization can occur, is increased by a sub-exponential factor in the number of qubits,
\begin{align}
	\tilde{M}_{\textrm{Constrained}} \sim O(\frac{1}{\tau J}\textrm{poly}(N))\tilde{M}_{\textrm{Unconstrained}}~.
\end{align}
Other than a shift in the overparameterization boundary, the given constraints do not appear to fundamentally affect the exponential infidelity convergence. We also note that even for relatively small system sizes of $N \leq 4$, several orders of magnitude deeper circuits than are typically classically simulated \cite{Kiani2020,Wiersema2020,Larocca2021}, with up to $O(10^{5})$ gates, are necessary to study such realistic systems.

\subsection{Noisy State Preparation}\label{subsec:noisy_state_preparation}
We now investigate the effects of local noise on Haar random pure state preparation. Here parameters are unconstrained according to experimental feasibility to ensure all observed phenomena are due to noise. Given our findings on the effects of constraints, to leading-order, imposing constraints should only shift the depth dependent results by a noise-independent factor.

\begin{figure}[h]
	\centering
	\begin{subfigure}[t]{\columnwidth}
		\captionsetup{singlelinecheck = false, justification=raggedright,margin={0pt,0pt},skip=0pt}
		\subcaption{}		
		\includegraphics[width=0.9\columnwidth]{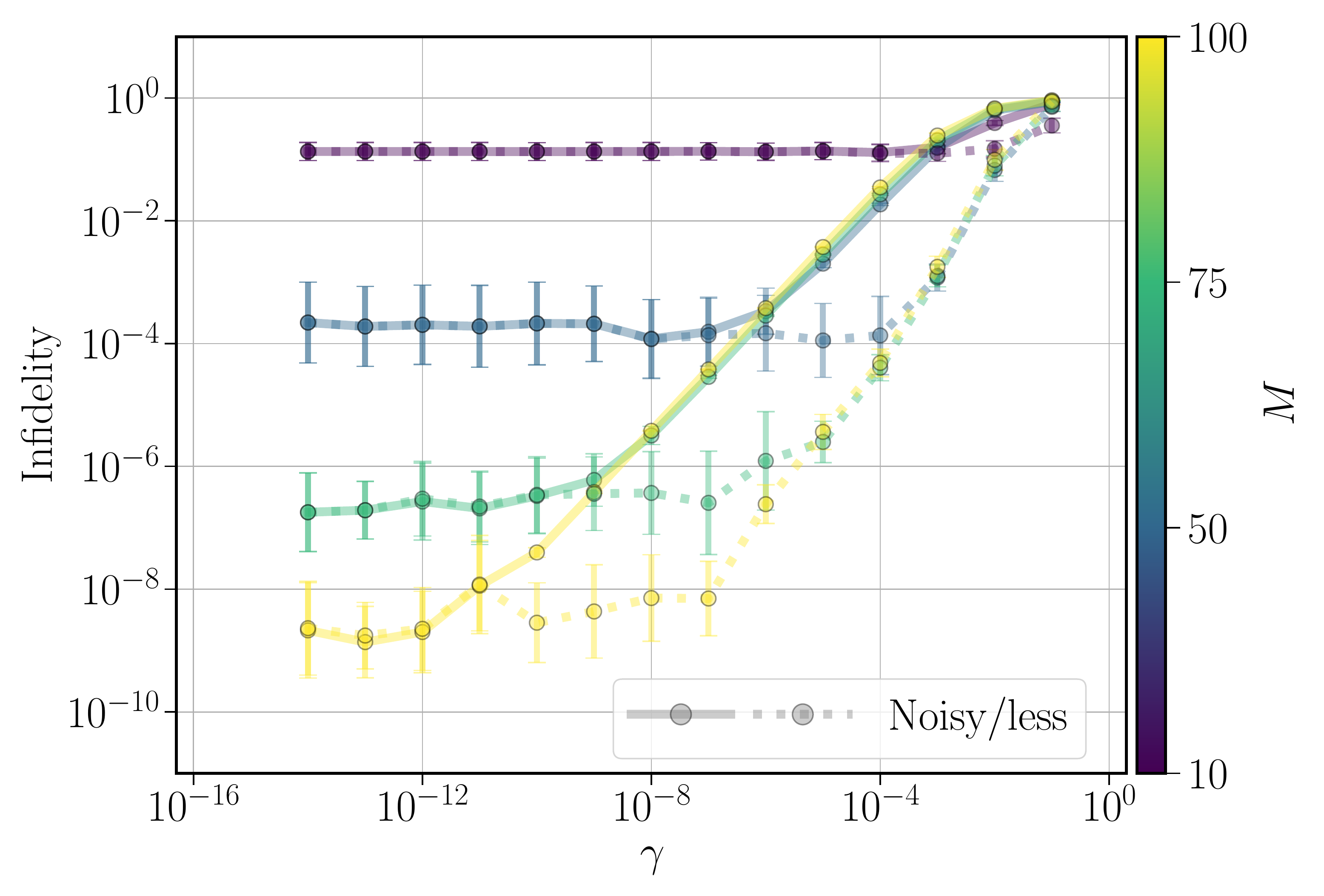}
		\label{fig:noise_noise}
	\end{subfigure}
	\hfill
	\begin{subfigure}[t]{\columnwidth}
		\captionsetup{singlelinecheck = false, justification=raggedright,margin={0pt,0pt},skip=0pt}
		\subcaption{}
		\includegraphics[width=0.93\columnwidth]{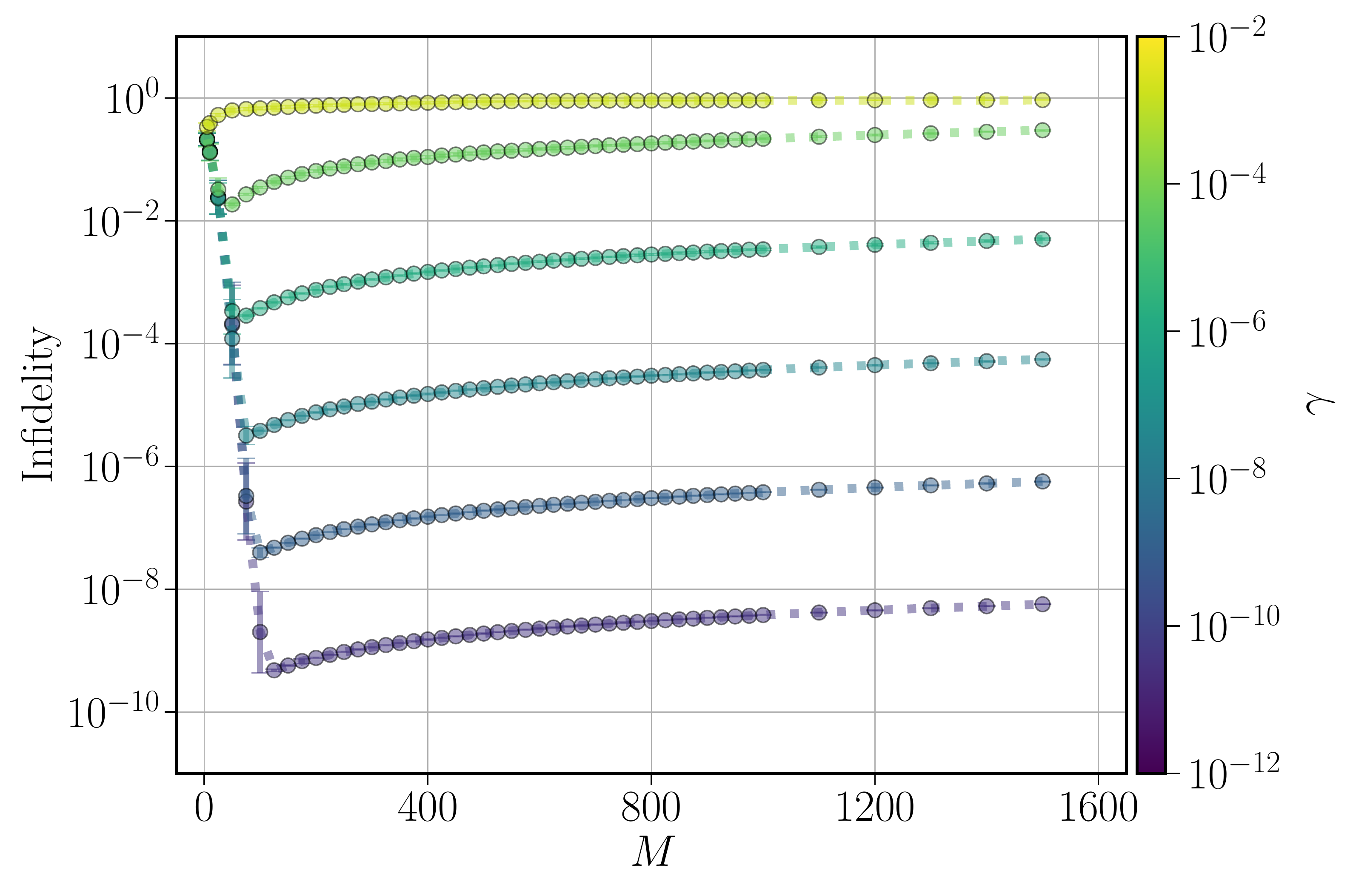}
		\label{fig:noise_M}
	\end{subfigure} 
	\captionsetup{justification=raggedright}	
	\caption{behavior of state preparation infidelity with respect to unital dephasing noise $\gamma$ and depth $M$ for the $N=4$ NMR ansatz. (a) Trained noisy infidelity (solid), and tested infidelity of noisy parameters in noiseless ansatz (dashed), with respect to noise scale, for various depths $M$ (colored/gradient). Infidelities are depth dependent and noise independent for small noise scales, before universally increasing polynomially with noise. Tested noiseless infidelities indicate that the underlying unitary dynamics are being learned resiliently. (b) Critical depth for noisy infidelity for various noise scales (colored/gradient). Infidelities improve exponentially with depth, up until a noise-induced critical depth, where entropic effects worsen infidelities polynomially.}
	\label{fig:noise}	
\end{figure}

To demonstrate the interplay between depth and noise scales on optimized infidelities, we plot infidelities with respect to each independent quantity in \cref{fig:noise}. Here we display unital dephasing noise, and other unital and non-unital noise models are shown to exhibit similar behavior in \cref{app:behavior_of_multiple_layer_noise_channels}. When varying the noise scale in \cref{fig:noise_noise}, for small noise scales, the average optimal infidelity (solid lines in \cref{fig:noise_noise}) strictly scales as expected with depth. Increasing the noise scale past a depth-dependent critical noise scale $\gamma_{M}$ causes infidelities to increase polynomially, between linearly and quadratically, until the infidelities plateau at their maximum value.

We also investigate trends when inserting parameters learned in the noisy setting into an identical, but noiseless unitary ansatz, yielding infidelities (dashed lines in \cref{fig:noise_noise}). Such noiseless, tested infidelities are superior than their corresponding trained noisy infidelities, in terms of their greater critical noise scale $\gamma_{M_{\textrm{Noiseless}}} > \gamma_{M_{\textrm{Noisy}}}$: the noiseless infidelities increase in the divergent regime at much larger noise scales. This suggests the optimization is learning about the underlying unitary dynamics, and is not just preparing another mixed state that happens to be close to the target pure state. 

The error bars of both the noisy trained and noiseless tested infidelities are equal in the convergent regime with $\gamma < \gamma_{M}$ (overlapped markers in the left of \cref{fig:noise_noise}), and they are orders of magnitude smaller in the divergent regime with $\gamma > \gamma_{M}$. Overall, the trends in error suggests the optimization and dynamics are most uncertain within the regime around and past the noise scale $\gamma_{M}$, indicating increased complexity near this transition.

The noiseless behavior can partly be explained from parameter shift rules for gradients of parameterized channels with constant noise, as derived in \cref{app:behavior_of_multiple_layer_noise_channels}. The noisy state is a convex combination of parameterized pure states, each with identical gradient directions to the noiseless case, albeit with magnitudes that are scaled by polynomials of the noise scale. Therefore, the trajectory of the gradient-based optimization remains similar at small noise scales in both noisy and noiseless cases. 

When varying the depth, we observe a critical noise-dependent depth $M_{\gamma}$ that occurs in \cref{fig:noise_M}. Previously decreasing infidelities with depth in a convergent regime for $M < M_{\gamma}$, increase uniformly with depth in a divergent regime for $M > M_{\gamma}$. Beyond this depth, the increase in expressiveness of the ansatz to prepare arbitrary states from increasing the number of variable parameters, is outweighed by the accumulated noise from the increased sources of error. 
From a trainability standpoint, beyond this critical depth, noise-induced barren plateaus may be occurring, leading to a decrease in trainability where the gradients are unable to find the ideal trajectory to reach optimality. Alternatively, from an expressiveness standpoint, noise potentially increases in an uncontrolled manner the number of directions permitted to be explored in the objective landscape \cite{Garcia-Martin2023}. Future work should investigate the density of mixed states that are perturbatively away from a given pure state \cite{Sharma2020,Kattemolle2023}. As discussed below, through analytical calculations, we offer complementary interpretations into exactly how entropic effects begin to dominate infidelity behaviors.

\subsection{Universal Effects of Classical and Quantum Sources of Noise}
\begin{figure}[h]
	\centering
	\begin{subfigure}[t]{\columnwidth}
		\includegraphics[width=0.95\columnwidth]{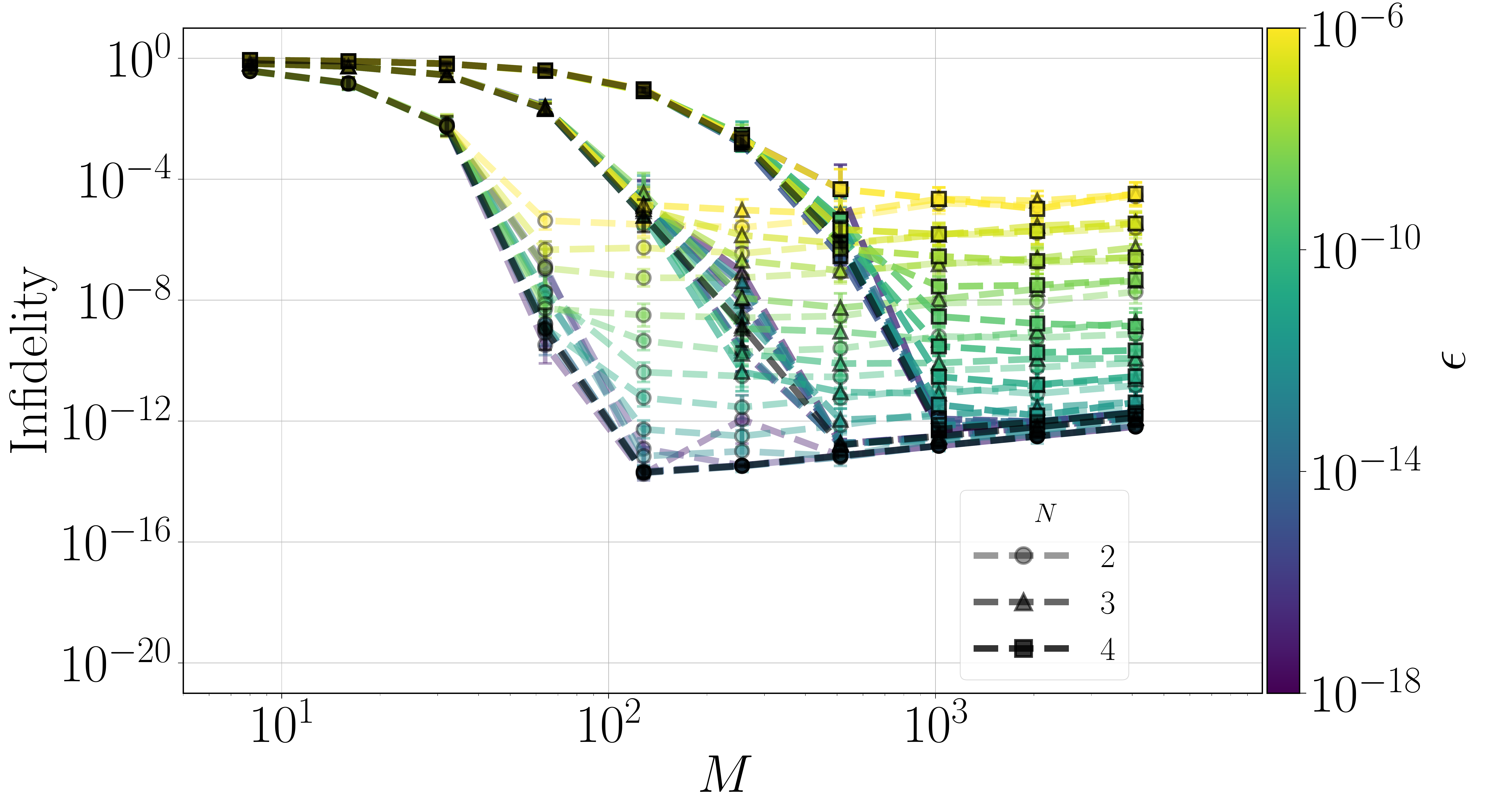}
	\end{subfigure} 
	\captionsetup{justification=raggedright}	
	\caption{behavior of unitary compilation infidelity with respect to depth $M$ and classical floating point noise scale $\epsilon$ (colored/gradient), for various $N$ NMR ansatz, relative to the noiseless case (black). Decreasing error scale verifies the classical noise model's suitability, and estimates the architecture dependent machine precision of $\varepsilon \sim O(10^{-16})$. Classical noise is also shown to exhibit a critical depth $M_{\epsilon}$ and divergent regime.}
	\label{fig:noise_classical}	
\end{figure}

These studies bring to mind the question of whether noise-induced critical depth phenomena can be attributed to strictly quantum, or potentially classical noise phenomena. For very deep, ideal noiseless ansatz with $M>O(1000)$, infidelities approach machine precision, and start to increase with depth. This suggests floating point errors accumulate for large numbers of simulated operations. As per the noise models derived in \cref{app:classical_and_quantum_error_analysis}, we insert artificial classical floating point error of different scales into simulations, as per \cref{fig:noise_classical}. Adding zero-mean random errors with standard deviation proportional to error scales, to results of floating point operations, shows similar trends in noise-induced convergent and divergent regimes. Since unitary compilation tasks have quadratically more degrees of freedom than state compilation tasks, their infidelities decrease slower with depth. Furthermore, infidelity curves reach their critical depth at earlier depths for smaller system sizes due to the exponentially smaller spaces to search. 

Infidelities are shown to increase, and enter the divergent regime when they reach a scale proportional to that of the error scale. This error scale may be approximately upper bounded by the derived deviations of the noisy infidelities from their noiseless counterparts. By decreasing the artificial floating point error, the curves also converge to the supposedly noiseless case, offering an estimate for the machine precision of $\varepsilon \sim O(10^{-16})$. These trends open many questions on the viability of large-scale simulations close to machine precision with finite floating point architectures. Can arbitrarily large systems be accurately simulated, without resorting to inefficient arbitrary precision arithmetic, or error mitigation or correction approaches \cite{Bharti2022,Wang2021a}?

\section{Discussion}\label{sec:discussion}
From numerical experiments, we find for a given scale of local noise $\gamma$ that there is a critical depth of circuit $M_{\gamma}$, beyond which optimal infidelities increase with depth due to an accumulation of noise. From fitting procedures discussed in \cref{app:behavior_of_multiple_layer_noise_channels}, we are able to determine the critical depth to be logarithmic in the noise scale
\begin{align}
	M_{\gamma} \sim&~ \log{1/\gamma}~.
\end{align}
The optimal infidelity is therefore approximately
\begin{align}
	\mathcal{L}_{\theta^{*}\gamma|M_{\gamma}} \sim&~ e^{-\alpha M_{\gamma}} \sim \gamma^{\alpha}~,
\end{align}
with $1 \leq \alpha \leq 2$, confirming previous conjectures of linear, or quadratic scaling of infidelity with noise \cite{Fontana2021}. The interpretation of a noisy channel being a binomial distribution of $k$-error channels also suggests that parameterized quantum channels can mitigate approximately
\begin{align}
	\bar{K}_{\gamma} \sim&~ \gamma \log{1/\gamma}
\end{align}
errors. Determining whether the optimization is finding a parameterization that explicitly performs error mitigation \cite{Bharti2022,Cai2020,Ravi2022}, or even error correction through a parameterized encoding \cite{Johnson2017}, would constitute important future contributions \cite{Wang2021a,Niroula2023}. The presence of a noise-induced depth is also reminiscent of weak measurement-induced phase transitions \cite{Skinner2018}. However, these noise-induced effects intuitively should be apparent at all system sizes, and do not seem to be related to typical indicators of phase transitions such as scale invariance.

As derived in \cref{app:behavior_of_multiple_layer_noise_channels}, the Bloch representation allows explicit leading-order scaling of quantities with respect to depth and noise to be derived. Here, we represent states as $\rho = (1/d)(I + \lambda \cdot \omega)$, with Bloch coefficients $\lambda$ associated with a set of $d^{2}-1$ non-identity, trace orthogonal basis operators $\omega$. Channels $\Lambda$ may then be represented as affine transformations $\lambda \to \Gamma\lambda + \upsilon$. Parameterized noisy channels $\Lambda_{\theta\gamma}$ with $K$ possible errors may be decomposed into strictly unitary $u_{\theta}$, and unital $u_{\theta\gamma}$ and non-unital $\eta_{\theta\gamma}$ noise dependent components
\begin{align}
	\Gamma_{\theta\gamma} =&~ (1-\gamma)^{K}u_{\theta} + (1-(1-\gamma)^{K})u_{\theta\gamma}\\
	\upsilon_{\theta\gamma} =&~ (1-(1-\gamma)^{K})\eta_{\theta\gamma}~.
\end{align}
In this Bloch representation, we may then express a parameterized noisy state as
\begin{align}
	\rho_{\theta\gamma} =&~ (1-\gamma)^{K}\rho + (1-(1-\gamma)^{K})\epsilon_{\theta\gamma} + \Delta_{\theta\gamma}~.
\end{align}
This decomposition expresses the interplay of the parameterized unitary and noise-induced non-unitary components of the channel. The unitary component rotates what we refer to as the pure component of the state. This component consists of a superposition of the pure target state $\rho$, with associated coefficients $\lambda$, and an orthogonal pure state with orthogonal associated coefficients $\zeta \perp \lambda$, represented within the traceless deviation term $\Delta_{\theta\gamma}$. The noise component of the channel scales the pure component of the state with the noise scale $1-\gamma$, plus it shifts the state by what we refer to as the mixed component of the state $\epsilon_{\theta\gamma}$, with associated Bloch coefficients $\varepsilon_{\theta\gamma}$. In the limit of the optimization reaching optimality in the noisy setting $\theta \to \theta^{*}_{\gamma}$, the pure component of the state approaches the pure target state, the deviation term $\Delta_{\theta^{*}_{\gamma}\gamma} \to 0$ approaches zero, and there only remains an inherent noise-dependent mixed component $\epsilon_{\theta^{*}_{\gamma}\gamma}$. We may then assume that optimality is reached in this noisy setting, and channel-dependent quantities may be expanded in the number of errors and noise scale.

At optimality, we find our quantities of interest analytically scale similarly to leading-order in $K,\gamma$, namely
\begin{align}
	\mathcal{L}_{\theta\gamma}^{\rho} \sim&~ K\gamma \frac{d-1}{d}\left( 1 - \frac{\lambda \cdot \varepsilon_{\theta\gamma}}{\lambda^2}\right) ~+~ O(\tbinom{K}{2}\gamma^2) ~,\\
	\mathcal{I}_{\theta\gamma} \sim&~ 2K\gamma \frac{d-1}{d}\left(1 -  \frac{\lambda \cdot \varepsilon_{\theta\gamma}}{\lambda^{2}}\right) ~+~ O(\tbinom{K}{2}\gamma^2) ~,\\
	\mathcal{S}_{\theta\gamma} \sim&~ O(K\gamma)  \quad , \quad \mathcal{D}_{\theta\gamma}^{\rho} \sim  O(K\gamma)~.
\end{align}
As derived in \cref{app:behavior_of_multiple_layer_noise_channels}, the Bloch representation allows exact leading-order terms to be derived for quantities that are strictly polynomial functions of the Bloch coefficients. Quantities that are more complex, for example logarithmic, functions of the Bloch coefficients require knowledge of the algebras governing the specific choice of basis $\omega$, yielding in principle calculable \cite{Sarkar1971}, but unintuitive forms. Importantly, when the mixed components $\varepsilon_{\theta\gamma} \parallel \lambda$ become pure, or aligned with the pure target coefficients, quantities differ from their noiseless values by strictly higher-order terms. This purification of the mixed component could be due to the lack of noise, or the specific combination of parametrizations and noise models forcing the system towards a pure state, and it describes general error mitigation. The use of Bloch representations thus allows simplified, and occasionally ansatz-independent depictions of noise-induced phenomena.

\begin{figure}[h]
	\centering
	\begin{subfigure}[t]{\columnwidth}
		\includegraphics[width=0.95\columnwidth]{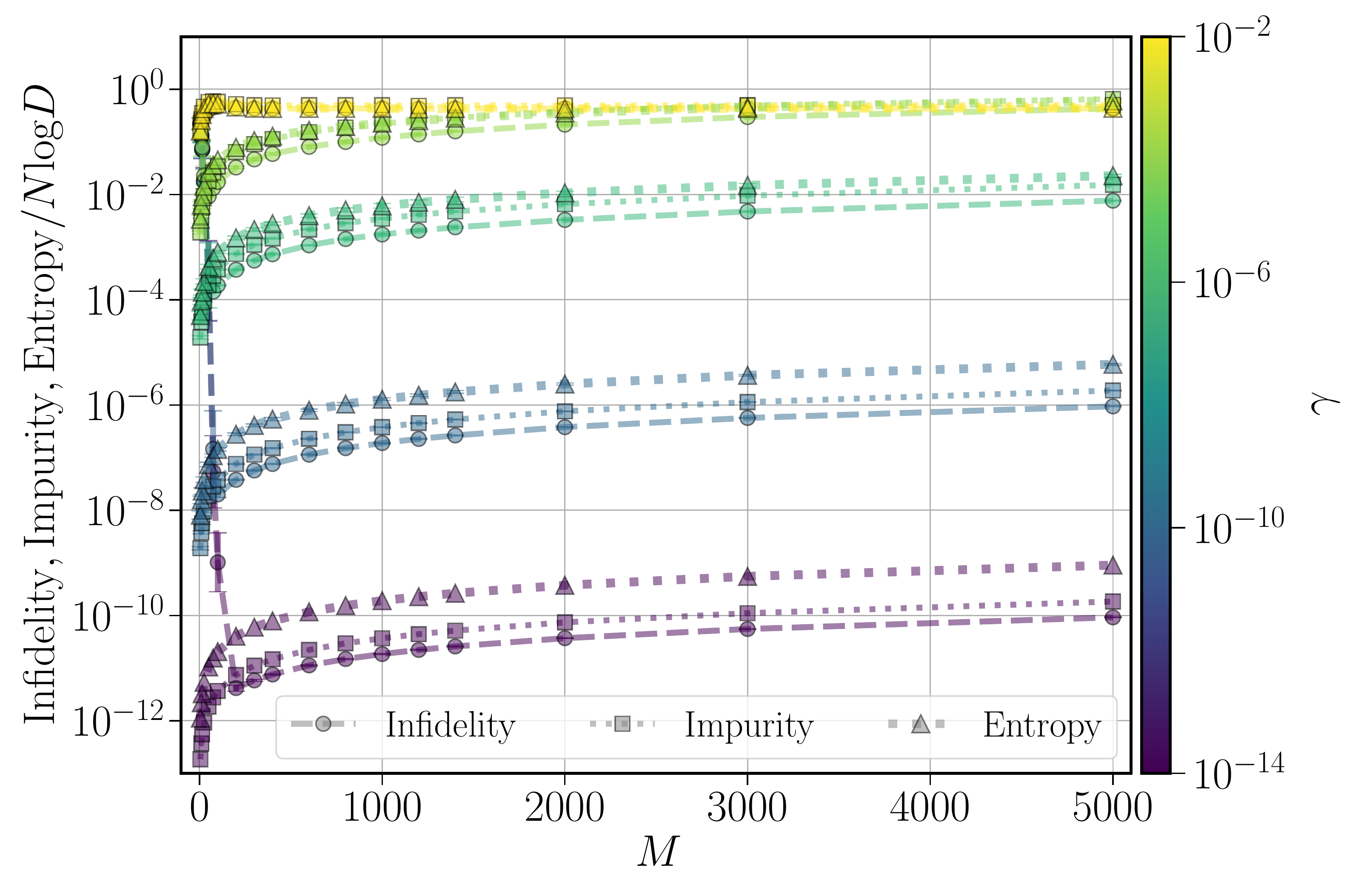}
	\end{subfigure}
	\captionsetup{justification=raggedright}	
	\vspace{-12pt}	
	\caption{behavior of infidelity, entropy, and impurity with respect to depth $M$ and non-unital amplitude damping noise $\gamma$ (colored/gradient), for the $N=4$ NMR ansatz. At small noise scales, quantities scale identically linearly with depth and with noise in the divergent, entropic driven regime. At large noise scales, unlike unital noise, non-unital noise infidelities decrease polynomially with depth, once the parameterized unitary aligns the state towards the target pure state.}
	\label{fig:quantity}
	\vspace{-12pt}
\end{figure}

Beyond the critical depth, infidelities appear both analytically and numerically to be linear functions of entropy and impurity, and they scale with the overlap of the pure target state with the mixed component. In particular, at low noise scales in the divergent noisy regime, analytical predictions and numerical simulations of the discussed quantities all correspond precisely, as demonstrated in \cref{fig:objective_purity_entropy_divergence_similarity} in \cref{app:behavior_of_multiple_layer_noise_channels}. Further, from \cref{fig:quantity}, all quantities appear to collapse together with increasing noise scales. Reasoning that noise phenomena dominate at $M=M_{\gamma}$, when the scale of the optimization-driven decreasing optimized infidelity $\mathcal{L}_{\theta^{*}_{\gamma}\gamma}^{\rho}$, reaches the scale of the entropic-driven increasing analytical infidelity $\mathcal{L}_{\theta^{*}\gamma}^{\rho}$,
\begin{align}
	\mathcal{L}_{\theta^{*}_{\gamma}\gamma}^{\rho} \sim e^{-\alpha M}\vert_{M_{\gamma}} \approx&~ \mathcal{L}_{\theta^{*}\gamma}^{\rho} \sim NM\gamma \vert_{M_{\gamma}}~,
\end{align}
and we recover our numerically predicted noise-induced critical depth $M_{\gamma} \sim \log{1/\gamma}$.

We note that bi-partite (entanglement) entropy, between a system and its environment, is generally bounded strictly by the system size $N = K/M$. However, noisy quantities are polynomials of $k$-error factors $p_{K\gamma}(k) \sim O(K^k \gamma^k)$. The additional degree of freedom representing the strength of system-environment interactions $\gamma$, appears to suppress the higher-order polynomial factors.

Universal behaviors in the divergent regime can be attributed to entropy-increasing phenomena, once the parameterized channel has rotated the state to within a depth-dependent distance from the target state. However, at large noise scales, there are important distinctions between unital versus non-unital types of noise.

For unital noise, such as dephasing noise in \cref{fig:noise}, a potential explanation is that the combination of the potentially close to Haar random parameterized unitaries, and the accumulated noise, induces depolarization. Entropy is shown to increase linearly with depth, at practically all noise scales, and dominates the infidelity behaviors.

For non-unital noise, such as amplitude damping noise in \cref{fig:quantity}, at small noise scales, the behavior appears qualitatively similar to unital noise, of linearly increasing infidelity, along with impurity and entropy. However, non-unital noise appears to have fundamentally different, non-universal behavior at large noise scales. Non-unital noise forces the state into a specific (pure) state, which appears to improve the infidelities. Unlike in the overparameterized regime, however, infidelities appear to decrease polynomially with depth. Adequately parameterized unitaries at depths far beyond the typical overparameterization bound appear necessary to rotate some components of this forced state towards the target state. 

Ultimately, optimization routines in a noisy setting are shown to be capable of rotating the pure components of states towards target pure states, and they exhibit overparameterization phenomena. Once objectives approach a noise-induced entropy-dependent scale, entropic effects then dominate objective behavior with increasing depth. Finally, fundamental differences between unital and non-unital noise at large depths and large noise scales are relevant when considering NISQ applications.

\vspace{-16pt}
\section{Conclusion}\label{sec:conclusion}
Through this work's classical simulation and analytical treatments, overparameterization phenomena for quantum systems are shown to be robust under realistic settings. Infidelities decrease exponentially with depth in the convergent regime, before increasing polynomially with depth and with noise in the divergent regime. These scalings provide essential data for the experimental design of variational quantum algorithms.

When assessing the relevance of this work, given we simulate specifically NMR systems with depths $M \in [10,5000]$, and noise scales $\gamma \in [10^{-18},10^{-2}]$, we must assess the depth and noise scales of other implementations, such as trapped ions, super-conducting qubits, and neutral atoms in \cref{tab:implementations}. Given our derived logarithmic dependence with noise scales of the noise-induced critical depth, we conjecture that NMR systems' robustness deteriorates at depths $M_{\gamma} \sim O(200)$, and noise scales of $\gamma_{M} < O(10^{-3})$ if we require infidelities $\mathcal{L} < O(10^{-4})$. Although the full span of considered depths and noise scales exceeds currently experimentally feasible regimes for NISQ implementations, the derived limits where robustness deteriorates are still approaching currently feasible scales. Other implementations are expected to show identical phenomena, with some ansatz-dependent shift in these depth and noise scales.

It should also be noted that non-unital noise appears to allow for re-improved infidelities in the divergent regime at exceptionally large depths of $M \sim O(5000)$ and large noise $\gamma \sim O(10^{-2})$. There may be intriguing non-trivial, noise-type induced emergent phenomena at these large depths, even if these regimes are currently impractical experimentally. This work further serves as studies of general noise phenomena, which are likely to be encountered when existing implementations are scaled to address practical problems. The conclusions drawn, therefore, support the necessity of quantum error correction, and challenge aspirations \cite{Schuld2022a,Abbas2023} of existing NISQ applications being scaled to thousands of qubits and gates.

Finally, we remark that entropic effects appear to dominate only beyond a critical number of errors in the system. General parameterized systems are thus shown to be capable of suppressing entropic behavior imposed by their environment. By locating the noise-induced critical depth, problems can be optimized to their best-case objectives across all depths, for example in coveted quantum control problems \cite{Magann2021}. This opens up intriguing applications \cite{Endo2018} for variational ansatze, both classical and quantum, and we are excited about their potential.

\begin{acknowledgments}
	The authors would like to thank Gerardo Ortiz, Hemant Katiyar, Zachary Mann, Andrew Jreissaty, and Roeland Wiersema, for their valuable insights into noisy quantum processes. MD and JC would like to acknowledge the support of the Natural Sciences and Engineering Research Council of Canada (NSERC). JC acknowledges support from the Shared Hierarchical Academic Research Computing Network (SHARCNET), Compute Canada, and the Canadian Institute for Advanced Research (CIFAR) AI chair program. Resources used in preparing this research were provided, in part, by the Province of Ontario, the Government of Canada through CIFAR, and companies sponsoring the Vector Institute \url{www.vectorinstitute.ai/#partners}. RL would like to thank Mike and Ophelia Lazaridis. Research at the Perimeter Institute is supported in part by the Government of Canada through the Department of Innovation, Science and Economic Development Canada and by the Province of Ontario through the Ministry of Economic Development, Job Creation and Trade.
\end{acknowledgments}

\bibliography{main}

\appendix
\onecolumngrid
\newpage

\section{Background}\label{app:background}
In these appendixes, we elaborate on the specific parameterized quantum channel ansatze studied in this work. We discuss the unitary and non-unitary components of the channel, and their approximations and derivations via Trotterization. These Trotterized forms correspond to quantum circuit models of continuous evolution of quantum systems up to a specified order of precision, and they are used for classical simulation.

\subsection{Unitary Evolution}\label{subsec:unitary_evolution}
The unitary evolution operator takes the form of the time ordered matrix exponential of the Hamiltonian
\begin{align}
	U_{\theta} =&~ \Tau e^{-i \int_{0}^{T} dt ~H_{\theta}^{(t)}}~.
\end{align}
Here the time-dependent Hamiltonian driving the evolution takes the parameterized form
\begin{align}
	H_{\theta}^{(t)} =&~ \sum_{\mu}H_{\mu}^{(t)} ,
\end{align}
with a set of $\textrm{poly}(N)$ parameters $\theta^{(t)}$, at each time $t \in [0,T]$.
Each term in the Hamiltonian is parameterized with fixed generators $\mathcal{G} = \{G_{\mu}\}$ as
\begin{align}
	H_{\mu}^{(t)} =&~ \theta_{\mu}^{(t)} G_{\mu}~.
\end{align}
The set of operators $\{G_{\mu}\}$ may contain local or non-local operators, and generally is at least partially non-commuting.

This continuous evolution generated by exponential maps of Hamiltonians must be discretized temporally and spatially across the space of subsystems for feasible classical simulation. This discretization further allows for comparison with the variational quantum circuit paradigm. Depending on the control problem of interest, there are several choices for the specific discretization scheme. Further, given constraints placed on the parameters, the explicitly optimized parameters may take various functional forms.

\subsection{Trotterization}\label{subsec:trotterization}
To classically simulate such time-dependent systems, unitaries are trotterized, both temporally and spatially across the space. To first-order in time,
\begin{align}
	U_{\theta} \approx &~ \prod_{m}^{M} U_{\theta}^{(m)} ~+~ O(\tau^2)
	\intertext{where the time has been discretized into $M$ time steps of size $\tau = T/M$, and evolution at each time step $m \in [M]$ is}
	U_{\theta}^{(m)} =&~ e^{-i \tau H_{\theta}^{(m)}}~.
\end{align}
Further, depending on the commutation relations between terms in the Hamiltonian, to $Q$-order in space across the qubits, at a given instance $m$ in time,
\begin{align}
	U_{\theta}^{(m)} \approx&~ \prod_{\mu}^{(Q)} U_{\mu}^{(m,Q)} ~+~ O(\comm{H_{\mu}^{(m)}}{H_{\nu}^{(m)}}_{Q+1} \tau^{Q+1})~.
\end{align}
The product $\prod_{\mu}^{(Q)}$ represents a product over some function of lower-order Trotterizations, denoted by $U_{\mu}^{(m,Q)}$, with the error being proportional to the $Q+1$-deep nested commutator of Hamiltonian terms at time $m$ \cite{Yang2022}. For $Q = 2$, the product is equal to the forward and backward ordering of first-order Trotterized operators
\begin{align}
	U_{\mu}^{(m,Q)} \overset{Q=2}{=}&~ {U_{\mu}^{(m)}}^{1/Q}~,
\end{align}
with a factor of $1/Q$ in the generators to ensure consistency, and
\begin{align}
	U_{\theta}^{(m)} \overset{Q=2}{\approx}&~ \prod_{\mu}^{\rightharpoonup} U_{\mu}^{(m)/Q} \prod_{\mu}^{\leftharpoonup} U_{\mu}^{(m)/Q} ~+~ O(\tau^{3})~.
\end{align}
The resulting evolution can be directly described by a parameterized quantum circuit, consisting of operators $U_{\mu}^{(m)}$ with locality of the corresponding Hamiltonian generator $G_{\mu}$. Therefore the following results, up to the precision of these discretizations, are relevant both in the discrete gate-based circuit model of quantum systems, and the continuous time evolution model, more generally seen in pulse-level and quantum control problems \cite{Magann2021,Egger2023}.

\subsection{Parameterized Quantum Channels}\label{subsec:parameterized_quantum_channels}
To understand the effects of noise on the evolution and abilities of such a system to represent various targets, we describe the evolution with parameterized quantum channels acting on states as
\begin{align}
	\Lambda_{\theta\gamma} =&~ \mathcal{N}_{\gamma} \circ \mathcal{U}_{\theta}
\end{align}
such that the evolved parameterized state from an initial state $\sigma$ is
\begin{align}
	\rho_{\theta\gamma} = \Lambda_{\theta\gamma}(\sigma)~.
\end{align}
The channel is composed of a noiseless unitary channel with variable parameters $\theta$, 
\begin{align} 
	\mathcal{U}_{\theta}(\cdot) =&~ U_{\theta} \cdot U_{\theta}^{\dagger}~,
\end{align}
and a noisy non-unitary Kraus operator channel with fixed noise parameters $\gamma$,
\begin{align}
	\mathcal{N}_{\gamma}(\cdot) =&~ \sum_{\alpha} K_{\gamma}^{(\alpha)} \cdot {K_{\gamma}^{(\alpha)}}^{\dagger}~.
\end{align}
All channels must be normalized such that they are trace-preserving.

Given the Trotterization of the continuous time evolution, we define the channel as a composition of $M$ layers of channels at each time step,
\begin{align}
	\Lambda_{\theta\gamma} =&~ \circ_{m}^{M} \left(\mathcal{N}_{\gamma}^{(m)} \circ \mathcal{U}_{\theta}^{(m)}\right)~.
\end{align}
We also make use of the notation for the composition of channels before or after an index $m$ as
\begin{align}
	\Lambda_{\theta\gamma}^{\lessgtr m} =&~ \circ_{l \lessgtr m} \left(\mathcal{N}_{\gamma}^{(l)} \circ \mathcal{U}_{\theta}^{(l)}\right)~.
\end{align}
We also write the $Q$-order spatial Trotterization of the unitary part of the channel as a composition over the Trotterized unitaries, with identical notation to the products of unitaries when deriving their Trotterization,
\begin{align}
	\mathcal{U}_{\theta}^{(m)} =&~ \circ_{\mu}^{(Q)} \mathcal{U}_{\mu}^{(m,Q)}~,
\end{align}
and we can similarly define the notation for composition of channels before or after an index $\mu$ as
\begin{align}
	\mathcal{U}_{\lessgtr \mu}^{(m)} =&~ \circ_{\nu \lessgtr \mu}^{(Q)} \mathcal{U}_{\nu}^{(m,Q)}~.
\end{align}
These notations may be combined for partial channels relative to before or after an index $(\mu,m)$, such that
\begin{align}
	\Lambda_{\theta\gamma} =&~ \Lambda_{\gamma}^{(>m)} \circ \mathcal{N}_{\gamma}^{(m)} \circ \mathcal{U}_{>\mu}^{(m)} \circ \mathcal{U}_{\mu}^{(m)} \circ \mathcal{U}_{<\mu}^{(m)} \circ \Lambda_{\gamma}^{(<m)}~.
\end{align}
The corresponding partially forward evolved state relative to a state $\sigma$ from the action of the channel before an index $(\mu,m)$ may then be denoted as
\begin{align}
	\rho_{<\mu\gamma}^{(< m)} =&~ \Lambda_{<\mu\gamma}^{(< m)}(\sigma)~,
\end{align}
and the corresponding backward evolved operator relative to an input operator $O$ from the adjoint action of the channel after an index $(\mu,m)$ may also be denoted as 
\begin{align}
	O_{> \mu \gamma}^{(>m)} =&~ {\Lambda_{>\mu\gamma}^{(> m)}}^{\dagger} (O)~.
\end{align}
This notation is used to understand parameter shift rules for channels, and to derive the scaling of objectives with noise and with depth. Here we have dropped the $(m,Q)$ superscript notation for the $Q$-order spatial Trotterization, in favor of $(m)$ superscript notation, for simplicity, and it is assumed that operators at spatial indices are implicit functions of the Trotterization scheme.

\newpage
\section{Learning Phenomena}\label{app:learning_phenomena}
In these appendixes, we give an overview of learning phenomena in quantum settings, in particular, overparameterization phenomena, and we discuss their relevance to this work. We derive a unitary version of the quantum Fisher information, and we show numerically that its rank acts as an indicator of overparameterization identically to the state quantum Fisher information studied in previous works. Finally, we use these results to confirm that overparameterization phenomena occur in realistic settings, as per established definitions.

A key aspect of this work involves understanding whether in realistic settings, indicators of overparameterization, in particular exponential convergence of the optimization with the number of parameters, still occur. We follow the approaches by Larocca \etal \cite{Larocca2021}, later followed up by Garcia-Martin \etal \cite{Garcia-Martin2023}, which set bounds on the rank of the quantum Fisher information to determine whether a quantum system is overparameterized, both in noiseless and then noisy settings.

Overparameterization in this context, as per the Fisher information definition \cite{Larocca2021}, refers to when there are adequate number $P$ parameters such that the model ansatz can span the space \cite{Kiani2020,Liu2020} of the dynamical Lie algebra $\mathcal{G}$ formed by its generators $\{G_{\mu}\}$. This parameterization generally translates to the optimization procedure being able to converge exponentially with the number of optimization iterations. This may occur due to a fundamental change in the objective landscape where it becomes much more convex in this regime. It may also be accompanied by what is known as lazy training, where the optimal parameters are negligibly different from their random initial values.

In the continuous time evolution, or gate based circuit formalisms, these generators are the non-commuting terms in the underlying Hamiltonian that drives the evolution, and the dynamical Lie algebra is formed by the Lie closure
\begin{align}
	\mathcal{G} =&~ \langle \{G_{\mu}\}\rangle_{\textrm{Lie}}~,
\end{align}
of all linearly independent nested commutators, of these generators.

This dynamical Lie algebra span forms a subspace $\mathcal{G} \subseteq \mathcal{H}$ of the full space $\mathcal{H}$ where the operators act, and it has dimension
\begin{align}
	G =&~ \abs{\mathcal{G}}~.
\end{align}
Within the context of spaces with dimension $d=D^{N}$, example algebras associated with the Lie closure that arise in quantum control settings include the special unitary algebra $\mathit{su}(D^{N})$ with dimension $\abs{\mathit{su}(D^{N})}=D^{2N}-1$, or the symplectic algebra $\mathit{sp}(2N)$ with dimension $\abs{\mathit{sp}(2N)} = 2N^{2} + N$. Generally, the dynamical Lie algebra dimension is either a polynomial or an exponential function of the system size $N$, and one-dimensional Lie algebras have been classified in \cite{Wiersema2023}. The number of parameters $P$, for a fixed periodic ansatz repeated for $M$ layers, generally scales as 
\begin{align}
	P \sim &~ O(\textrm{poly}(N)M)
\end{align}
and the depth itself may depend on the system size, depending on the ansatz. For overparameterization to occur, the number of parameters must be of at least similar order to this dimension,
\begin{align}
	P > \tilde{P} \sim O(G)
\end{align}
and in ideal settings, overparameterization occurs when exactly $\tilde{P} = G$.

A key indicator \cite{Larocca2021,Garcia-Martin2023} of overparameterization is whether the rank of the quantum Fisher information $\mathcal{F}_{\theta}$,
\begin{align}
	R_{\theta}^{\mathcal{F}} =&~ \textrm{rank}(\mathcal{F}_{\theta}) \leq G
\end{align}
saturates at this dimension. It can be shown that this saturation is independent of where you are in the objective or parameter landscapes, and it does not occur only at optimality.

Similar bounds for the Hessian $\mathcal{H}_{\theta}$ of the objective, being objective- and target-dependent, only occur at optimality,
\begin{align}
	R_{\theta^{*}}^{\mathcal{H}} =&~ \textrm{rank}(\mathcal{H}_{\theta}) \leq G
\end{align}
and otherwise the Hessian is generally full rank $R_{\theta}^{\mathcal{H}} = P$.

We should also note recent developments \cite{Garcia-Martin2023} in understanding the effects of noise on variational quantum circuits, in the context of the spectrum of the quantum Fisher information. Small amounts of local noise are shown to allow more directions to be searched \cite{Kattemolle2023}. These additional directions in the state space may increase or decrease the parameterized state's purity. However, for increasing noise scales, the system becomes exponentially less sensitive to its parameters, and it can search fewer and fewer directions. As shown in \cref{sec:results}, we also observe different regimes depending on the noise scales, which confirm this recent analysis. We observe that the quasi-linear scaling of infidelity with noise is at the boundary of the convergent and divergent regimes. There is finite noise that suppresses the system's abilities to achieve perfect infidelity, however not so much as to reach the divergent regime.

\subsection{Unitary Quantum Fisher Information}\label{subapp:unitary_Fisher_information}

We now study the form of the quantum Fisher information, which can be shown \cite{Meyer2021} to be proportional to the second-order correction to the Bures metric, also referred to as the Fubini-Study metric in the case of pure states. This quantity offers insight into which directions of the space can be reached, given the variational ansatz.

Here, we generalize the definition of the Fisher information, from states, to a state-independent definition that only depends on the underlying parameterized unitary, similar in form to another recent study of generalized metrics \cite{Haug2023}.

We now investigate the second-order term, or Fisher-like information of distance measures $\mathcal{L}_{\theta}^{U}$ between a parameterized unitary $U_{\theta}$ with parameters $\theta = \{\theta_{\mu}\}$, and a fixed reference unitary $U$, over a $d$-dimensional space. We define $\mathcal{L}^{U}_{\theta^{*}} = \mathcal{L}_{\theta}^{U}\vert_{U_{\theta} = U}$ to be the parameterized unitary evaluated exactly at the reference unitary.

Let a distance $\mathcal{L}_{\theta}^{U}$ between unitaries (that is not a proper distance metric as it does not satisfy the triangle inequality), be related to the absolute trace overlap
\begin{align}
	\mathcal{L}_{\theta}^{U} =&~ 1 - \frac{1}{d^{2}}\abs{\trace{U^{\dagger}U_{\theta}}}^{2}~,
\end{align}
with derivatives with respect to $\theta$ of
\begin{align}
	\partial_{\mu} \mathcal{L}^{U}_{\theta} =&~ -2 \frac{1}{d^{2}} \real{\trace{U_{\theta}^{\dagger}U}\trace{U^{\dagger}\partial_{\mu} U_{\theta}}}\\
	\partial_{\mu\nu} \mathcal{L}^{U}_{\theta} =&~ -2 \frac{1}{d^{2}} \real{\trace{U_{\theta}^{\dagger}U}\trace{U^{\dagger}\partial_{\mu\nu} U_{\theta}} - \trace{\partial_{\mu} U_{\theta}^{\dagger}U}\trace{U^{\dagger}\partial_{\nu} U_{\theta}}}~.
\end{align}
At optimality where $U_{\theta}=U$,
\begin{align}
	\mathcal{L}^{U}_{\theta^{*}} = 0
	\quad , \quad 		
	\partial_{\mu} \mathcal{L}^{U}_{\theta^{*}} = 0
	\quad ,&~ \quad 		
	\partial_{\mu\nu} \mathcal{L}^{U}_{\theta^{*}} = 2 \frac{1}{d^{2}} \real{d~\trace{\partial_{\mu} U_{\theta}^{\dagger}\partial_{\nu} U_{\theta}} - \trace{\partial_{\mu} U_{\theta}^{\dagger}U_{\theta}}\trace{U_{\theta}^{\dagger}\partial_{\nu} U_{\theta}}}~.
\end{align}
To define the Fisher information metric, we define it as the leading-order behavior of the objective given a perturbation of parameters $\theta \to \theta + \delta$, evaluated at $U=U_{\theta}$, yielding
\begin{align}
	\mathcal{L}_{\theta + \delta}^{U} =&~ \mathcal{F}^{U}_{\theta_{\mu\nu}}\delta_{\mu}\delta_{\nu} ~+~ O(\delta^3)~,
\end{align}
with the Fisher information metric being
\begin{align}
	\mathcal{F}^{U}_{\theta_{\mu\nu}} =&~ \frac{1}{d^{2}} \real{d~\trace{\partial_{\mu} U_{\theta}^{\dagger}\partial_{\nu} U_{\theta}} - \trace{\partial_{\mu} U_{\theta}^{\dagger}U_{\theta}}\trace{U_{\theta}^{\dagger}\partial_{\nu} U_{\theta}}}~.
\end{align}
This state-independent quantity, identical to other Fisher information metrics, contains a term that reflects the change in the ansatz, plus a corrective term to ensure gauge invariance, with additional dimensionality $d$ factors to reflect that the action of the ansatz with respect to specific states is not being considered.

We can also use a proper distance metric in terms of the Frobenius norm $\norm{A}^2 = \trace{A^{\dagger}A}$ \cite{Haah2023}, such that a proper objective is invariant up to phases between the operators
\begin{align}
	\tilde{\mathcal{L}}^{U}_{\theta} =&~ 1 - \sqrt{1 - \mathcal{L}^{U}_{\theta}} = \frac{1}{2}\frac{1}{d}\max_{\phi} \norm{U_{\theta} - e^{i\phi}U}^{2}~,
\end{align}
with derivatives with respect to $\theta$ of
\begin{align}
	\partial_{\mu} \tilde{\mathcal{L}}^{U}_{\theta} =&~ \frac{1}{2} \frac{1}{1 - \tilde{\mathcal{L}}^{U}_{\theta}} \partial_{\mu} \mathcal{L}^{U}_{\theta} \\
	\partial_{\mu\nu} \tilde{\mathcal{L}}^{U}_{\theta} =&~  \frac{1}{2}\frac{1}{1 - \tilde{\mathcal{L}}^{U}_{\theta}} \left(\partial_{\mu\nu} \mathcal{L}^{U}_{\theta} + 2 \partial_{\mu} \mathcal{L}^{U}_{\theta} \partial_{\nu} \mathcal{L}^{U}_{\theta}\right)~,
\end{align}
such that the optimal proper quantities are identical up to constant scalings to the improper quantities
\begin{align}
	\tilde{\mathcal{L}}^{U}_{\theta^{*}} = 0
	\quad , \quad 	
	\partial_{\mu} \tilde{\mathcal{L}}^{U}_{\theta^{*}} = 0
	\quad ,&~ \quad 	
	\partial_{\mu\nu} \tilde{\mathcal{L}}^{U}_{\theta^{*}} = \frac{1}{2} \partial_{\mu\nu} \mathcal{L}^{U}_{\theta^{*}}~.
\end{align}
Therefore, the proper Fisher information metric definition is identical up to a constant scaling to the improper definition
\begin{align}
	\tilde{\mathcal{F}}^{U}_{\theta_{\mu\nu}} =&~ \frac{1}{2}\mathcal{F}^{U}_{\theta_{\mu\nu}}~.
\end{align}
For example, when the trace is over $d=1$ states, $\ket{\rho_{\theta}} = U_{\theta}\ket{\sigma}$ from an initial fixed state $\ket{\sigma}$, or equivalently the unitaries are projected onto a $d=1$-dimensional subspace, the Fisher information reduces to the standard definition
\begin{align}
	\mathcal{F}^{\rho}_{\theta_{\mu\nu}} =&~ \real{\bra{\partial_{\mu}\rho_{\theta}}\ket{\partial_{\nu}\rho_{\theta}} - \bra{\rho_{\theta}}\ket{\partial_{\mu}\rho_{\theta}}\bra{\partial_{\nu}\rho_{\theta}}\ket{\rho_{\theta}}}~.
\end{align}



\subsection{Numerical Overparameterization}\label{subapp:numerical_overparameterization}
\begin{figure}[h]
	\centering
	\begin{subfigure}[t]{0.45\textwidth}
		\captionsetup{singlelinecheck = false, justification=raggedright,margin={0pt,0pt},skip=0pt}
		\subcaption{}
		\hspace{-1cm}
		\includegraphics[width=0.7\columnwidth]{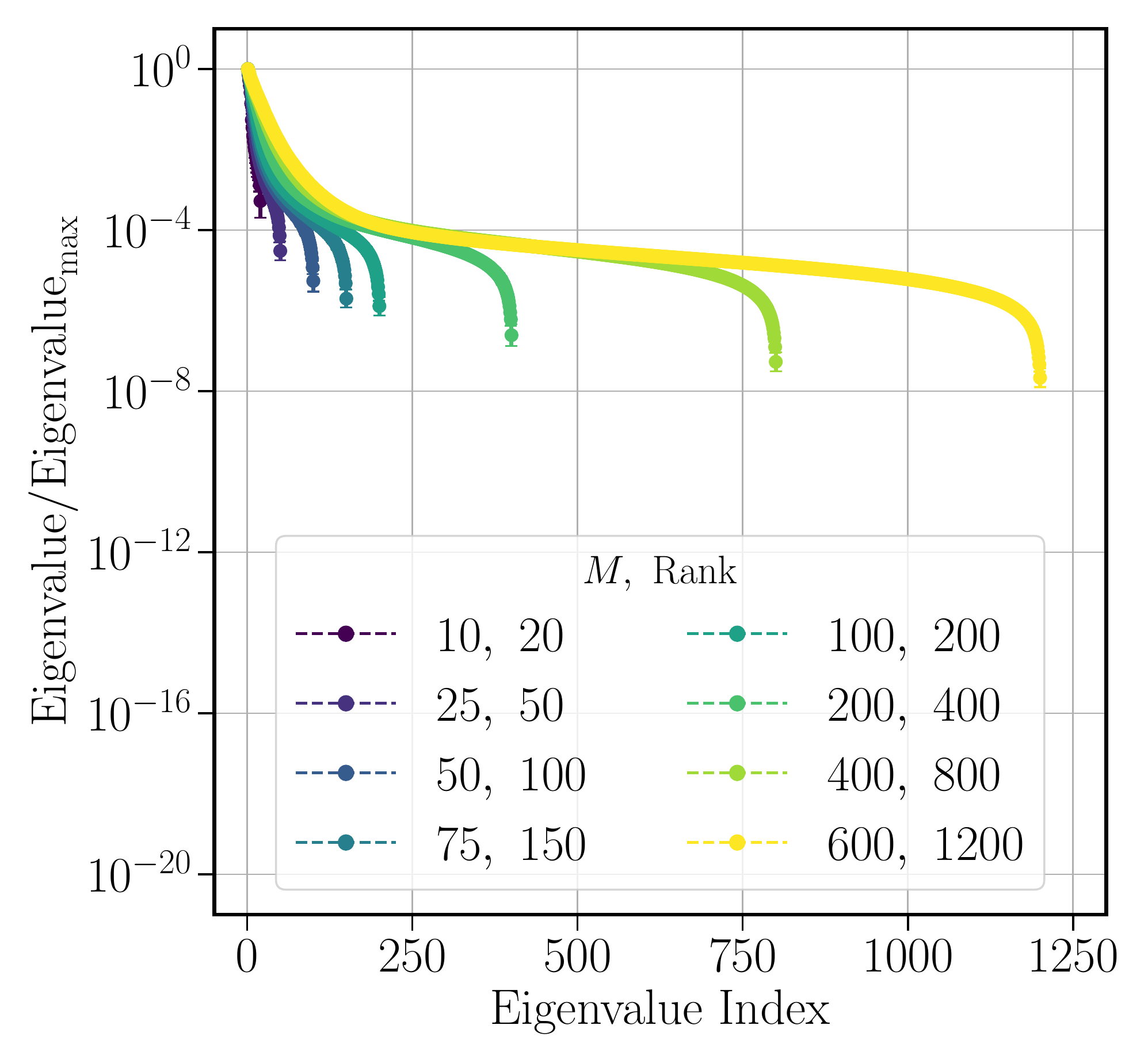}
		\vspace{-0.2cm}
		\label{fig:overparameterization_hessian}
	\end{subfigure}
	\hfill
	\begin{subfigure}[t]{0.45\textwidth}
		\captionsetup{singlelinecheck = false, justification=raggedright,margin={0pt,0pt},skip=0pt}
		\subcaption{}
		\hspace{-1cm}	
		\includegraphics[width=0.7\columnwidth]{figures/plot.None.fisher.eigenvalues.M.pdf}
		\vspace{-0.2cm}		
		\label{fig:overparameterization_fisher}
	\end{subfigure} 
	\captionsetup{justification=raggedright}	
	\caption{Metrics of overparameterization for constrained unitary compilation of an $N=4$ qubit NMR ansatz for various depths $M$ (colored/gradient). (a) Hessian spectrum of eigenvalues. The Hessian is shown to be full-rank at optimality for depths below the constrained overparameterization depth. Except at optimality in the overparameterized regime, all directions of the ansatz' span are able to be reached during optimization with respect to a target unitary. (b) Quantum Fisher information spectrum of eigenvalues. The quantum Fisher information is shown to be full-rank at optimality for depths below the overparameterization depth. It then saturates at the dynamical Lie algebra dimension for depths above the overparameterization depth, indicating the capabilities of the ansatz, independent of the specific target unitary.}
	\label{fig:overparameterization}
\end{figure}

We now investigate the effects of overparameterization numerically via the quantum Fisher information $\mathcal{F}_{\theta}$ and the objective Hessian $\mathcal{H}_{\theta}$. Here, we optimize $N=4$ qubit, constrained, noiseless unitary compilation tasks. For the NMR ansatz there are $D^{2}-2 = 2$ variable parameters per time step, meaning there are $P = 2M$ variable parameters per ansatz. Given the universal ansatz, $G = D^{N}-1$, we observe in \cref{fig:overparameterization} that the expected rank saturation occurs at $R = P = G = 255$ for $N=4$ for the Fisher information, and full rank saturation occurs at likely $R = P \gg G$ for the Hessian.

As per previous studies \cite{Larocca2021}, the Fisher metric exhibits saturation behavior and indicates overparameterization at any point in the objective or parameter landscape. However, the Hessian rank does not saturate, and remains full rank at this point in the landscape achieved by the optimizer. This point in the constrained landscape, even for $P \geq G$ is likely not optimal, and therefore not saturating the Hessian rank. This non-optimality is attributed to the previous studies indicating that $M \sim O(1000)$ depth is necessary for this constrained ansatz to achieve infidelities close to machine precision. Due to the quadratic scaling of the memory requirements for computing these $P \times P$ dimensional matrices, plus determining their spectrum, only $M \sim O(600)$ are currently feasible to compute in the current implementation.

We should note that for finite machine precision simulations, there is not a definitive method of determining the rank, or number of non-zero eigenvalues of matrices. Here we choose the heuristic when there is an obvious visual distinction between the set of zero and non-zero spectra. In the case in which there is not an obvious cutoff, we choose a relative precision of $\lambda/\lambda_{\textrm{max}} > P\varepsilon$ for $P$ parameters and machine precision $\varepsilon$. It remains an interesting question whether there is a more principled, and physics-informed approach for determining the rank.

\newpage
\section{behavior of Multiple Layer Noise Channels}\label{app:behavior_of_multiple_layer_noise_channels}
In these appendixes, we investigate properties of quantum channels consisting of parameterized unitary layers, interlaced with noise. We show that many local noise models form a binomial distribution, over the number of errors, or non-identity noise operations applied throughout the layers. We perform additional numerical studies to show differences in the behavior of infidelities with depth for unital versus non-unital noise. We extract from piecewise fitting, the scaling of the noise-induced critical depth for infidelities, to be logarithmic in the noise scale. Finally, we perform analytical calculations of the leading-order scaling, with depth and noise scale, of infidelities, impurities, entropies, and relative entropy divergences. These scalings match the numerical studies in this work exactly, and confirm that the divergent regime of optimization is driven by entropic effects.

There are many choices in the exact noise model $\mathcal{N}_{\gamma}$ used, and whether the noise acts globally or locally, both temporally and spatially. We assume that all noise acts independently in time and locally on each qubit in this work.
The total noise channel is chosen to take the form of temporally local noise
\begin{align}
	\Lambda_{\theta\gamma} =&~ \circ_{m}^{M} \Lambda_{\theta\gamma}^{(m)}  \\
	\intertext{where the unitary and non-unitary components are separable}
	\Lambda_{\theta\gamma}^{(m)} =&~ \mathcal{N}_{\gamma}^{(m)} \circ \mathcal{U}_{\theta}^{(m)}~.
\end{align}

For this work, we also consider spatially local noise models to represent the non-unitary components of our channels,
\begin{align}
	\mathcal{N}_{\gamma}^{(m)} =&~ \circ_{i}^{N}\mathcal{N}_{\gamma_{i}}^{(m)}~,
\end{align}
which act identically, and independently, on qubit $i$ at time step $m$, with noise scale $\gamma_{i}^{(m)} = \gamma$. 

We will discuss noise models in terms of the number of errors, or non-identity operations that they apply to the evolution. The temporally (and spatially across qubits) noise models in this work implies that there are
\begin{align}
	K =&~ NM
\end{align}
independent sites of possible errors at each site in time and in space.

As a key assumption for our analysis, we explicitly assume that the local non-unitary components of the channel can be written as a convex combination of an identity channel and a non-trivial (non-identity) error channel
\begin{align}
	\mathcal{N}_{\gamma} \equiv (1-\gamma)\mathcal{I} ~+~ \gamma\mathcal{K}_{\gamma}~.
\end{align}
Here, there is a scale $\gamma$ of a non-trivial operation $\mathcal{K}_{\gamma}$ applied, causing the evolution to deviate from being strictly unitary, resulting in what we refer to as an error. The non-trivial error channel is normalized identically as $\mathcal{N}_{\gamma}$. If the explicit noise model $\mathcal{N}_{\gamma}$ does not strictly contain an identity Kraus operator, for example if it is non-unital, then the non-trivial error is defined as
\begin{align}
	\mathcal{K}_{\gamma} =&~ \frac{1}{\gamma}\mathcal{N}_{\gamma} - \frac{1-\gamma}{\gamma}\mathcal{I}~.
\end{align}

This noise model for the non-unitary component of the channel forms a binomial distribution over $K$ possible errors defined by $\mathcal{K}_{\gamma}$. Errors can occur at spatial and temporal sites $i \in [N]$, and $m \in [M]$, and we use the multi-index ${\chi_{k}} = \{{\chi_{k}}_{i}^{m} \in [2] , k \in [K]\}$ as an indicator function for where the $\abs{{\chi_{k}}} = k \leq K$ errors occur across the sites. Given this binomial distribution description, and using the binomial expansion of the noise scale factors, the channel may be written as a convex combinations of $k$-error, or at most $k$-error channels
\begin{align}
	\Lambda_{\theta\gamma} =&~ \sum_{k}^{K}\tbinom{K}{k}\gamma^{k}(1-\gamma)^{K-k}\Lambda_{\theta\gamma_{k}} \\
	=&~ \sum_{k}^{K}\sum_{l}^{K-k}\tbinom{K}{k}\tbinom{K-k}{l}(-1)^{l}\gamma^{k+l}\Lambda_{\theta\gamma_{k}} \\
	=&~ \sum_{k}^{K}\sum_{l=k}^{K}\tbinom{K}{k}\tbinom{K-k}{l-k}(-1)^{l-k}\gamma^{l}\Lambda_{\theta\gamma_{k}} \\
	=&~ \sum_{k}^{K}\sum_{l}^{k}\tbinom{K}{k}\tbinom{k}{l}(-1)^{k-l}\gamma^{k}\Lambda_{\theta\gamma_{l}} \\
	=&~ \sum_{k}^{K}\tbinom{K}{k}\gamma^{k}\left[\sum_{l}^{k}\tbinom{k}{l}(-1)^{k-l}\Lambda_{\theta\gamma_{l}}\right] \\
	=&~ \sum_{k}^{K} \tbinom{K}{k}\gamma^{k} \Lambda_{\theta\gamma_{\leq k}} \\
	\intertext{We denote trace-preserving channels with $k$ non-trivial errors as the uniform convex combination of all possible locations of the $k \leq K$ errors,}
	\Lambda_{\theta\gamma_{k}} =&~ \frac{1}{\tbinom{K}{k}}\sum_{{\chi_{k}}} \Lambda_{\theta\gamma}^{{\chi_{k}}}~, \\
	\intertext{and we denote traceless operators with at most $k$ non-trivial errors as the non-uniform combination of all possible locations of the $l \leq k$ errors,}
	\Lambda_{\theta\gamma_{\leq k}} =&~ \sum_{l}^{k} (-1)^{k-l}\tbinom{k}{l} \Lambda_{\theta\gamma_{l}}~,
	\intertext{where specific $k$-error channels with $k$ non-trivial errors at indices ${\chi_{k}}$ are denoted as}
	\Lambda_{\theta\gamma}^{{\chi_{k}}} =&~ \circ_{m}^{M} \left[\left[\circ_{i}^{N} {\mathcal{K}_{\gamma_{i}}^{(m)}}^{{\chi_{k}}_{i}^{m}}\right] \circ \mathcal{U}_{\theta}^{(m)}\right]~.
\end{align}
Therefore, to leading-order, the noisy and noiseless channels differ as per the non-triviality $\mathcal{K}_{\gamma_{i}}^{(m)} \neq \mathcal{I}_{i}$ of $K$ possible local errors,
\begin{align}
	\Lambda_{\theta\gamma} - \Lambda_{\theta} =&~ \left(\sum_{m}^{M}\sum_{i}^{N}~\mathcal{U}_{\theta}^{(>m)}\circ{\mathcal{K}_{\gamma_{i}}^{(m)}} - \mathcal{I}_{i}
	\circ \mathcal{U}_{\theta}^{(\leq m)}\right)\gamma ~+~ O(\tbinom{K}{2}\gamma^2)~.
\end{align}

\subsection{Noise Models}\label{subapp:noise_models}
In this work, we consider several noise models $\mathcal{N}_{\gamma} = (1-\gamma)\mathcal{I} ~+~ \gamma\mathcal{K}_{\gamma}$ that are relevant to existing quantum devices, namely independent, local, dephasing, amplitude damping, and depolarizing noise. These noise models all belong to the class of unital or non-unital Pauli noise, that transform local Pauli operators $\mathcal{P}_{D}$. For the case of $D=2$ qubits, $\mathcal{P}_{2} = \{I,Z,X,Y\}$. The forms of the models are identified by their non-trivial error component $\mathcal{K}_{\gamma}$ and are described as follows.

Unital local dephasing noise for all inputs may be written as
\begin{align}
	{\mathcal{K}_{\gamma}}_{}^{\textrm{dephase}} (\cdot) =&~ Z_{} \cdot Z_{}^{\dagger}~,
\end{align}
where the non-trivial channel is unitary.

Unital local depolarizing noise for all inputs may be written as
\begin{align}
	{\mathcal{K}_{\gamma}}_{}^{\textrm{depolarize}}(\cdot) =&~ \frac{\trace{\cdot}}{D} I_{}~,
\end{align}
where the non-trivial channel is maximally depolarizing.

Non-unital amplitude damping noise for Pauli inputs $P_{} \in \mathcal{P}_{2}$ may be written as
\begin{align}
	{\mathcal{K}_{\gamma}}_{}^{\textrm{amplitude}}(P_{}) =&~ \left\{ \begin{array}{cc}
	P_{} + \gamma Z_{} & P_{} \in \{I_{}\}\\
	(1-\gamma) P_{} & P_{} \in \{Z_{}\}\\
	\sqrt{1-\gamma} P_{} & P_{} \in \mathcal{P}_{2}\backslash\{I_{},Z_{}\}
	\end{array}\right.~,
\end{align}
and the non-trivial channel is complicated by the lack of an inherent identity Kraus operator.

In the case of the non-trivial components of the noise channel being single unitary operators $V$ such that $\mathcal{K}_{\gamma}(\cdot) = V \cdot V^{\dagger}$, each $k$-error channel is unitary, and the original unitary channel is interlaced with unitaries $V$ at indices $\chi_{k}$ where errors occur. 

In the case of depolarizing noise where $\mathcal{K}_{\gamma}(\cdot) = (\trace{\cdot}/D)\mathcal{I}$, partial traces remove any information about the state at the local indices in $\chi_{k}$ where errors occur.

\subsection{Probabilistic Interpretation of Noise Channels}\label{subapp:probabilistic_interpretation_of_noise_channels}
Within this formalism, the channel can be represented as an expectation over a distribution of the number of $k \leq K$ possible errors, 
\begin{align}
	\Lambda_{\theta\gamma} =&~ \expval{\Lambda_{\theta\gamma_{k}}}_{k \sim p_{K\gamma}}
\end{align}
where for moments of channels with $k$ non-trivial errors,
\begin{align}
	\Lambda_{\theta\gamma_{k}} =&~ \frac{1}{\tbinom{K}{k}}\sum_{{\chi_{k}}} \Lambda_{\theta\gamma}^{{\chi_{k}}}~.
\end{align}
The exact distribution with this local noise model used in this work is the binomial distribution
\begin{align}
	p_{K\gamma}(k) =&~ p_{K\gamma}^{\textrm{Binomial}}(k) = \tbinom{K}{k}\gamma^{k}(1-\gamma)^{K-k}
\end{align}
which has mean and variance
\begin{align}
	\mu_{K\gamma} = K \gamma \quad &, \quad \Sigma_{K\gamma} = K\gamma(1-\gamma)~.
\end{align}
Such interpretations are also used in error mitigation approaches \cite{Endo2018}.
In the limit $K \to \infty$, this distribution tends to the Gaussian distribution
\begin{align}
	p_{K\gamma}(k) \to&~ p_{K\gamma}^{\textrm{Gaussian}}(k) = \sqrt{\frac{1}{2\pi \Sigma_{K\gamma}}}e^{-\frac{1}{2}(k - \mu_{K\gamma})\Sigma_{K\gamma}^{-1}(k - \mu_{K\gamma})}~,
\end{align}
which can be shown through the De Moivre - Laplace theorem \cite{Fischer2011}, a central-limit version of the standardized binomially distributed variables $\Sigma_{K\gamma}^{-1/2}(k-\mu_{K\gamma})$.

In the limit $K \to \infty, \gamma \to 0$, and the finite limit $K \gamma \to \lambda_{K\gamma}$, the distribution tends to the Poisson distribution
\begin{align}
	p_{K\gamma}(k) \to&~ p_{K\gamma}^{\textrm{Poisson}}(k) = \frac{1}{k!}(K \gamma)^{k} e^{-K \gamma}~,
\end{align}
which can be shown from using Sterling's approximation for the binomial coefficient, and equating the generating functions for the distributions.

\subsection{Noisy State Preparation}\label{subapp:noisy_state_preparation}
\begin{figure}[h]
	\centering
	\begin{subfigure}[t]{0.49\columnwidth}
		\captionsetup{singlelinecheck = false, justification=raggedright,margin={0pt,0pt},skip=0pt}
		\subcaption{}
		\captionsetup{singlelinecheck = false, justification=raggedright}
		\includegraphics[width=\columnwidth]{figures/plot.noise.parameters.objective.M.dephase.pdf}
		\label{fig:noises_noise_dephasing}
	\end{subfigure}
	\hfill
	\begin{subfigure}[t]{0.49\columnwidth}
		\captionsetup{singlelinecheck = false, justification=raggedright,margin={0pt,0pt},skip=0pt}
		\subcaption{}
		\includegraphics[width=1.02\textwidth]{figures/plot.M.objective.noise.parameters.dephase.pdf}
		\label{fig:noises_M_dephasing}
	\end{subfigure}
	\vfill
	\begin{subfigure}[t]{0.49\columnwidth}
		\captionsetup{singlelinecheck = false, justification=raggedright,margin={0pt,0pt},skip=0pt}
		\subcaption{}
		\includegraphics[width=\columnwidth]{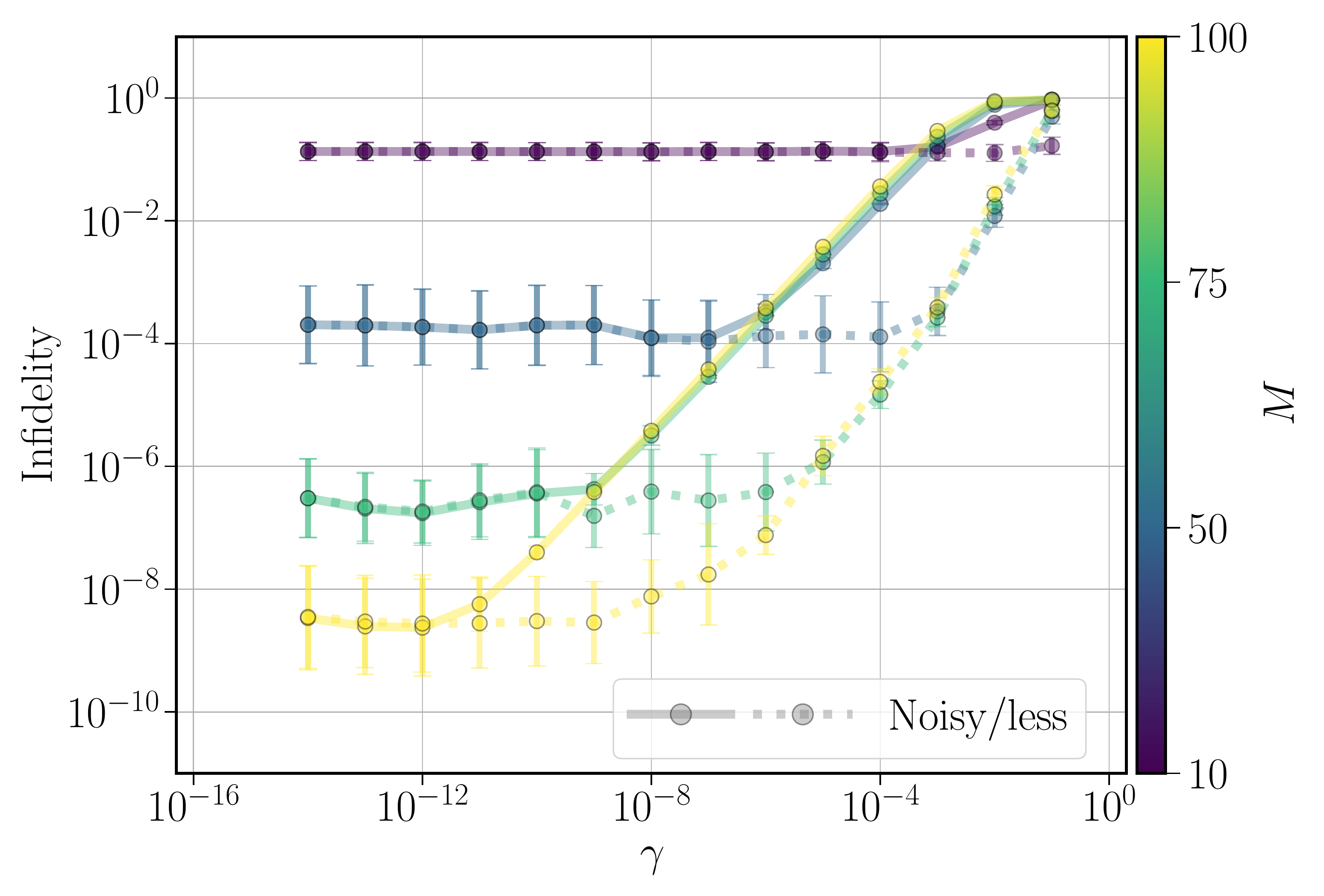}
		\label{fig:noises_noisy_depolarizing}
	\end{subfigure}
	\hfill
	\begin{subfigure}[t]{0.49\columnwidth}
		\captionsetup{singlelinecheck = false, justification=raggedright,margin={0pt,0pt},skip=0pt}
		\subcaption{}
		\includegraphics[width=1.02\textwidth]{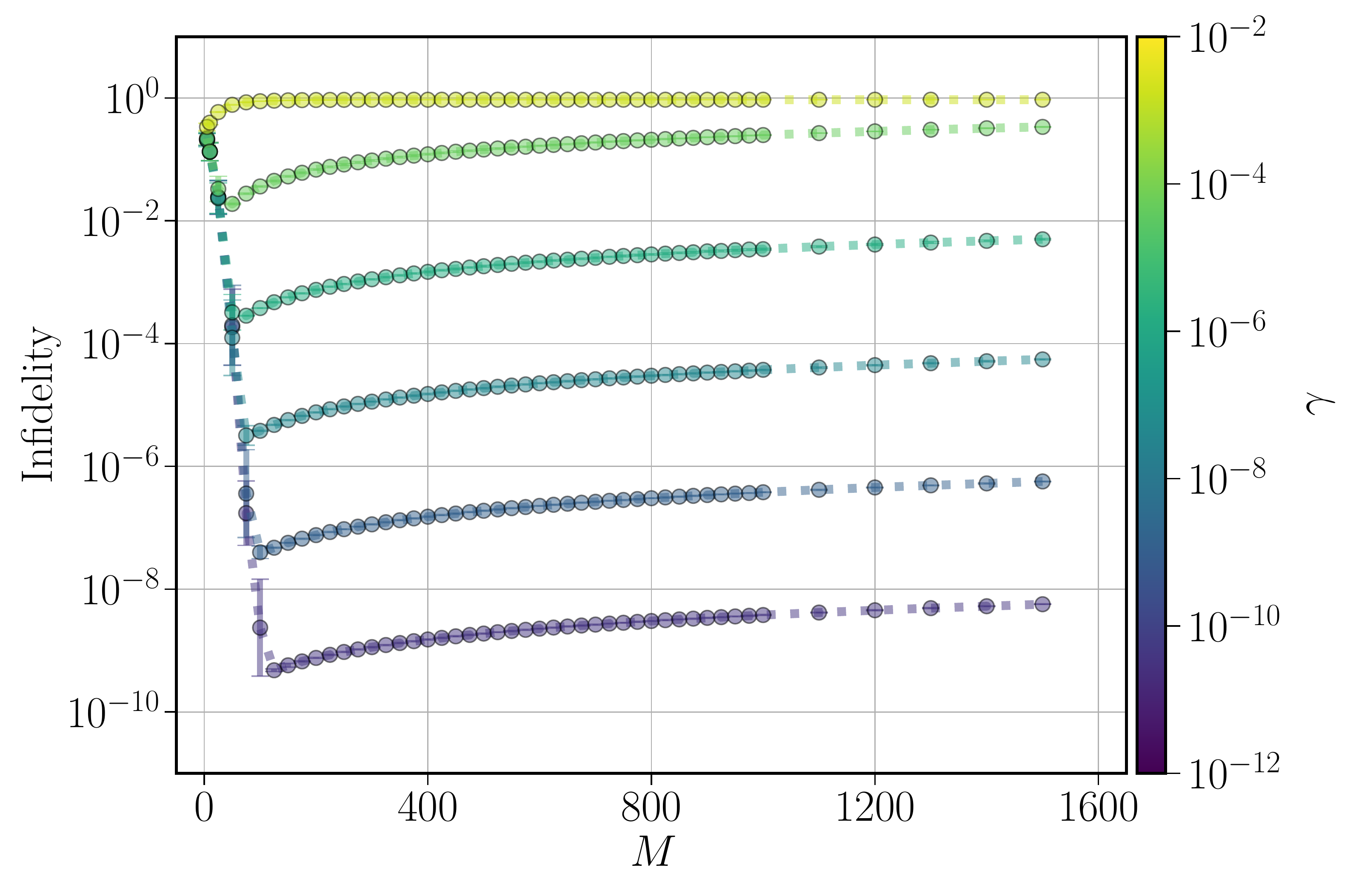}
		\label{fig:noises_M_depolarizing}
	\end{subfigure}
	\vfill
	\begin{subfigure}[t]{0.49\columnwidth}
		\captionsetup{singlelinecheck = false, justification=raggedright,margin={0pt,0pt},skip=0pt}
		\subcaption{}
		\includegraphics[width=\columnwidth]{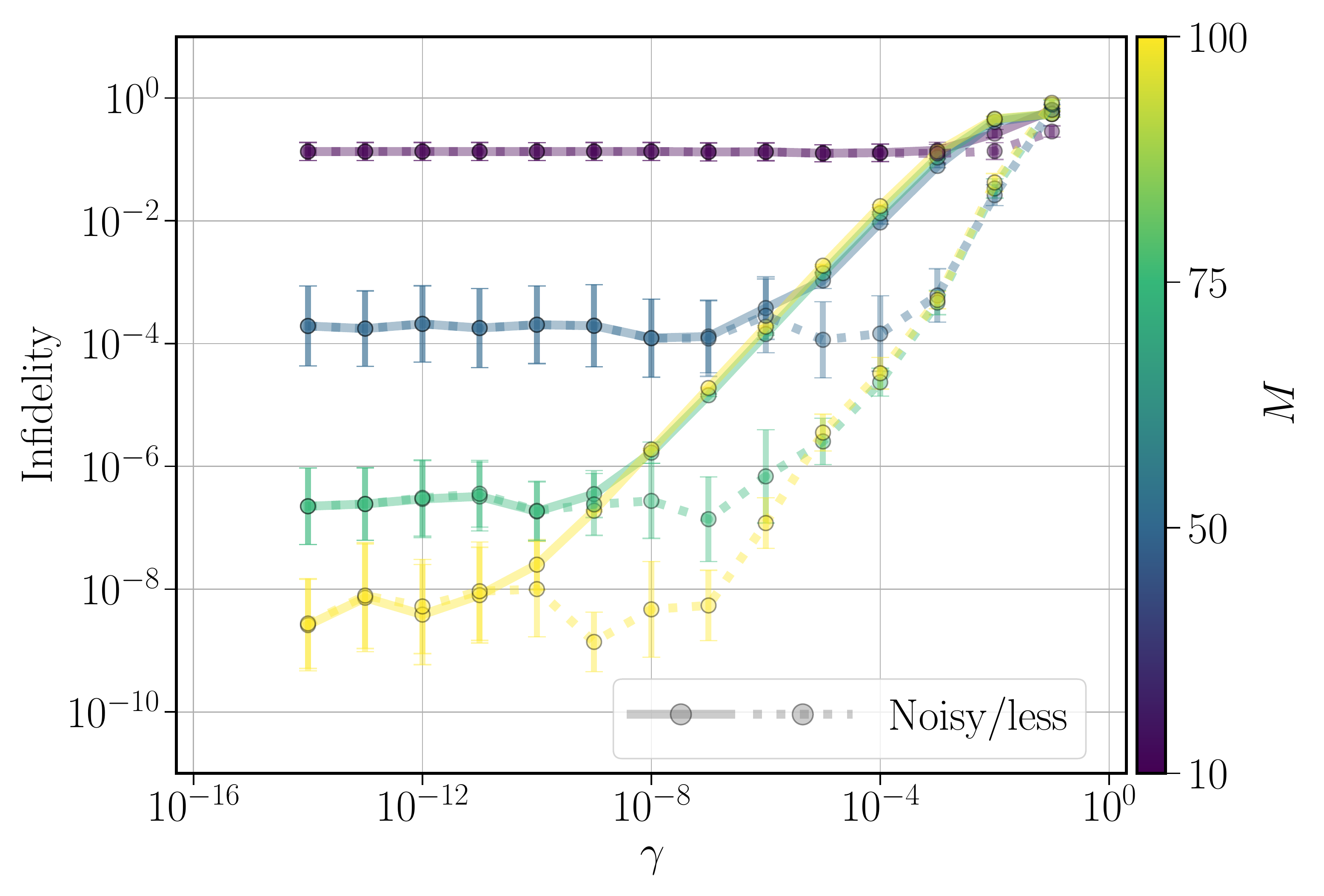}
		\label{fig:noises_noisy_amplitude}
	\end{subfigure}
	\hfill
	\begin{subfigure}[t]{0.49\columnwidth}
		\captionsetup{singlelinecheck = false, justification=raggedright,margin={0pt,0pt},skip=0pt}
		\subcaption{}
		\includegraphics[width=1.02\textwidth]{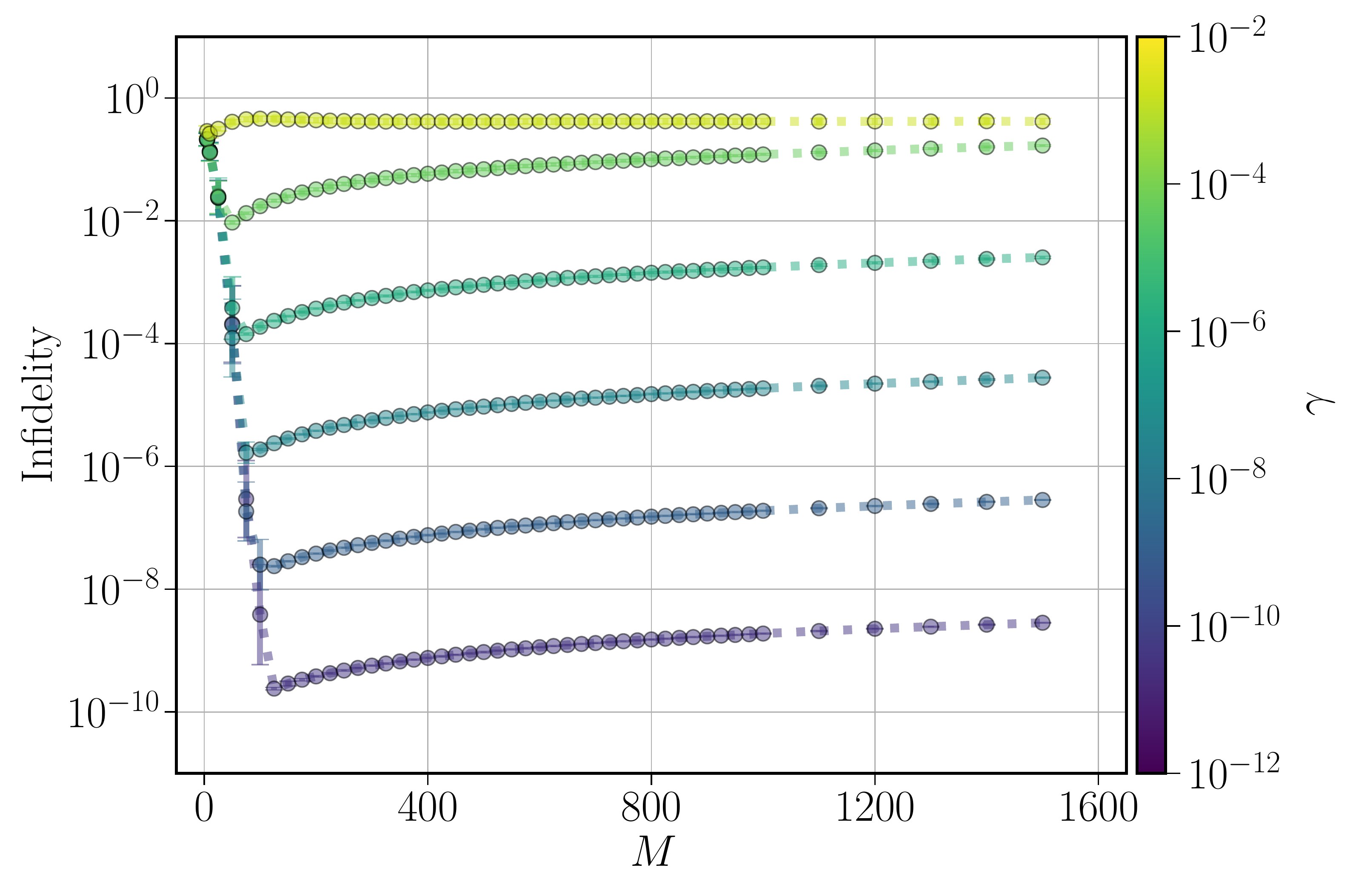}
		\label{fig:noises_M_amplitude}
	\end{subfigure}
	\vfill	
	\captionsetup{justification=raggedright}	
	\caption{behavior of infidelity objectives with respect to noise and depth for the $N=4$ NMR ansatz. (a),(b) Dephasing, (c),(d) Depolarizing, (e),(f) Amplitude Damping. Left)  Trained noisy infidelity (solid), and tested infidelity of noisy parameters in noiseless ansatz (dashed) for various depths $M$ (colored/gradient). Infidelities of noisy ansatz are shown to be depth dependent and noise independent for small noise scales, before universally increasing polynomially with noise. Inserting parameters learned in the noisy setting into an identical noiseless ansatz indicates that the underlying unitary evolution is being learned, and is resilient to noise. Right) Critical depth for noisy infidelity for various noise scales $\gamma$ (colored/gradient). Infidelities are shown to improve exponentially with depth, up until a noise-induced critical depth, where entropic effects worsen infidelities polynomially with depth. Non-unital noise is also shown to be less dominated be entropic effects at large noise scales.}
	\label{fig:noises}	
\end{figure}

We conduct studies in \cref{fig:noises} of the noise and depth dependence on the optimal infidelities for each of the unital dephasing and depolarizing, and non-unital amplitude damping noise models. For unital noise, we observe that the critical depth phenomena seem to be consistent across noise models at all noise scales, and entropic effects dominate. For non-unital noise, we observe that the critical depth phenomena seem to be consistent with unital noise models, at all small noise scales. However, for large noise scales, non-unital noise forces the state into a specific (pure) state, which dominates over entropic effects. The infidelities then appear to converge to non-unity values or potentially decrease polynomially to zero. The unitary component of the channel appears to be able to slowly rotate this noise-induced pure state, towards the correct target pure state. In both unital and non-unital noise models, inserting parameters learned in the noisy setting, into the corresponding noiseless unitary ansatz indicates that the noiseless infidelities have greater resilience to increasing noise scales. This suggests the underlying unitary transformations are being learned in the noisy settings, however with greater variance.

\subsection{Noise-induced Critical Depth}\label{subapp:noise_induced_critical_depth}
\begin{figure}[h]
	\centering
	\begin{subfigure}[t]{0.49\textwidth}
		\captionsetup{singlelinecheck = false, justification=raggedright,margin={0pt,0pt},skip=0pt}
		\subcaption{}
		\includegraphics[width=\columnwidth]{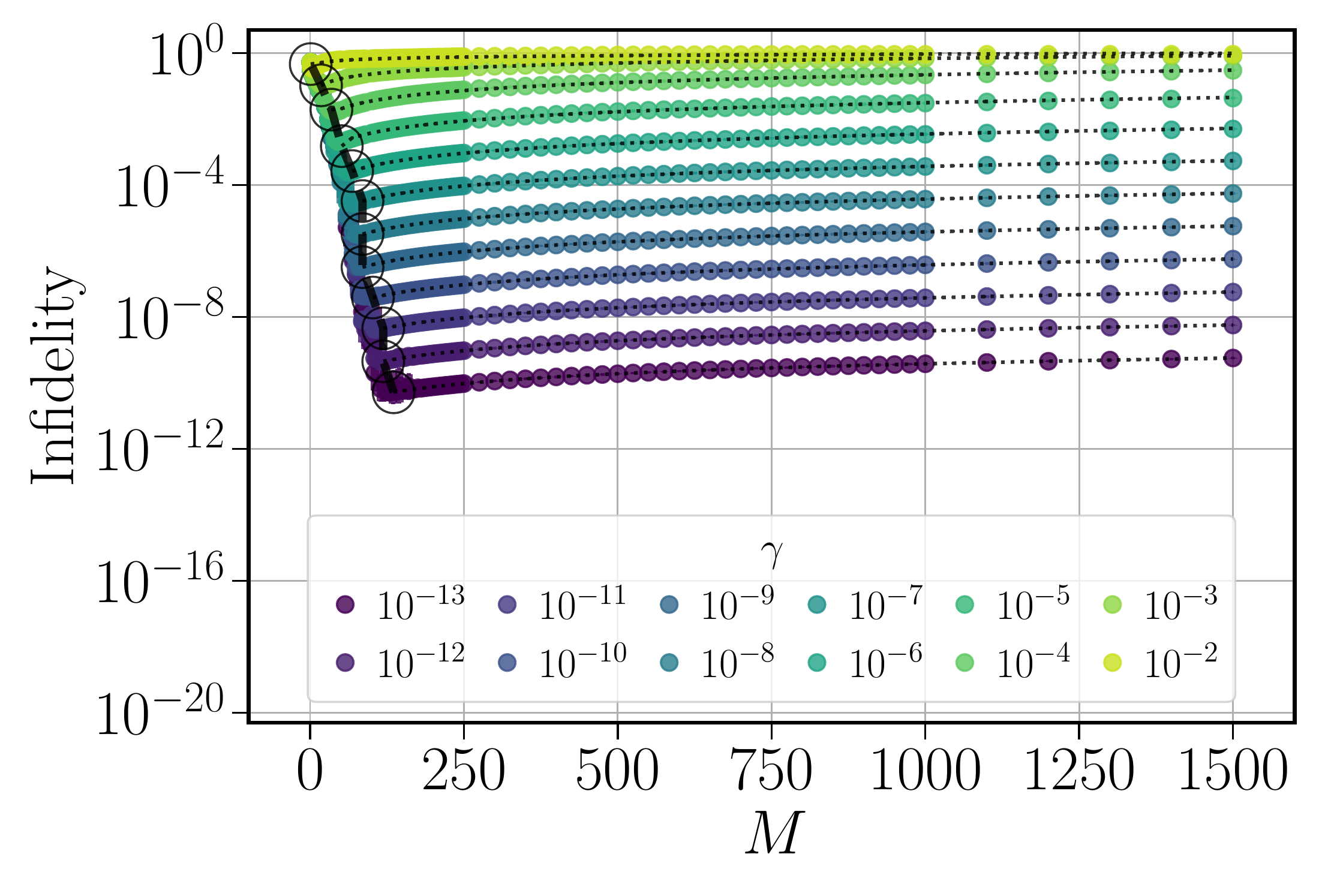}
		\label{fig:critical_depth_data}
	\end{subfigure} 
	\hfill
	\begin{subfigure}[t]{0.48\textwidth}
		\captionsetup{singlelinecheck = false, justification=raggedright,margin={0pt,0pt},skip=0pt}
		\subcaption{}
		\includegraphics[width=\columnwidth]{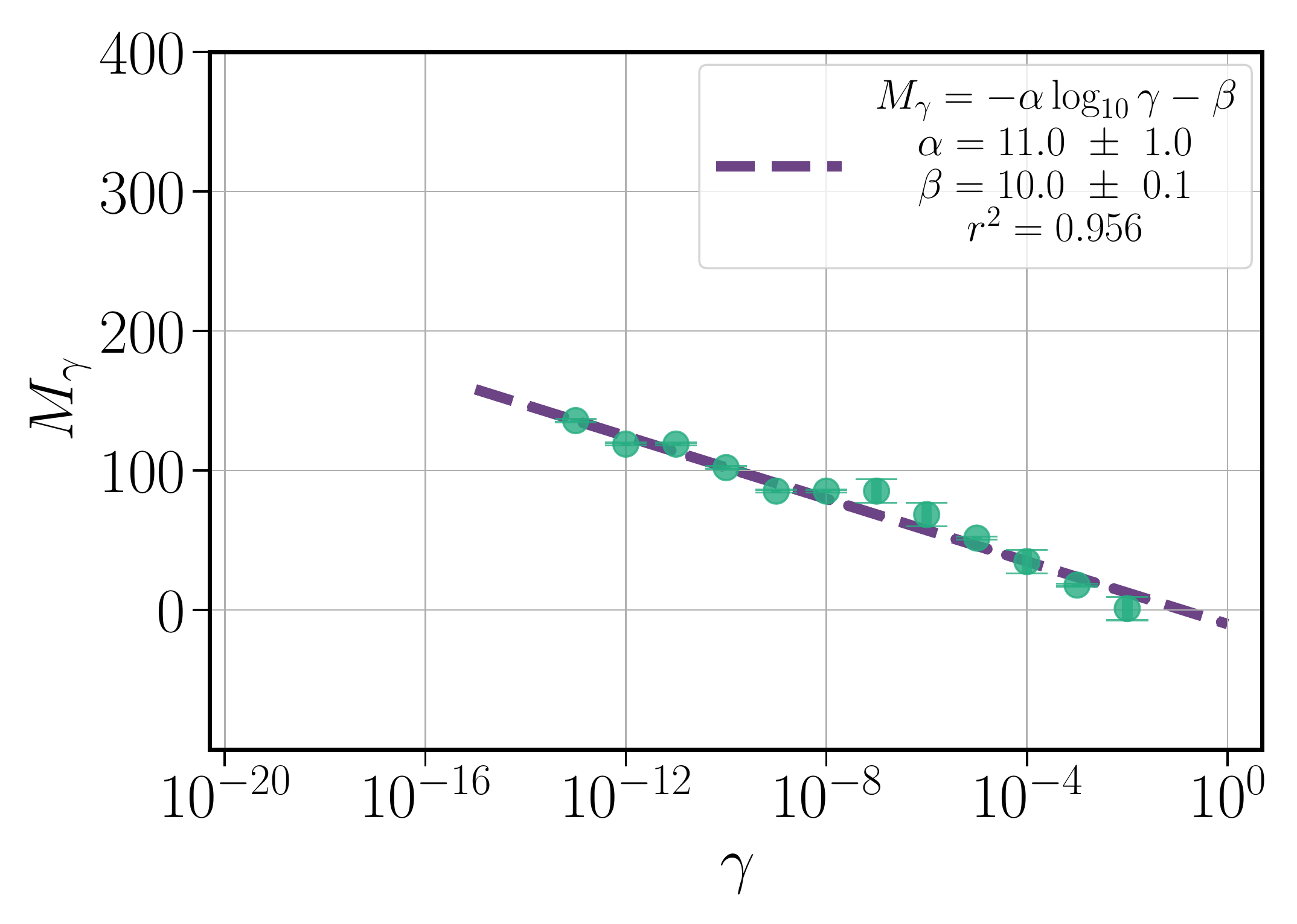}
		\label{fig:critical_depth_fit}
	\end{subfigure} 
	\captionsetup{justification=raggedright}	
	\caption{Fit of noise-induced critical depth. (a) Piecewise fits for various noise scales $\gamma$ (colored/gradient). Objectives decrease exponentially with depth, for depths less than the critical depth (black circles), and objectives increase polynomially with depth for depths greater than the critical depth. (b) Uncertainty propagation is used to estimate the error of the linear-log fit (dashed line) between the estimated critical depths where the piecewise curves intersect, and the associated noise scales.}
	\label{fig:critical_depth}		
\end{figure}

To develop a relationship between noise and an induced critical depth beyond which infidelities no longer converge exponentially, we perform piecewise fits, as per \cref{fig:critical_depth}. For each noise scale, we perform exponential, and then polynomial fits, respectively, before and after an approximate location of the critical depth suggested by the finite amount of data points available. We are then able to approximate where the piece wise curves intersect, indicating the location of the critical depth as a function of noise. Plotting this relationship suggests a logarithmic relationship between noise and critical depth. This leads to the optimal infidelity scaling approximately polynomially, between linearly and quadratically with noise, and in agreement with previous conjectures about these relationships \cite{Fontana2021}.

\subsection{Parameter Shift Rules for Quantum Channels}\label{subapp:parameter_shift_rules_for_quantum_channels}
Here, we derive parameter shift rules for the gradients of multiple layer quantum channels. This generalizes previous results, and provides some explanation for why the optimization routines in noisy settings generally appear to converge to the noiseless optima at small noise scales. Given our composite channel definition over $M$ layers, written decomposed into channels before and after an index $(\mu,m)$, our state preparation objectives of the trace infidelity with respect to a pure state $\rho$, and initial state $\sigma$ may be written as
\begin{align}
	\mathcal{L}_{\theta\gamma} =&~ 1 - \trace{\mathcal{U}_{\mu}^{(m)}\left(\sigma_{<\mu\gamma}^{(<m)}\right) \mathcal{U}_{>\mu}^{(m)\dagger}\left(\mathcal{N}_{\gamma}^{(m)\dagger}\left(\rho_{>\mu\gamma}^{(>m)}\right)\right)}~.
\end{align}
These objectives are crucially linear in the states, and therefore gradients of the objectives with respect to parameters at the index $(\mu,m)$ are linear functions of the gradients of the state. 

Let $\mathcal{U}_{\theta}$ be a single parameter unitary channel, with the corresponding unitary operator $U_{\theta} = e^{-i\theta G}$, where the hermitian generator is involutory up to a factor $G^{2} = \zeta^2 I$. The gradient of such a unitary channel for an arbitrary input follows the parameter shift rule
\begin{align}
	\partial_{\theta} \mathcal{U}_{\theta} =&~ -i\comm{G}{\mathcal{U}_{\theta}} = \zeta(\mathcal{U}_{\theta+\varphi} - \mathcal{U}_{\theta-\varphi})
\end{align}
where we denote $\varphi = \pi/4\zeta$, and other choices of $\varphi$ are also possible depending on experimental feasibility. 
This can be extended to the $k$-order gradient of this channel as the $k$-order nested commutator
\begin{align}
	\partial_{\theta}^{k} \mathcal{U}_{\theta} =&~ (-i)^{k}\comm{G}{\mathcal{U}_{\theta}}_{k} = \zeta^{k}\sum_{l}^{k}(-1)^{l}\tbinom{k}{l}\mathcal{U}_{\theta+(k-2l)\varphi}~.
\end{align}
Unitary channels with generators with more complicated spectra can be expressed as a linear combination of unitary channels, with perturbative angles $\varphi$, weighted by coefficients $\alpha_{\varphi}$
\begin{align}
	\partial_{\theta}^{k} \mathcal{U}_{\theta} =&~ \sum_{\varphi}\alpha_{\varphi}\mathcal{U}_{\theta+\varphi}~.
\end{align}
Thus gradients of general parameterized channels for constant noise scales $\gamma$ have the form
\begin{align}
	\partial_{\mu}\Lambda_{\theta\gamma}=&~ \sum_{\varphi}\alpha_{\varphi_{\mu}}\Lambda_{\theta+\varphi_{\mu}~\gamma}~,
\end{align}
and similarly any linear objectives such as $\partial_{\mu}^{(m)}\mathcal{L}_{\theta\gamma}$ have an identical form.

\subsection{Relationship between Infidelity, Impurity and von-Neumann Entropy}\label{subapp:relationship_between_fidelity_impurity_and_von_neumann_entropy}
To better understand the phenomena dictating the behavior of infidelities with respect to noise scales and depth, we also investigate the scaling of the impurity and entropy of the noisy parameterized states. For completeness, we restate the impurity, von-Neumann entropy, conditional entropy, and relative entropy divergence, between a $d$-dimensional state $\rho$ and a pure state $\rho^{\prime}$, as
\begin{align}
	\mathcal{L}_{\rho}^{\prime} =&~1- \trace{\rho^{\prime}\rho} \\
	\mathcal{I}_{\rho} =&~1- \trace{\rho^{2}} \\
	\mathcal{S}_{\rho} =&~- \trace{\rho\log{\rho}}/\log{d} \\
	\mathcal{S}_{\rho}^{\rho^{\prime}} =&~- \trace{\rho^{\prime}\log{\rho}}/\log{d} \\
	\mathcal{D}_{\rho}^{\rho^{\prime}} =&~ \mathcal{S}_{\rho}^{\rho^{\prime}} - \mathcal{S}_{\rho^{\prime}}^{} ~.
\end{align} 
We also note that when we use the definitions of parameterized states $\rho_{\theta\gamma}$ and target reference states $\rho$ in place of $\rho$ and $\rho^{\prime}$, we instead use the subscripts $\theta\gamma$ and superscripts $\rho$ for these quantities.

For this analysis, we use the Bloch representation \cite{Crowder2015}, which describes operators in terms of a trace orthogonal basis $\mathcal{P}_{d}$, such that $\trace{\alpha^{\dagger}\beta} = d\delta_{\alpha\beta}$ for all operators $\alpha,\beta \in \mathcal{P}_{d}$. This basis contains the identity, is of size $\abs{\mathcal{P}_{d}}=d^{2}$, and may represent an algebra, with structure constants defining their commutation relations. Let $\omega = \{P ~:~ P \in \mathcal{P}_{d}\backslash\{I\}\}$ represent the vector of all non-identity operators such that quantum states with unit traces have the form
\begin{align}
	\rho =&~ \frac{I + \lambda \cdot \omega}{d}~.
\end{align}
Here the $\abs{\mathcal{P}_{d}}-1$ Bloch coefficients $\lambda$ fully describe quantum states. These coefficients are constrained to represent positive-semidefinite operators, and their magnitude is bounded by the pure-state boundary described by
\begin{align}
	\lambda^{2} \leq&~ d-1~.
\end{align}

The action of arbitrary trace-preserving channels on quantum states can further be described by the affine linear transformation on the coefficients,
\begin{align}
	\Lambda~:~ \lambda \to \Gamma\lambda + \upsilon~.
\end{align}
This affine linear transformation represents a rotation and scaling of $\lambda$ by $\Gamma$, plus a translation to a different axis by $\upsilon$, and it can be thought of as a transformation on the generalized $d$-dimensional Bloch sphere of radius $d-1$. The primary constraints on the transformations are to remain within this boundary such that $(\Gamma\lambda + \upsilon)^2 \leq d-1$. 

If the channel is unital, then the transformation is strictly linear and $\upsilon = 0$, otherwise the channel is non-unital. Finally, if the channel is unitary, $\Gamma = u$ is an orthogonal transformation that preserves the length of $\lambda^{2}$.

We also note a useful identity when $\lambda$ is associated with a pure state with zero entropy, given the expansion for the matrix logarithm
\begin{align}
	\log{I + \lambda \cdot \omega} =&~ \sum_{k>0}\frac{(-1)^{k+1}}{k} (\lambda \cdot \omega)^{k} 
\end{align}
and the tracelessness of $\lambda \cdot \omega$, meaning
\begin{align}
	\frac{1}{d\log{d}}\sum_{k>1}\frac{(-1)^{k}}{k(k-1)} \trace{(\lambda \cdot \omega)^{k}} =&~ 1~.
\end{align}
This identity avoids complicated expressions for powers of Bloch vectors \cite{Kimura2003}.

We may now define quantities in terms of the Bloch coefficients. Similarity between quantum states $\rho,\rho^{\prime}$ can be described either by their trace inner products
\begin{align}
	\trace{\rho\rho^{\prime}} =&~ \frac{1 + \lambda \cdot \lambda^{\prime}}{d} = 1 - \mathcal{L}_{\rho}^{\rho^{\prime}}~,
\end{align}
or by their cosine similarity with an angle $\phi_{\rho}^{\rho^{\prime}}$ between their coefficients
\begin{align}
	\cos{\phi_{\rho}^{\rho^{\prime}}} =&~ \frac{\lambda \cdot \lambda^{\prime}}{\sqrt{\lambda^{2}\lambda^{\prime2}}} = \frac{d(1 - \mathcal{L}_{\rho}^{\rho^{\prime}}) - 1}{\sqrt{(d(1 - \mathcal{I}_{\rho}^{}) - 1)(d(1 - \mathcal{I}_{\rho^{\prime}}^{}) - 1)}}~.
\end{align}
Similarly for other functions of interest with respect to a pure state $\rho^{\prime}$,
\begin{align}
	\mathcal{L}_{\rho}^{\rho^{\prime}} =&~ 1 - \frac{1 + \lambda \cdot \lambda^{\prime}}{d} \\
	\mathcal{I}_{\rho} =&~ 1 - \frac{1 + \lambda^{2}}{d} \\
	\mathcal{S}_{\rho} =&~ 1 - \frac{1}{d\log{d}}\sum_{k>1}\frac{(-1)^{k}}{k(k-1)} \trace{(\lambda \cdot \omega)^{k}} = 1 - \frac{1}{2}\frac{1}{\log{d}} \lambda^{2} ~+~ O(\lambda^{3})\\
	\mathcal{D}_{\rho}^{\rho^{\prime}} =&~ 1 - \frac{1}{d\log{d}}\sum_{k>0}\frac{(-1)^{k+1}}{k} \trace{(I + \lambda^{\prime} \cdot \omega)(\lambda \cdot \omega)^{k}} = 1 - \frac{1}{2}\frac{1}{\log{d}}\left(2\frac{\lambda \cdot \lambda^{\prime}}{\lambda^{2}} - 1\right)\lambda^{2} ~+~ O(\lambda^3)~.
\end{align}
We can also define the generalized $\alpha$-Renyi infidelities, impurities, entropies, and divergences for $\alpha \geq 1/2$,
\begin{align}
	\mathcal{L}_{\rho}^{\rho^{\prime}(\alpha)} =&~ \trace{(\rho^{\prime\frac{1-\alpha}{2\alpha}}\rho\rho^{\prime\frac{1-\alpha}{2\alpha}})^{\alpha}} \\
	\mathcal{I}_{\rho}^{(\alpha)} =&~ 1 - \trace{\rho^{\alpha}} \\
	\mathcal{S}_{\rho}^{(\alpha)} =&~ \frac{1}{\alpha-1}\frac{1}{\log{d}}\log{1-\mathcal{I}_{\rho}^{(\alpha)}} \\
	\mathcal{D}_{\rho}^{\rho^{\prime}(\alpha)} =&~ \frac{1}{\alpha-1}\frac{1}{\log{d}}\log{1-\mathcal{L}_{\rho}^{\rho^{\prime}(\alpha)}}~.
\end{align}
Using these definitions we can show \cite{Nuradha2023} that the entropies and divergences are monotonically increasing with decreasing $\alpha$, and we can identify $\mathcal{D}_{\rho}^{\rho^{\prime}}$ with $\alpha \to 1$, and related divergences and infidelities with $\mathcal{D}_{\rho}^{\rho^{\prime}(1/2)} = -2\log{1-\mathcal{L}_{\rho}^{\rho^{\prime}}}/\log{d}$. Based on these relationships, we have the bounds relating the conditional entropy and the infidelity
\begin{align}
	\mathcal{S}_{\rho}^{\rho^{\prime}} \geq \frac{2}{\log{d}} \mathcal{L}_{\rho}^{\rho^{\prime}}~,
\end{align}
however we have not found any known similar bounds relating entropy and infidelity or impurity.

Given our previous expressions for channels consisting of layers of non-unitary noise with scale $\gamma$, interlaced by unitary channels with parameters $\theta$, resulting in a binomial distribution over states with at most $K$ errors, we may represent such parameterized quantum channels in this formalism as
\begin{align}
	\Lambda_{\theta\gamma} ~:~ \Gamma_{\theta\gamma} = (1-\gamma_{K})u_{\theta} + \gamma_{K}u_{\theta\gamma} \quad,\quad \upsilon_{\theta\gamma} = \gamma_{K}\eta_{\theta\gamma}~.
\end{align}
Here, we have explicitly separated the noiseless unitary rotation from the other non-unitary and non-unital transformations, and we defined the error-dependent noise scale as
\begin{align}
	(1-\gamma_{K}) =&~ (1-\gamma)^{K}~.
\end{align}
Therefore the transformed coefficients, given an initial state $\sigma$ with coefficients $\xi$, are transformed by unitary and noise components of the transformation
\begin{align}
	\lambda_{\theta\gamma} =&~ (1-\gamma_{K})u_{\theta}\xi + \gamma_{K}u_{\theta\gamma}\xi + \gamma_{K}\eta_{\theta\gamma} \\
	=&~ (1-\gamma_{K})\lambda_{\theta|\gamma} + \gamma_{K}\varepsilon_{\theta\gamma}~.
\end{align}
Here, we decompose the transformed state into what we refer to as the pure and mixed components. The pure component of the state may be written in terms of a reference pure state $\rho$ with coefficients $\lambda$, and a pure state with orthogonal coefficients $\zeta \perp \lambda$ as 
\begin{align}
	\lambda_{\theta|\gamma} =&~ u_{\theta}\xi = (1-\alpha_{\theta|\gamma})\lambda + \beta_{\theta|\gamma}\zeta~,
\end{align}
where the coefficients implicitly contain a dependence on the noise from optimization, and they are constrained such that $(1-\alpha_{\theta|\gamma})^{2} + \beta_{\theta|\gamma}^{2} = 1$. The mixed components of the state can be written in terms of its unital and non-unital components,
\begin{align}
	\varepsilon_{\theta\gamma} =&~ u_{\theta\gamma}\xi + \eta_{\theta\gamma}~.
\end{align}
This decomposition allows us to understand how the unitary and noise components transform the state. The unitary component, independent of the noise, rotates the state into the pure components parallel and perpendicular to $\rho$. The noise component then scales both the pure and mixed components with $\gamma_{K}$, and performs an affine translation with $\eta_{\theta\gamma}$. In the limit that $\gamma \to 0$, and assuming converged optimization to the optimal $\theta = \theta^{*}$ such that $\rho_{\theta^{*}} \to \rho$, the coefficients should then reduce to $\alpha_{\theta^{*}},\beta_{\theta^{*}} \to 0$, and the non-unital affine translation should also vanish, $\eta_{\theta} \to 0$.

Due to the non-trivial optimization, we do not have a general closed-form expression depicting the $\theta,\gamma,K$ dependence of the pure state components $\alpha_{\theta|\gamma}, \beta_{\theta|\gamma}$ or the mixed component $\varepsilon_{\theta\gamma}$. However, by expanding out the quantities of interest in terms of these expressions for the coefficients $\lambda_{\theta\gamma}$, we can obtain the leading-order scaling of quantities in terms of $K,\gamma$.

The inner products of the coefficients are
\begin{align}
	\lambda_{\theta\gamma}^{2} =&~ (1-\gamma_{K})^{2}\lambda^{2} + \gamma_{K}^{2}\varepsilon_{\theta\gamma}^{2} + 2\gamma_{K}(1-\gamma_{K})^{2}((1-\alpha_{\theta|\gamma})\lambda + \beta_{\theta|\gamma}\zeta) \cdot \varepsilon_{\theta\gamma} \\
	=&~ \lambda^{2} - 2K\gamma \left(1 -  (1-\alpha_{\theta|\gamma})\frac{\lambda \cdot \varepsilon_{\theta\gamma}}{\lambda^{2}} - \beta_{\theta|\gamma}\frac{\zeta \cdot \varepsilon_{\theta\gamma}}{\lambda^{2}}\right)\lambda^{2} ~+~ O(\tbinom{K}{2}\gamma^2)\\
	\lambda_{\theta\gamma}\cdot \lambda =&~ (1-\gamma_{K})(1-\alpha_{\theta|\gamma})\lambda^{2} + \gamma_{K}\lambda \cdot \varepsilon_{\theta\gamma}\\
	=&~ \lambda^{2} - \alpha_{\theta|\gamma}\lambda^{2} - K\gamma \left(1-\alpha_{\theta|\gamma} - \frac{\lambda \cdot \varepsilon_{\theta\gamma}}{\lambda^{2}}\right)\lambda^{2} ~+~ O(\tbinom{K}{2}\gamma^2)~.	
\end{align}
Powers of Bloch vectors in terms of the target coefficients, even when $\alpha_{\theta|\gamma}=\beta_{\theta|\gamma} = 0$, scales with $\gamma$ as
\begin{align}
	(\lambda_{\theta\gamma} \cdot \omega)^{k} =&~ (1-\gamma_{K})^{k}(\lambda \cdot \omega)^{k} ~+~ O(\gamma)~.
\end{align}
All powers $k$ of Bloch vectors, such as those that occur in infinite series expansions for logarithmic functions appearing in entropies, therefore contribute leading-order terms in $K,\gamma$. It does not then suffice to retain only some $O(\lambda^{k})$-order terms in expansions of such functions, to capture the coefficients of their leading-order behavior with $\gamma$. Obtaining these coefficients to all orders, for all dimensions $d$, is possible, however it requires opaque, recursive expressions for products $\trace{(\lambda \cdot \omega)^{k}}$ in terms of the structure constants, such as those found for the Pauli basis by Sarkar \cite{Sarkar1971}.

Therefore, to leading-order in $K,\gamma$, the quantities of interest show similar scaling with overlaps of the parameterized state with the target pure state,
\begin{align}
	\mathcal{L}_{\theta\gamma}^{\rho} =&~ \frac{d-1}{d}\alpha_{\theta|\gamma} ~+~  K\gamma \frac{d-1}{d}\left( (1-\alpha_{\theta|\gamma}) - \frac{\lambda \cdot \varepsilon_{\theta\gamma}}{\lambda^2}\right) ~+~ O(\tbinom{K}{2}\gamma^2)\\
	\mathcal{I}_{\theta\gamma} =&~ 2K\gamma \frac{d-1}{d}\left(1 -  (1-\alpha_{\theta|\gamma})\frac{\lambda \cdot \varepsilon_{\theta\gamma}}{\lambda^{2}} - \beta_{\theta|\gamma}\frac{\zeta \cdot \varepsilon_{\theta\gamma}}{\lambda^{2}}\right) ~+~ O(\tbinom{K}{2}\gamma^2)\\
	\mathcal{S}_{\theta\gamma} =&~  K\gamma\frac{d-1}{\log{d}} \left(1 -  (1-\alpha_{\theta|\gamma})\frac{\lambda \cdot \varepsilon_{\theta\gamma}}{\lambda^{2}} - \beta_{\theta|\gamma}\frac{\zeta \cdot \varepsilon_{\theta\gamma}}{\lambda^{2}}\right) ~+~ O(\lambda_{\theta\gamma}^{3}) ~\sim~ O(K\gamma)\\
	\mathcal{D}_{\theta\gamma}^{\rho} =&~ \frac{d-1}{\log{d}}\alpha_{\theta|\gamma} ~+~ K\gamma \frac{d-1}{\log{d}}\left( \alpha_{\theta|\gamma}(1 + \frac{\lambda \cdot \varepsilon_{\theta\gamma}}{\lambda^2}) - \beta_{\theta|\gamma}\frac{\zeta \cdot \varepsilon_{\theta\gamma}}{\lambda^2}\right) ~+~ O(\lambda_{\theta\gamma}^{3}) ~\sim~ O(K\gamma)~.
\end{align}

If the unitary component of the channel transforms the pure component of the state to exactly the target state such that $\alpha_{\theta|\gamma}=\beta_{\theta|\gamma} = 0$, then expressions simplify further. A hierarchy between quantities can be partially observed to leading-order in terms of the overlap $\lambda \cdot \varepsilon_{\theta\gamma}$ between the mixed component of the state and the target state
\begin{align}
	\mathcal{L}_{\theta\gamma}^{\rho} =&~ K\gamma \frac{d-1}{d}\left( 1 - \frac{\lambda \cdot \varepsilon_{\theta\gamma}}{\lambda^2}\right) ~+~ O(\tbinom{K}{2}\gamma^2)\\
	\mathcal{I}_{\theta\gamma} =&~ 2K\gamma \frac{d-1}{d}\left(1 -  \frac{\lambda \cdot \varepsilon_{\theta\gamma}}{\lambda^{2}}\right) ~+~ O(\tbinom{K}{2}\gamma^2)\\
	\mathcal{S}_{\theta\gamma} =&~  O(K\gamma)\\
	\mathcal{D}_{\theta\gamma}^{\rho} =&~ O(K\gamma)~.
\end{align}

\begin{figure}[h]
	\centering
	\begin{subfigure}[t]{0.49\columnwidth}
		\centering
		\captionsetup{singlelinecheck = false, justification=raggedright,margin={0pt,0pt},skip=0pt}
		\subcaption{}
		\includegraphics[width=\textwidth]{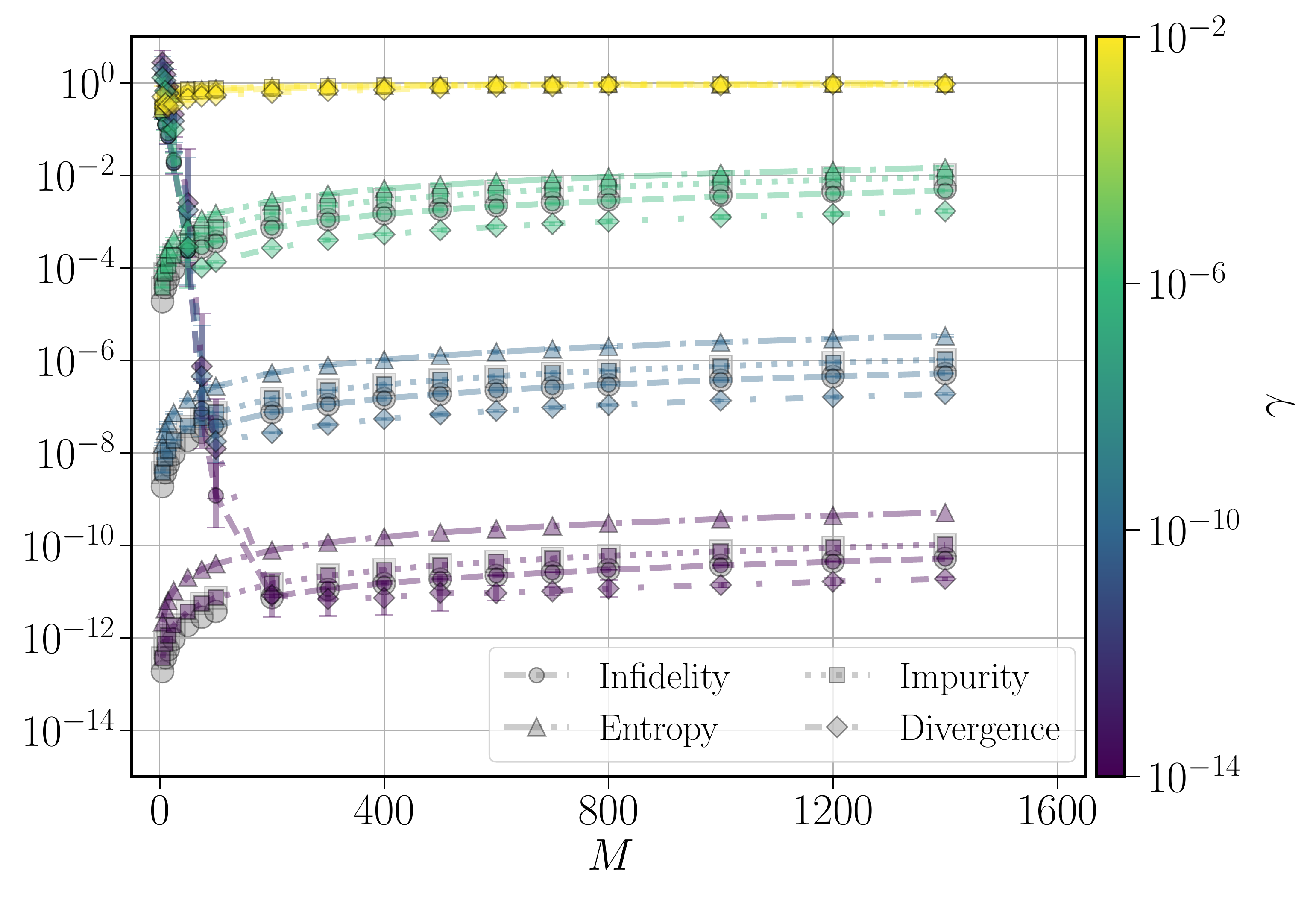}
		\label{fig:objective_purity_entropy_divergence_similarity_objective_purity_entropy_divergence}
	\end{subfigure}
	\hfill	
	\begin{subfigure}[t]{0.49\columnwidth}
		\centering
		\captionsetup{singlelinecheck = false, justification=raggedright,margin={0pt,0pt},skip=0pt}
		\subcaption{}
		\includegraphics[width=\textwidth]{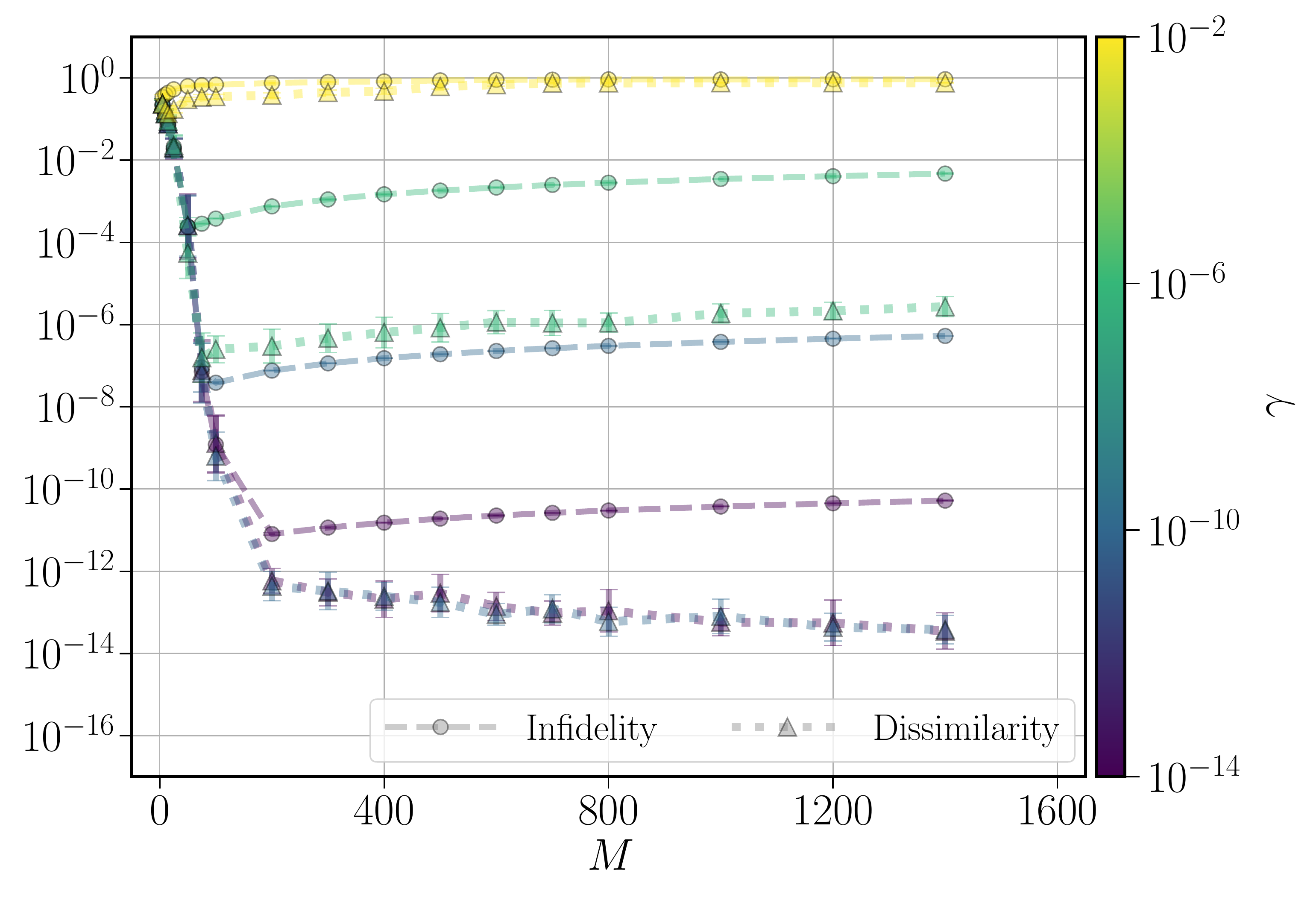}
		\label{fig:objective_purity_entropy_divergence_similarity_similarity}
	\end{subfigure}
	\captionsetup{justification=raggedright}	
	\caption{behavior of infidelity, impurity, entropy, relative entropy divergence, and cosine dissimilarity of parameterized states relative to pure target states, with respect to dephasing noise $\gamma$ (colored/gradient) and depth for the $N=4$ NMR ansatz. (a) Impurity, entropy, relative entropy divergence. Once the optimized parameterized ansatz achieves infidelities or relative entropy divergences of the order of the impurity and entropies, the system is dominated by entropic effects. In this divergent regime, all quantities (colored markers) scale identically, and leading-order infidelity and impurity analytical values (enlarged gray markers) show exact agreement. (b) Cosine dissimilarity. The unitary component of the channel is able to align the parameterized state with the target state. For sufficiently low noise scales, this dissimilarity remains constant and infidelities increase in the divergent regime strictly due to being scaled by the depth-dependent noise scale.}
	\label{fig:objective_purity_entropy_divergence_similarity}	
\end{figure}

Plotting each of these quantities from numerical optimizations in \cref{fig:objective_purity_entropy_divergence_similarity} for unital dephasing noise, we observe that once the infidelities or divergences are less than the impurities or entropies, there is a transition from the convergent to the divergent regime. This transition occurs at a critical depth, before which increasing depth allows the dominant unitary component of the channel to rotate the state to converge to the target state. Beyond this critical depth, the noise component of the channel dominates, and the parameterized state scales towards the mixed state. This divergence of all quantities is shown numerically at small noise scales to be linear in depth and noise scale. We also plot the derived analytical leading-order scaling of the infidelity and impurities with gray enlarged markers, which are in excellent agreement with the numerical results. In this divergent regime at small noise scales, we also observe the numerical hierarchy of the quantities
\begin{align}
	\mathcal{D}_{\theta\gamma}^{\rho} \leq \mathcal{L}_{\theta\gamma}^{\rho} \leq \mathcal{I}_{\theta\gamma} \leq \mathcal{S}_{\theta\gamma}~,
\end{align}
which is in agreement with our infidelity and impurity analytical results. All quantities also numerically appear to converge together as noise increases. In the case of unital noise, all quantities diverge to the worst-case unity values as the noise and depth increases. In the case of non-unital noise, quantities potentially converge to optimality polynomially with depth as the noise and depth increases. This contrasting behavior to the unital noise case suggests the non-unital terms $\eta_{\theta\gamma}\neq 0$ dominate the leading-order scaling in this regime.

We also note that the relative entropy divergences show identical convergent and divergent overparameterized regimes of optimization as the infidelities. The numerically found hierarchy of quantities also suggests that infidelity is potentially lower bounded by the divergence. This appears to be reasonable due to both quantities reflecting distances between distributions corresponding to the quantum states. Given that the divergence is used in classical learning tasks, it is also suitable as an objective function for quantum fidelity-based tasks, although it is more demanding to compute.

To assess the validity of this assumption of the unitary component aligning the pure component of the state with the target state, we note that for small noise scales, the cosine dissimilarity
\begin{align}
	1-\abs{\cos{\phi_{\theta\gamma}^{\rho}}} \sim O(\abs{\alpha_{\theta|\gamma}})	
\end{align}
is of order of the alignment of the pure component with the target state. Plots of the cosine dissimilarity in \cref{fig:objective_purity_entropy_divergence_similarity} indicate that for sufficiently low noise scales, the dissimilarity remains constant, or even decreases to machine precision scales. This suggests that divergences in the infidelities at large depths are strictly due to the noise component of the channel scaling the pure component by the depth-dependent noise scales. At larger noise scales, the cosine dissimilarity is larger, but still orders of magnitude smaller than the infidelities, and diverges similarly to the infidelities. \\

\newpage

\section{Classical and Quantum Error Analysis}\label{app:classical_and_quantum_error_analysis}
In these appendixes, we derive and compare the noise-induced bias of the parameterized noisy states from their noiseless values, for both classical and quantum noise. We show both noise types have biases that scale polynomially with the noise scale, and exponentially with the number of errors induced by the noise. Finally, we discuss the implications of classical floating point error on the viability of large scale simulations without error mitigation.

To bound generally independent errors in simulations, we use Schatten norms $\norm{A} = \norm{A}_{p}~,~ p \in [1,\infty]$ for $d$-dimensional matrices $A,B \in \mathcal{M}(d)$. Such norms satisfy the convenient properties of monotonicity $\norm{A}_{p} \leq \norm{A}_{q} ~,~ q \leq p$, sub-additivity $\norm{A+B}_{p} \leq \norm{A}_{p}+\norm{B}_{p}$, sub-multiplicativity $\norm{AB}_{p} \leq \norm{A}_{p}\norm{B}_{p}$, and being invariant under unitaries $\norm{UAV^{\dagger}}_{p} = \norm{A}_{p}$ for $U,V \in \mathcal{U}(d)$. Unitary matrices have constant norm $\norm{U}_{p} = d^{1/p}$ for $U \in \mathcal{U}(d)$, and we denote the $\lim_{p \to \infty} \norm{A}_{p} = \lambda(A)$ norm as the largest singular value. Finally, Schatten norms satisfy Holder's $\norm{AB}_{s} \leq \norm{A}_{p}\norm{B}_{q} ~,~ 1/p + 1/q = 1/s$, and von-Neumann $\abs{\Tr[AB]} \leq \norm{A}_{p}\norm{B}_{q} ~,~ 1/p + 1/q = 1$ inequalities \cite{Mirsky1975}.

\subsection{Classical Error Analysis}\label{subapp:classical_error_analysis}

We first investigate the effect of classical floating point errors on scalar operations, generalizing the results of \cite{Wilkinson1965} to the case of an arbitrary number of successive matrix multiplications.  

\subsubsection{Scalar Floating Point Error}\label{subsubapp:scalar_floating_point_error}
Let us assume that we are performing binary floating point operations, with the exact, ideal operations denoted by $\circ \in \{+,-,\times,\mathbin{/}\}$. We denote floating point representations of any exact operations $\circ$ with subscripts $\circ_{\epsilon}$.\\

For operations between scalars, and scalar values themselves, we assume they may only be represented with a relative error, upper bounded by $\varepsilon$, sometimes referred to as machine precision.\\

For vectorized operations comprised of vectors of dimension $d$, let the relative error be upper bounded by $\epsilon = \epsilon(d,\varepsilon)$. This dependence of the error on the number of operations $d$ and the machine precision $\varepsilon$ may be polynomial, or even exponential. This general definition includes the case of more sophisticated computer architectures, that perform operations such as fused-addition-multiplications. \\

The scalar product between scalars $x,y$, when $d=1$, therefore has error
\begin{align}
	x \circ_{\epsilon} y - x \circ y =&~ (x \circ y)\epsilon~,
	\intertext{and in this scalar case, the error is generally proportional to the machine precision}
	\epsilon =&~ \varepsilon~.
\end{align}
It should be noted that this error could be considered as deterministic error, where a fixed magnitude error $\epsilon$ is associated with the operations. Instead, stochastic error could be considered,
\begin{align}
	x \circ_{\epsilon} y - x \circ y =&~ (x \circ y)\xi~,
	\intertext{where the error is assumed to be the random variable}
	\xi \sim&~ \Omega_{\epsilon}
\end{align}
from a distribution $\Omega_{\epsilon}$, generally with zero mean and generally a standard deviation that is proportional to the scale $\epsilon$.\\

For successive additions, the order of the operations affects which terms have greater error, as initial terms accumulate more error over the course of successive additions. The sum of $d$ scalars $\{a_{\mu}\}$ has error
\begin{align}
	\sideset{}{_{\epsilon}}\sum_{\mu}^{d}a_{\mu} -  \sum_{\mu}^{d} a_{\mu} =&~ \sum_{\mu}^{d} a_{\mu}\epsilon_{\mu}~,
	\intertext{where the accumulated error from addition is}
	\epsilon_{\mu} =&~ (1 + \varepsilon)^{d-\mu} - 1~.
\end{align}\\

The error of summations of $d$ scalars $a = \{a_{\mu}\}$ can be represented as inner products with a ones vectors $1$, and error vectors $\epsilon = \{\epsilon_{\mu}\}$ to account for the accumulated error from addition
\begin{align}
	{1 \cdot_{\epsilon} a} - 1 \cdot a =&~ \epsilon \cdot a~.
\end{align}

\subsubsection{Matrix Floating Point Error}\label{subsubapp:matrix_floating_point_error}
Generalizing beyond scalars, the deterministic error of inner products of $d$-dimensional matrices $A,B$ can therefore be represented as an inner product with respect to an error matrix 
\begin{align}
	{AB}_{\epsilon} - AB  =&~ A \Sigma B~.
\end{align}
Similarly, the stochastic error of inner products of $d$-dimensional matrices $A,B$ can therefore be represented as an inner product with respect to an error matrix,
\begin{align}
	{AB}_{\epsilon} - AB  =&~ A \Xi B~
	\intertext{where the error is assumed to be the random variable}
	\Xi \sim&~ \Omega_{\epsilon}^{d \times d}~,
\end{align}
from a distribution $\Omega_{\epsilon}$, generally with zero mean and generally a standard deviation that is proportional to the scale $\epsilon$.\\

The relative error matrix is arbitrary depending on the model of computation, and we choose it to be diagonal,
\begin{align}
	\Sigma =&~ \{\epsilon_{\mu} = (1 + \varepsilon)^{d-\mu} - 1\}~.,
\end{align}
with norm
\begin{align}
	\norm{\Sigma}_{p} =&~ \left(\sum_{\mu}^{d} ((1+\varepsilon)^{d-\mu}-1)^p\right)^{1/p} \hspace{-0.4cm} \approx ~ \left(\sum_{\mu}^{d}(d-\mu)^{p}\right)^{1/p}\hspace{-0.4cm}\varepsilon ~\approx ~O(d^{\frac{p+1}{p}})\varepsilon ~\sim~ O\left(\textrm{poly}(d)\right)\varepsilon ~\equiv~ \epsilon^{(p)} ~\leq~ \epsilon ~,
\end{align}
which we define to be upper bounded by a $p$-norm independent error $\epsilon$.

This can be extended to a product of $k$ matrices $\prod_{\mu}^{k}A_{\mu}$. We assume the error acts recursively, propagating from the initial to final matrix multiplications. Therefore. the matrix multiplication operation becomes a sum over all possible locations of the error matrix interlaced within the products, denoted with a multi-index $\chi_{l} = \{ {\chi_{l^{}}^{}}_{\mu} \in [2], \mu \in [k]\}$, with the number of errors $\abs{\chi_{l}} = l \leq k$. We assume that errors also exist on the representation of the matrix elements, yielding an extra error matrix factor at the end of the product, and errors from additions of the error terms are a secondary effect. 

The error of the product of $k$ matrices $\prod_{\mu}^{k}A_{\mu}$ therefore takes the form
\begin{align}
	{\sideset{}{_{\epsilon}}\prod_{\mu}^{k}A_{\mu}} - \prod_{\mu}^{k}A_{\mu}  =&~ \sum_{l>0}^{k} \sum_{\chi_{l}} \prod_{\mu}^{k} A_{\mu}\Sigma^{{\chi_{l}}_{\mu}}~.
\end{align}

We then use norm inequalities, and define an upper bound on the norms values $\norm{A_{\mu}}_{p} \leq \norm{A}_{p}$, to bound the norm of the absolute matrix multiplicative error,
\begin{align}
	\norm{{\sideset{}{_{\epsilon}}\prod_{\mu}^{k}A_{\mu}} - \prod_{\mu}^{k}A_{\mu}}_{s} =&~ \norm{\sum_{l>0}^{k} \sum_{\chi_{l}} \prod_{\mu}^{k} A_{\mu}\Sigma^{\chi_{l_{\mu}}}}_{s} \\
	\leq&~ \sum_{l>0}^{k} \sum_{\chi_{l}} \norm{\prod_{\mu}^{k} A_{\mu}\Sigma^{\chi_{l_{\mu}}}}_{s} \\
	\leq&~ \sum_{l>0}^{k} \sum_{\chi_{l}} \min_{\sum_{\mu}^{k}\frac{1}{p_{\mu}} + \frac{\chi_{l_{\mu}}}{q_{\mu}} = \frac{1}{s}}\prod_{\mu}^{k}  \norm{A_{\mu}}_{p_{\mu}} \norm{\Sigma}_{q_{\mu}}^{\chi_{l_{\mu}}} \\
	\leq&~ \sum_{l>0}^{k} \min_{\frac{k}{p} + \frac{l}{q} = \frac{1}{s}} \tbinom{k}{l}  \norm{A}_{p}^{k} \norm{\Sigma}_{q}^{l} \\
	\leq&~ \sum_{l>0}^{k} \tbinom{k}{l}  \lambda(A)^{k} \epsilon^{l} \\
	=&~ \lambda(A)^{k}((1+\epsilon)^{k}-1)~.
\end{align}
Therefore, the relative multiplicative error is upper bounded for a general norm by
\begin{align}
	\frac{\norm{{\sideset{}{_{\epsilon}}\prod_{\mu}^{k}A_{\mu}} - \prod_{\mu}^{k}A_{\mu}}_{}}{\norm{\prod_{\mu}^{k}A_{\mu}}_{}} \leq&~ \frac{\lambda(A)^{k}}{\norm{\prod_{\mu}^{k}A_{\mu}}_{}}\left(\vphantom{\frac{}{}}(1+ \epsilon^{})^{k}-1\right) ~ \overset{A_{\mu} \in \mathcal{U}(d)}{\to} ~ \frac{1}{\textrm{poly}(d)}\left(\vphantom{\frac{}{}}(1+ \epsilon^{})^{k}-1\right)~,
\end{align}
where in the case of unitary matrices with unit singular values, the bound is simplified.


\subsection{Classical Error Scaling}\label{subapp:classical_error_scaling}
Let us define a unitary evolution as a product of $k$, $d$-dimensional unitaries $U = \prod_{\mu}^{k} U_{\mu}$ that transforms an initial state as $\sigma \to \rho = U\sigma U^{\dagger}$.  Suppose the evolution is subject to classical floating point error $U \to U_{\epsilon}$ and $\sigma \to \rho_{\epsilon}$. An interesting interpretation is that this noisy evolution with classical floating point error can be represented by unormalized Kraus-like operators, $\{I,\Sigma\}$, where $I + \Sigma^{\dagger}\Sigma \neq I$, meaning the operation is not trace-preserving.

The composite adjoint action of unitaries with $k$ matrix multiplications with classical error is therefore
\begin{align}
	\rho_{\epsilon} =&~ U_{\epsilon} \sigma U_{\epsilon}^{\dagger} 
	= \rho_{} + \delta_{\epsilon}~,
\end{align}
where the perturbation is
\begin{align}	
	\delta_{\epsilon} =&~ \sum_{l+l^{\prime}>0}^{k} \sum_{{\chi_{l^{}}^{}},{\chi_{l^{\prime}}^{\prime}}} \left[\prod_{\mu}^{k} U_{\mu}\Sigma^{{\chi_{l^{}}^{}}_{\mu}}\right] \sigma \left[\prod_{\mu^{\prime}}^{k}{\Sigma^{{\chi_{l^{\prime}}^{\prime}}_{\mu^{\prime}}}}^{\dagger}{U_{\mu^{\prime}}}^{\dagger}\right]~.
\end{align}

The absolute norm error, given the non-unitary classical noise has norm $\norm{\Sigma} = \epsilon$, and the operators are unitaries with $\norm{U_{\mu}} = \textrm{poly}(d)$, is therefore bounded by
\begin{align}
	\norm{\rho_{\epsilon} - \rho_{}} \leq&~ \sum_{l+l^{\prime}>0}^{k} \sum_{{\chi_{l^{}}^{}},{\chi_{l^{\prime}}^{\prime}}} \norm{\left[\prod_{\mu}^{k} U_{\mu}\Sigma^{{\chi_{l^{}}^{}}_{\mu}}\right] \sigma \left[\prod_{\mu^{\prime}}^{k}{\Sigma^{{\chi_{l^{\prime}}^{\prime}}_{\mu^{\prime}}}}^{\dagger}{U_{\mu^{\prime}}}^{\dagger}\right]} \\
	\leq&~ \sum_{l+l^{\prime}>0}^{k} \sum_{{\chi_{l^{}}^{}},{\chi_{l^{\prime}}^{\prime}}} \left[\prod_{\mu}^{k}\norm{U_{\mu}\Sigma^{{\chi_{l^{}}^{}}_{\mu}}}\right] \norm{\sigma} \left[\prod_{\mu^{\prime}}^{k}\norm{{\Sigma^{{\chi_{l^{\prime}}^{\prime}}_{\mu^{\prime}}}}^{\dagger}U_{\mu^{\prime}}^{\dagger}}\right] \\
	=&~ \sum_{l+l^{\prime}>0}^{k} \sum_{{\chi_{l^{}}^{}},{\chi_{l^{\prime}}^{\prime}}} \left[\prod_{\mu}^{k}\norm{\Sigma^{{\chi_{l^{}}^{}}_{\mu}}}\right] \norm{\sigma} \left[\prod_{\mu^{\prime}}^{k}\norm{{\Sigma^{{\chi_{l^{\prime}}^{\prime}}_{\mu^{\prime}}}}^{\dagger}}\right] \\	
	\leq&~ \sum_{l+l^{\prime}>0}^{k} \sum_{{\chi_{l^{}}^{}},{\chi_{l^{\prime}}^{\prime}}} \norm{\sigma}\prod_{\mu,\mu^{\prime}}^{k}\norm{\Sigma}^{{\chi_{l^{}}^{}}_{\mu}+{\chi_{l^{\prime}}^{\prime}}_{\mu^{\prime}}} \\	
	=&~ \sum_{l+l^{\prime}>0}^{k} \tbinom{k}{l}\tbinom{k}{l^{\prime}} \norm{\sigma}\epsilon^{l+l^{\prime}} \\	
	=&~ \norm{\sigma}\left[\left(\sum_{l}^{k} \tbinom{k}{l} \epsilon^{l}\right)^{2} - 1\right] \\	
	=&~ \norm{\sigma}\left((1 + \epsilon)^{2k}-1\right)~. 
\end{align}
Given the norm of pure input and noiseless output states is $\norm{\rho} = \norm{\sigma} = 1$, the relative error for pure states is
\begin{align}
	\frac{\norm{\rho_{\epsilon} - \rho_{}}}{\norm{\rho_{}}} \leq&~ \abs{1-(1+\epsilon)^{2k}}~. 
\end{align}

We may then calculate functions of classical noisy states, such as the infidelity with a (pure) state $\rho$, as
\begin{align}
	\mathcal{L}_{\epsilon} =&~ 1 - \trace{\rho\rho_{\epsilon}} = \mathcal{L}_{} - \trace{\rho\delta_{\epsilon}}~.
\end{align}

Using the Cauchy-Schwartz inequality, we may bound the bias of the classical noisy infidelities as
\begin{align}
	\abs{\mathcal{L}_{\epsilon} - \mathcal{L}_{}} =&~ \abs{\trace{\rho\delta_{\epsilon}}}  \leq \min_{\frac{1}{p} + \frac{1}{q} = 1}\norm{\rho}_{p}\norm{\delta_{\epsilon}}_{q}  \leq \min_{\frac{1}{p} + \frac{1}{q} = 1}\norm{\rho}_{p}\norm{\sigma}_{q}\abs{1-(1+\epsilon)^{2k}}~,
\end{align}
and this noisy linear objective deviation, for pure states, scales as
\begin{align}
	\abs{\mathcal{L}_{\epsilon} - \mathcal{L}_{}} \leq&~ \abs{1-(1+\epsilon)^{2k}}~.
\end{align}

\begin{figure}[h]
	\centering
	\includegraphics[width=0.7\columnwidth]{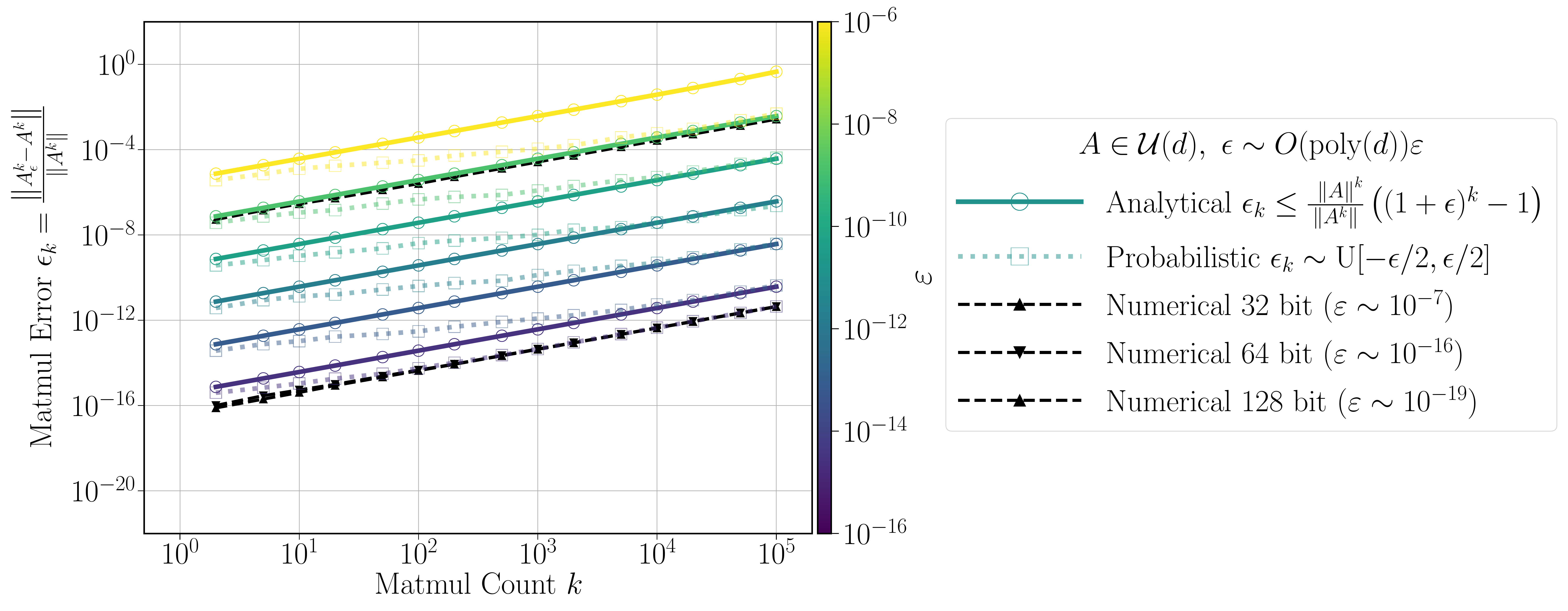}
	\captionsetup{justification=raggedright}	
	\caption{Matrix multiplication floating point error scaling with number of multiplications $k$, with analytical (circles/solid lines), numerical probabilistic (squares/dotted lines) and numerical data-type (triangles/dashed lines) models scaling for $d=2^2$-dimensional random unitaries $A$. Simulated floating point error is shown to increase polynomially with the number of floating point operations. Exact floating point data-types are in exact agreement with the analytical upper bounds, with the exception of simulated $128$-bit data-type, whose error appears to be bounded by the $64$-bit precision, likely attributed to intermediate backend calculations being cast to $64$-bit precision.}
	\label{fig:classical_floating_point_error}
\end{figure}

An example simulation of classical floating point error for $k$ successive matrix multiplications, by artificially including noise with different scales $\varepsilon$, is shown in \cref{fig:classical_floating_point_error}. To investigate this classical error, we compare analytical results (exact upper bounds on the difference between noisy and noiseless matrix multiplications), probabilistic results (numerically adding zero-mean uniformly random errors $U[-\varepsilon/2,\varepsilon/2]$ to results of floating point operations), and numerical results (numerically performing matrix multiplications without artificial noise for standard datatypes: single point $32$-bit precision with $\varepsilon \approx 10^{-7}$, double point $64$-bit precision with $\varepsilon \approx 10^{-16}$, and quadruple point $128$-bit precision with $\varepsilon \approx 10^{-19}-10^{-24}$, depending on whether the precision is only simulated to $\sim80$-bit for double precision architectures).

We observe good agreement across all models, and we note the drift in the probabilistic results at a large number of matrix multiplications $k$, to an effective $\varepsilon$ an order of magnitude smaller than simulated. This suggests the built-in floating point operation algorithms may be cacheing results or eliminating some sources of error. We repeat such experiments across $S=10$ samples, and we also observe negligible errorbars within plot markers. This classical error also shows similar trends to the quantum noise case of noise-induced convergent and divergent regimes. 

Finally, we note that it is difficult to precisely and consistently define the machine precision $\varepsilon$, or the number of decimal points of accuracy of a simulation, even for a recognized floating point data-type. Apart from the often non-linear dependence of the true machine precision value on the number of operations and dimensionality of the problem, there are additionally software effects. These effects include how basic floating point operations may be vectorized or fused together. Similarly, there may be hardware effects, such as how the exact memory layout of different computing architectures can affect the way in which operations are performed. These studies, in particular the exact agreement between analytical and $32$ and $64$-bit data-type numerical results, however confirm the appropriateness of our derived models, and they offer valuable insight into upper bound estimates for a range of machine precision values.

The simulation package developed for this work \cite{Duschenes2022}, as discussed in \cref{app:classical_simulation_and_optimization}, can perform single $\varepsilon \sim O(10^{-7})$, double $\varepsilon \sim O(10^{-16})$, or (simulated) quadratic $\varepsilon \sim O(10^{-19})$ floating point arithmetic, and each data-type is useful in understanding the sources of floating point error. However, quadruple floating point arithmetic is unable to currently be compiled efficiently for large systems, in addition to the overhead of the simulated precision. Automatic differentiation computations also empirically have less than double precision accuracy, typically $\varepsilon \sim O(10^{-12})$. Therefore current large scale simulations are infeasible with this data-type.

\subsection{Quantum Error Scaling}\label{subapp:quantum_error_scaling}
Let us define a unitary evolution as a product of $k$, $d$-dimensional unitaries $U = \prod_{\mu}^{k} U_{\mu}$ that transforms an initial state as $\sigma \to \rho = U\sigma U^{\dagger}$. Suppose the evolution is subject to quantum error $U \to U_{\gamma}$ and $\sigma \to \rho_{\gamma}$. This noisy evolution with quantum error can be represented by the normalized Kraus operators $\{\sqrt{1-\gamma}I, \sqrt{\gamma}\Sigma\}$, where $(1-\gamma)I + \gamma\Sigma^{\dagger}\Sigma = I$, meaning the operation is trace-preserving.

The composite adjoint action of unitaries with $K = NM$ matrix multiplications with quantum error is therefore
\begin{align}
	\rho_{\gamma} =&~ U_{\gamma} \sigma U_{\gamma}^{\dagger} 
	= (1-\gamma)^{K}\rho_{} + \delta_{\gamma}~,
\end{align}
where the perturbation is
\begin{align}	
	\delta_{\gamma} =&~ \sum_{l>0}^{K} \sum_{\chi_{l}} \gamma^{l}(1-\gamma)^{K-l} \left[\prod_{\mu}^{K} U_{\mu}\Sigma^{{\chi_{l^{}}^{}}_{\mu}}\right] \sigma \left[\prod_{\mu^{\prime}}^{K}{\Sigma^{{\chi_{l^{\prime}}^{\prime}}_{\mu^{\prime}}}}^{\dagger}{U_{\mu^{\prime}}}^{\dagger}\right]~.
\end{align}

The absolute norm error, given the unitary quantum noise has norm $\norm{\Sigma} = \textrm{poly}(d)$, and the operators are unitaries with $\norm{U_{\mu}} = \textrm{poly}(d)$, is therefore bounded by
\begin{align}
	\norm{\rho_{\gamma} - \rho_{}} \leq&~ ((1-\gamma)^{K} - 1)\norm{U\sigma U^{\dagger}} + \sum_{l>0}^{K} \sum_{\chi_{l}} \gamma^{l}(1-\gamma)^{K-l} \norm{\left[\prod_{\mu}^{K} U_{\mu}\Sigma^{{\chi_{l^{}}^{}}_{\mu}}\right] \sigma \left[\prod_{\mu^{\prime}}^{K}{\Sigma^{{\chi_{l^{\prime}}^{\prime}}_{\mu^{\prime}}}}^{\dagger}{U_{\mu^{\prime}}}^{\dagger}\right]} \\
	=&~ ((1-\gamma)^{K} - 1)\norm{\sigma} + \sum_{l>0}^{K} \sum_{\chi_{l}} \gamma^{l}(1-\gamma)^{K-l} \norm{\sigma} \\
	=&~ ((1-\gamma)^{K} - 1)\norm{\sigma} + \sum_{l>0}^{K} \tbinom{K}{l} \gamma^{l}(1-\gamma)^{K-l} \norm{\sigma} \\
	=&~ 2\norm{\sigma}((1-\gamma)^{K} - 1)~.	
\end{align}
Given the norm of pure input and noiseless output states is $\norm{\rho} = \norm{\sigma} = 1$, the relative error for pure states is
\begin{align}
	\frac{\norm{\rho_{\epsilon} - \rho_{}}}{\norm{\rho_{}}} \leq&~ 2\abs{1 - (1-\gamma)^{K}}~. 
\end{align}

We may then calculate functions of quantum noisy states, such as the infidelity with a (pure) state $\rho$ as
\begin{align}
	\mathcal{L}_{\gamma} =&~ 1 - \trace{\rho\rho_{\gamma}} = \mathcal{L}_{} - \trace{\rho\delta_{\gamma}}~.
\end{align}

Using the Cauchy-Schwartz inequality, we may bound the bias of the quantum noisy infidelities 
\begin{align}
	\abs{\mathcal{L}_{\gamma} - \mathcal{L}_{}} =&~ \abs{\trace{\rho\delta_{\gamma}}}  \leq \min_{\frac{1}{p} + \frac{1}{q} = 1}\norm{\rho}_{p}\norm{\delta_{\gamma}}_{q} \leq \min_{\frac{1}{p} + \frac{1}{q} = 1} 2\norm{\rho}_{p} \norm{\sigma}_{q}\abs{1 - (1-\gamma)^{K}}~,
\end{align}
and this noisy linear objective deviation, for pure states, scales as
\begin{align}
	\abs{\mathcal{L}_{\gamma} - \mathcal{L}_{}} \leq&~ 2\abs{1 - (1-\gamma)^{K}}~.
\end{align}

\subsection{Classical versus Quantum Error Scaling}\label{subapp:classical_versus_quantum_error_scaling}

We may relate the classical and quantum error scales by identifying the dimension dependent classical scale
\begin{align}
	\epsilon =&~ \epsilon(d)
\end{align}
and the number of matrix multiplication operations for $M$, $\textrm{poly}(N)$ local unitaries being
\begin{align}
	k =&~ O(K) = O(\textrm{poly}({N})M)~.
\end{align}
The error scaling of classical and quantum noisily evolved states may then be summarized as
\begin{align}
	\norm{\rho_{\epsilon} - \rho_{}} \leq&~ \hspace{5pt}\abs{1-(1+\epsilon)^{2k}} \\
	\norm{\rho_{\gamma} - \rho_{}} \leq&~ 2\abs{1 - (1-\gamma)^{K}}~,
\end{align}
and the linear functions of the perturbed states are also perturbed with identical scaling
\begin{align}
	\abs{\mathcal{L}_{\epsilon} - \mathcal{L}_{}} \leq&~ \hspace{5pt}\abs{1-(1+\epsilon)^{2k}} \\
	\abs{\mathcal{L}_{\gamma} - \mathcal{L}_{}} \leq&~ 2\abs{1 - (1-\gamma)^{K}}~.
\end{align}

A subtle point to keep in mind is that we are comparing infidelities at fixed variable parameters values. If the parameters have been optimized in a noisy setting, then the associated noiseless infidelity $\mathcal{L}_{\theta_{\gamma}}$, evaluated with these parameters, likely differs from the true optimal noiseless infidelity $\mathcal{L}_{\theta}$ with parameters optimized in a noiseless setting. If we happen to be in an overparameterized regime, where the phenomenon of lazy training occurs \cite{Shirai2021}, parameters may change negligibly from their initial values. In this regime, the noiseless infidelities, with noisy and noiseless trained parameters, possibly approximately coincide
\begin{align}
	\theta^{*}_{\gamma} \approx \theta^{*}~.
\end{align}
In this setting, functions of the parameters such as the infidelities may also approximately coincide,
\begin{align}
	\mathcal{L}_{\theta^{*}_{\gamma}} \approx \mathcal{L}_{\theta^{*}}~,
\end{align}
and they can be used in place of each other. Bounds such as those above comparing noisy and noiseless fidelities may be relevant, however future studies should investigate the noise-induced bias in optimization.

\newpage
\section{Nuclear Magnetic Resonance Ansatz}\label{app:nuclear_magnetic_resonance_ansatz}
In these appendixes, we include details about the nuclear magnetic resonance (NMR) ansatz studied in this work. We describe the model Hamiltonian, and we document all parameter scales
\begin{align}
		H_{\theta}^{(t)} =&~ \sum_{i}{\theta_{i}^{x}}^{(t)}X_{i} ~+~ \sum_{i}{\theta_{i}^{y}}^{(t)} Y_{i} ~+~ \sum_{i}h_{i}Z_{i} ~+~ \sum_{i<j} J_{ij} Z_{i}Z_{j}~.
\end{align}
Here we have control over the variable time-dependent transverse $X$ and $Y$ fields, with constant time-independent longitudinal $Z$ and $ZZ$ fields. In these units, the parameters in \cref{tab:parameters} are used for this ansatz. We note that the non-local coupling scale is $J \ll h,\theta$, and there are time step sizes such that $\tau J \ll 1$. To generate a finite angle ($\pi/4$ for $ZZ$) rotation necessary to implement a single entangling gate, depths of order $M \sim 1/\tau J \sim O(10^4)$ are necessary. Furthermore, the number of entangling gates necessary to compile Haar random unitaries is generally exponential $O(D^{N})$ in the number of qubits \cite{Ashhab2022}. Therefore, these parameter scales, even for relatively small system sizes of $N \leq 4$, indicate that in general an even greater total number, possibly super-exponential $M > O(D^{N})$, of physical gates are necessary in practice for these tasks of interest.
\begin{table*}
\captionsetup{justification=raggedright}	
\caption{Constrained parameters for NMR ansatz \cite{Peterson2020a}. All parameters are chosen to be experimentally relevant. Some parameters however, in particular the time step of $100~\si{\micro\s}$ (unitary compilation), and $75~\si{\micro\s}$ (state preparation), are chosen to be on the maximum end of what is experimentally feasible, to allow for reasonably efficient simulation. 
}
\label{tab:parameters}
\begin{tabular}{|c|l|l|}
	\hline	
	Parameter & Description & Value \\ \hline
	$N$      & Number of qubits & $1-4$           \\
	$M$      & Number of time steps & $5-5000$  \\
	$\tau$ & Trotterization time step & $75-100~\si{\micro\s}$          \\
	$T$   & Evolution time & $375~\si{\micro\s}-500~\si{\milli\s}$  \\
	$Q$ & Spatial Trotterization order & $2$ \\
	$P$ & Number of parameters & $MN(N+5)/2$ \\
	$J_{i<j}$      & Constant longitudinal coupling & $\pi/2 ~\{72.4, -130.0, 50.0, 210.0, 20.0, -190.0, -30.0, 60.0, 90.0, -60.0\}~\si{\hertz}$ \\
	$h_{i}$      & Constant longitudinal field &$\pi/2 ~\{-10.0, 0, -1.0, 29.0, -20.0\}~\si{\kilo\hertz}$   \\
	$\bar{\theta}$ & Variable transverse field & $\pi/2 ~\si{\mega\hertz}$ \\
	$\gamma$ & Noise scale & $10^{-14} - 10^{-1}$  \\
	$\tilde{\theta}$ & Constraints & \makecell[l]{Constrained and Shared Transverse Fields \\Zero-Field Dirichlet Boundary Conditions in Time \\ $\theta^{x,y\hspace{1pt}(m)}_{i} = \theta^{x,y\hspace{1pt}(m)}~,~ \abs{\theta^{x,y\hspace{1pt}(m)}} \leq \bar{\theta}~	,~ \theta^{x,y\hspace{1pt}(0,M)} = 0$} \\
	\hline
\end{tabular}
\end{table*}

We also compare gate times, decoherence times, experimentally tested depths, and effective depths based on these times for various NISQ implementations in \cref{tab:implementations}. We note that each implementation has better and worse properties. NMR in this work has the largest $2$-qubit gate times $T_{U2} = O(1/J)$, which translates to NMR having the largest effective depth $T_{U2}/T_{U1} \sim O(100)$ required for each $2$-qubit gate, and having the smallest effective maximum depth $T_{\gamma}/T_{U2} \sim O(10^{2})$ before coherence. We emphasize, however, that tabulated values taken from a range of values in recent literature are very experimental apparatus- and algorithm- specific \cite{Linke2017}.

\begin{table*}
\captionsetup{justification=raggedright}	
	\caption{Comparison of $1$-qubit gate times $T_{U1}$, $2$-qubit gate times $T_{U2}$, decoherence times $T_{\gamma}$, experimentally achieved number of qubits $N_{U}$, experimentally achieved depths of $2$-qubit gates $M_{U}$, ratio of $2$-qubit to $1$-qubit gate times $M_{U12} = T_{U2}/T_{U1}$, and the ratio of decoherence to $2$-qubit gate times $M_{\gamma U2} = T_{\gamma}/T_{U2}$, for various experimental quantum computing implementations. The ratio of $2$ to $1$-qubit gate times can be interpreted as the effective depth $M_{U12}$ necessary to implement $2$-qubit gates. The ratio of decoherence times $T_{\gamma}$ to $2$-qubit gate times can be interpreted as the effective maximum depth $M_{\gamma U2}$ before decoherence. For NMR, $1$-qubit gate times $T_{U1} = O(\tau)$ are approximately the pulse time steps, $2$-qubit gate times $T_{U2} = O(1/J)$ are approximately inversely related to the $2$-qubit coupling, and decoherence times $T_{\gamma} = O(T_{2})$ are approximately the dephasing times. Tabulated values are experimental apparatus and algorithm specific \cite{Linke2017}, taken from a range of values in recent literature.}
	\label{tab:implementations}
\begin{tabular}{|l|c|c|c|c|c|c|c|}
	\hline
	Implementation & 
		$T_{U1}$ & 
		$T_{U2}$ & 
		$T_{\gamma}$ & 
		$N_{U}$ &
		$M_{U}$ &
		$M_{U12} = T_{U2}/T_{U1}$ & 
		$M_{\gamma U2} = T_{\gamma}/T_{U2}$ \\ \hline
	NMR \cite{Peterson2020a,Peterson2020,Majidy_Wilson_Laflamme_2024} & 
		$~~75~\si{\micro\s}~~$ & 
		$~~5000~\si{\micro\s}~~$ & 
		$~~1000~\si{\milli\s}~~$ &
		$~~12~~$ &
		$~~1000~~$ &
		$~~70~~$ & 
		$~~2 \times 10^{2}~~$ \\ \hline
	Trapped Ions \cite{Bruzewicz2019,Low2020,Monroe2021,Zhao2023,Majidy_Wilson_Laflamme_2024}  & 
		$~~1~\si{\micro\s}~~$ & 
		$~~50~\si{\micro\s}~~$ & 
		$~~10000~\si{\milli\s}~~$ &
		$~~12~~$ &
		$~~64~~$ &
		$~~50~~$  & 
		$~~1 \times 10^{6}$ \\ \hline
	Super-Conducting \cite{Kim2023,Werninghaus2021,Majidy_Wilson_Laflamme_2024} & 
		$~~0.02~\si{\micro\s}~~$ & 
		$~~0.1~\si{\micro\s}~~$ & 
		$~~1~\si{\milli\s}~~$ & 
		$~~127~~$ &
		$~~60~~$ &
		$~~5~~$ & 
		$~~1 \times 10^{4}~~$ \\ \hline
	Neutral Atoms \cite{Wintersperger2023,Bluvstein2021,Bluvstein2023} & 
		$~~2~\si{\micro\s}~~$ & 
		$~~0.5~\si{\micro\s}~~$ & 
		$~~1000~\si{\milli\s}~~$ &
		$~~280~~$ &
		$~~516~~$ &
		$~~0.25~~$ & 
		$~~2 \times 10^{6}~~$ \\
	\hline
\end{tabular}
\end{table*}

\newpage
\section{Classical Simulation and Optimization}\label{app:classical_simulation_and_optimization}
In these appendixes, we discuss details of the classical simulation and optimization of the quantum systems studied in this work. We explain our hyperparameter choices, optimization routines, and we list all optimization settings.

The quantum systems in this work are simulated and optimized classically using a compiled, automatic-differentiation library using the Python JAX backend, developed as a general differentiable exponentially deep circuit simulator \cite{Duschenes2022}. This library is optimized for noisy density matrix simulations with few qubits $N<6$, and state-of-the-art large depth circuits with $k \sim O(10^{5})$ gates per circuit instance. 

For example, in this work's NMR ansatz, circuits consist of $Q=2$ order spatially Trotterized evolution, consisting of fully connected two-body gates. Local noise on all qubits after each layer requires up to $M \sim 5 \times 10^{3}$ circuit layers, and the circuits are simulated roughly $10^{9}$ times throughout the optimization routines and loops over ansatz settings. All optimizer hyperparameters are shown in \cref{tab:hyperparameters}.

When performing gradient-based optimization, the gradients of these objectives can be expressed analytically with parameter shift rules, however for the $N\leq4$ system sizes considered in this work, we use automatic differentiation for efficiency. An exception is when computing the Hessian and quantum Fisher information, where computing analytical gradients is most memory-efficient. For small system sizes, the unitary and quantum noise operations, and infidelities, specifically the traces over the full space, can also be calculated exactly. Sampling methods, which introduces forms of shot-noise, and other quantum algorithms, such as the Hilbert-Schmidt test, and other forms of process tomography, are not necessary to be implemented. The effects of estimating such infidelities with sampling and approximate operators for larger system sizes are important studies to be conducted in future works.

For computing statistics such as the mean and variance across $S$ independent optimizations, we sample any inputs from specific distributions. We generally sample the initial states $\sigma$, target unitaries $U$, and target states $\rho$ according to the Haar measure, to avoid any biases in targeting a specific subspace \cite{Holmes2021}. Expectation values of parameterized functions of the initial and target states $\mathcal{F}_{\theta}(\sigma,\rho)$, such as infidelities, may then be computed as
\begin{align}
	\expval{\mathcal{F}_{\theta}}^{(S)} =&~ \frac{1}{S}\sum_{\substack{s\\\tilde{\theta}^{(s)} \sim \textrm{Uniform}\\\sigma^{(s)} \sim \textrm{Haar}~,~\rho^{(s)} \sim \textrm{Haar}}}^{S} \mathcal{F}_{\theta^{(s)}}(\sigma^{(s)},\rho^{(s)})~.
\end{align}
For artificially simulating floating point error, we add random matrices to each successive distinct matrix multiplication in a calculation. Plotted quantities are also calculated at the optimal parameterization, which is not necessarily at the last optimization iteration. We also uniformly randomly sample initial parameters $\theta = \tilde{\theta}$, and we then smooth them over the $M$ time steps with cubic interpolation to be experimentally implementable.

In this work, we define the parameters as $\theta = \theta(\phi)$, functions of explicit variables $\phi$ that are explicitly optimized. In the unconstrained case, all parameters for each qubit are independent, and are not constrained in magnitude. However in the constrained case, we impose that the fields are constrained, and coupled to act uniformly across all sites for each operator. Therefore, $\theta^{x,y}_{i} = \theta^{x,y}$ for each qubit $i \in [N]$. We also bound all transverse field magnitudes $\abs{\theta^{x,y}_{i}} \leq \bar{\theta}$. We finally impose Dirichlet boundary conditions in time, such that the initial and final fields are approximately zero, ${\theta^{x,y}}^{(0)} = {\theta^{x,y}}^{(M)} = 0$.

To perform the classical simulation and optimization of the parameterized channels, we perform first-order gradient based optimization routines to minimize the objectives $\mathcal{L}_{\theta\gamma}$ with respect to the variable parameters $\theta$. We may then compute gradients of objectives $\zeta = \partial \mathcal{L}_{\theta}$. 

Due to the high dimensionality of the problem, more effective variants of classical gradient descent must be performed. Here we choose a variant of the first-order conjugate gradient scheme, where the search direction $\xi$ is updated iteratively \cite{Nocedal2006} at iteration $0 \leq l \leq L$:
\begin{align}
	\theta^{(l+1)} =&~ \theta^{(l)} + \alpha^{(l)} \xi^{(l)} ~,\\
	\xi^{(l+1)} =&~ -\zeta^{(l+1)} + \beta^{(l)} \xi^{(l)} ~,
\end{align}
given initial conditions of $\theta^{(0)}$ and $\xi^{(0)} = -\zeta^{(0)}$.

The learning rates $\alpha,\beta$ must be chosen to satisfy the Wolfe convergence conditions \cite{Nocedal2006}, which guarantee the parameters and search directions are updated such that the objective is monotonically decreasing.

For the parameter learning rate $\alpha$, a line search is conducted, that involves at most $L_{\alpha}$ objective calls per iteration $l$, and it ensures the objective maximally decreases. For the search learning rate $\beta$, rates that obey
\begin{align}
	\abs{\beta} \leq&~ \bar{\beta} = \frac{\zeta^{(l+1)} \cdot \zeta^{(l+1)}}{\zeta^{(l)} \cdot \zeta^{(l)}}
\end{align}
ensure convergence \cite{Nocedal2006}. Various forms for this parameter include the standard Fletcher-Reeves rate $\beta_{\textrm{FR}}^{(l)} = \bar{\beta}^{(l)}$. However, for the range of problems in this work, we find that it and many definitions lead to the optimizer immediately getting stuck in local minima. We find the Hestenes-Stiefel rate, however, to be quite effective,
\begin{align}
	\beta_{\textrm{HS}}^{(l)} =&~ \frac{\zeta^{(l+1)} \cdot (\zeta^{(l+1)} - \zeta^{(l)})}{\xi^{(l)} \cdot (\zeta^{(l+1)} - \zeta^{(l)})}~.
\end{align}
The hyperparameters are chosen after heuristic, manual searches, that indicated stability and adequately fast convergence for noisy and noiseless state preparation and unitary compilation tasks. Additional heuristic stopping conditions are also implemented to avoid unnecessary iterations.

\begin{table}[h]
\begin{tabular}{|c|l|c|}
	\hline	
	Hyperparameter & Description & Value  \\ \hline
	Optimizer & Optimizer routine & Conjugate gradient \\
	Line Search & Line search routine & Wolf conditions \\
	Conjugate Search & Conjugate search routine & Hestenes-Stiefel \\
	$p_{\tilde{\theta}}$ & Initial parameters distribution & \textrm{Uniform}/\textrm{Cubic Smoothing}\\
	$p_{U,\rho}$ & Objectives distribution & \textrm{Haar}\\
	$p_{\sigma}$ & Initial state distribution & \textrm{Haar}\\
	$S$ & Number of sample objectives & 50 \\
	$L$ & Maximum number of optimization iterations & 500 \\
	$L_{<}$ & Minimum number of optimization iterations & 50 \\
	$L_{\alpha}$ & Maximum number of line search iterations per iteration & 2500 \\
	$\epsilon_{\mathcal{L}}$ & Minimum objective stopping condition & $10^{-16}$ \\
	$\epsilon_{\Delta\mathcal{L}}$ & Minimum absolute difference in objective per iteration stopping condition & $0$ \\
	$\epsilon_{\delta\mathcal{L}}$ & Maximum relative increase in objective per iteration stopping condition & $10^{-3}$ \\
	$\epsilon_{\partial\mathcal{L}}$ & Minimum gradient norm stopping condition & $0$ \\
	$\epsilon_{\Delta\partial\mathcal{L}}$ & Minimum absolute difference in gradient norm per iteration stopping condition & $0$ \\
	$\epsilon_{\delta\partial\mathcal{L}}$ & Maximum relative increase in gradient norm per iteration stopping condition & $\infty$ \\	
	
	$\alpha$ & Initial parameter learning rate & $10^{-4}$ \\
	$\beta$ & Initial search learning rate & $10^{-4}$ \\
	$\alpha_{\lessgtr}$ & Bounds on parameter learning rate before reset to $\alpha$ & $[0,\infty]$ \\
	$\beta_{\lessgtr}$ & Bounds on search learning rate before reset to $\beta$ & $[10^{-10},10^{10}]$ \\
	$c_{1}$ & Wolf objective parameter & $10^{-5} - 10^{-4}$ \\
	$c_{2}$ & Wolf gradient parameter & $10^{-1} - 9 \times 10^{-1}$ \\
	\hline	

\end{tabular}
	\captionsetup{justification=raggedright}	
	\caption{Optimization hyperparameters. All settings are selected from manual parameter searches, and are only heuristically shown to reasonably guarantee optimization convergence across all system settings.}
	\label{tab:hyperparameters}
\end{table}

\end{document}